\documentclass[notitlepage,showpacs,aps,floatfix,prd] {revtex4-1}
\input revstyle.tex
\input epsf
\usepackage{graphicx}
\usepackage{caption}\usepackage{subcaption}
\begin{document} 
\title{Kaluza-Klein theory for type $II \ b$ supergravity on the warped
  deformed conifold}    
\author{M. Chemtob} \email{marc.chemtob@ipht.fr}
\affiliation{Universit\'e Paris-Saclay, CNRS, CEA, Institut de physique
th\'eorique, 91191, Gif-sur-Yvette, France} 
\thanks {\it Supported by 
Direction  de la RechercheFondamentale et aux  Energies  Alternatives
Saclay} \date{\today} 
\begin{abstract} 
We discuss  Kaluza-Klein theory   for   type $II\ b $ supergravity on  
the warped deformed conifold  using  
a  large radial distance  limit of Klebanov-Strassler   solution  where
the radial coordinate  separates  from angle coordinates for a background
asymptotic to   $ AdS _5 \times T^{1,1}$ spacetime.  
The  decomposition of  field  fluctuations  
on   harmonics  of  the base manifold $T^{1,1}$ and plane waves
of $M_4= \dh (AdS _5 )$  is examined for the metric  tensor components along $M_4= \dh (AdS _5 )$, the axio-dilaton and the 4-form potential components along $T^{1,1}$.  Semi-classical
methods  are used to compute  the  mass spectra, wave functions
and interactions for a set of  modes in   low dimensional representations of the isometry  group.  Deformations of the  background  solution     due to compactification effects are also considered.  The information
on warped modes properties   is  utilized to explore the  thermal evolution   of a cosmic component of metastable modes  after  exit from brane inflation. 
\end{abstract}  
\maketitle \renewcommand{\thefootnote}{alph{footnote}}
\tableofcontents\addtocontents{toc}{\protect\setcounter{tocdepth}{1}}

\section{Introduction}

The   discussion  of string theory  flux
compactifications~\cite{dougkach05,granarev05} in the context of    gauge-string duality~\cite{malda99}  has  opened novel
perspectives to    particle physics    models invoking   warped extra  space  dimensions~\cite{rs9905,rs9906}. The attractive  synthesis  of  
proposals to anchor  the models in string theory~\cite{becker96,acharya99,verlinde99,dasgupta99,becker00,greene00,becker01},  realized through the     Giddings, Kachru and Polchinski (GKP)~\cite{gkp01} 
construction of  type $ II \ b$ superstring  theory compactifications,
has led to  progress  on  both 
formal~\cite{dealvisflux03,gidd05,firouz06,dougba07,dougtorr08,shiu08,frey08,douglas09,freyroberts13} and 
phenomenological~\cite{kofman05,chen06,berndsen07,kofman08,freydacline09,chemtob16}  grounds. 

Pursuing  along these lines,  we  examine   in this work
the Kaluza-Klein  theory  for  10-d supergravity theory   reduced on a   warped  deformed conifold throat~\cite{candelas90} glued  to a conic Calabi-Yau orientifold  $X_6$.  To ease computations,  we consider 
replacing  the  spacetime metric in  Klebanov-Strassler  solution~\cite{klebstrass00}   by an approximate factorizable ansatz. 
Instead  of  modifying  the      deformed  conifold $\calc _6$ radial  section
to  copies of a   sub-manifold     of  $S^3\times  S^2$  geometry,    as  proposed in~\cite{firouz06},   we consider  a   large radial distance limit where    the   conifold   radial coordinate    $\tau  $  separates   from the  angular   coordinates $\T _\a ,\ [a=1,\cdots , 5] $ of the  sub-manifold  $T^{1,1}$  leading to  
 a  geometry      asymptotic to  $ AdS _5 \times T^{1,1}$  spacetime.  The  dimensional reduction can then be performed  along standard
lines~\cite{kiroman84,ceresole99,ceresoII99},  easing 
the  comparison with the dual  Klebanov-Witten  gauge theory~\cite{klewit98}.

The   decomposition of  supergravity   multiplet   fields  on harmonic  functions   of    $ T^{1,1}$ yields a field theory
in Minkowski spacetime  $ M_4 = \dh (AdS _5) $  with  towers of  modes   whose   masses   and  radial wave functions  satisfy
a Sturm-Liouville  boundary eigenvalue  problem. 
We   derive  the wave equations   for warped modes   descending  from   the  metric tensor field along $M_4$, the 4-form potential field along $T^{1,1}$ and the axio-dilaton field and compute  their   properties (mass  spectra, wave functions   and local couplings).   We present  results in the  throat domination regime ignoring  the  throat-bulk interface and selecting   normalizable modes along the  conifold radial  direction. The semi-classical WKB (Wentzel-Kramers-Brillouin) method is applied to  obtain  the  wave  functions and mass parameters    for a set  of  singlet and  charged modes under the conifold isometry group with the view to   
identify candidates for the lightest charged Kaluza-Klein particle (LCKP). 
We also   obtain predictions for the
cubic and higher order   self  couplings of bulk  modes
and   for their  couplings to  $D3$-branes  embedded near the   conifold apex. 

The information on warped modes motivates us to
examine  their   cosmological   impact  on  the
universe  reheating in the $(D-\bar D)$-brane inflation
scenario~\cite{kklt03,kklmmt03,baumann06,baumann08}. 
We examine the possibility   that  the  interacting gas   of
 Kaluza-Klein particles   produced  in the inflationary  throat    
 could   reach thermal equilibrium   before   decaying
or could     tunnel  out  to a neighboring      throat  
hosting    Standard Model branes~\cite{barnaby04,frey05,chialva05,kofman05,chen06,kofman08,chialva12}.
We  also   explore in these two cases the possibility  that a 
fraction of  metastable modes    might survive as a
 cold thermal relic  that  could decay at later times~\cite{berndsen07,freydacline09,chen09}.      
 
The contents of this work  are organized  into four sections.  In
Section~\ref{sect1}, we review  the   constuction  of  
GKP flux vacua of type $II \ b $ supergravity on the spacetime $ M_4\times X_6 $   with an attached  deformed conifold throat   described  by  
 Klebanov-Strassler solution. In Section~\ref{sect2}, we examine
the Kaluza-Klein reduction of bosonic field components of the supergravity  multiplet   at  large radial distances inside $\calc _6$
 where the  background warped  asymptotes $ AdS _5
\times T^{1,1}$ spacetime.   The harmonic decomposition of fields
is  applied in Subsections~\ref{sect2sub1} and~\ref{sect2sub2} 
to   derive the wave equations  for  fluctuations of the (unwarped)  metric
tensor $\tilde g _{\mu \nu } $  along $ M_4$, 
the   real scalar from   the 4-form potential $C_{abcd}(X) $ 
along $T^{1,1} $ and the    axio-dilaton $\tau (X) $.  Numerical
results  are presented  in Subsection~\ref{sect3sub2} 
for the  mass spectra   and wave functions  of  a selected set  of warped
modes in low dimensional representations of the conifold isometry
group.  

The central issue in this work   concerning    
warped modes interactions     is discussed in Section~\ref{sect2p}. 
Subsection~\ref{sect2psub1} deals with the  mutual couplings of
graviton modes, Subsection~\ref{sect2psub3}  with 
compactification effects on warped modes couplings   
using the perturbative AdS/CFT duality
approach of~\cite{gandhi11} and Subsection~\ref{sect2psub2}  with 
trilinear  couplings  between graviton  and scalar  modes. 
In Section~\ref{sect4} we consider   a
cosmic population of warped modes produced after brane inflation
and  examine    both its thermal evolution and   ability to  leave  a
cold  thermal relic. 
Subsection~\ref{secT0} discusses  general assumptions, 
Subsection~\ref{subT1}  a single throat scenario  and  
Subsection~\ref{subT2}  a double  throat scenario.
In Section~\ref{sect5} we present main conclusions.
An introductory review of the deformed conifold is  presented   in   
Appendix~\ref{appwdc}.   
Subsection~\ref{appwdcsub1}   discusses   the algebraic
properties, Subsection~\ref{appwdcsubII1}    the harmonic analysis~\cite{pufu10},  Subsection~\ref{appwdcsub2}  the 
approximate separable version of Klebanov-Strassler
metric involving  a  cone over a base manifold of geometry $ S^2
\times S^3 $~\cite{firouz06} and   Subsection~\ref{appwdcsub3}
the  approximate  analytic  formalism  bridging  between the
deformed  and undeformed conifold cases.

\section{Type $II\ b $ supergravity   theory on the conifold} 
\label{sect1}  

\subsection{Warped background  spacetime for 10-d supergravity} 
\label{sect1.sub1} 

Our    discussions   will mostly   concentrate on  the
classical  bosonic action  of 10-d type $II\ b $  supergravity   theory in  Einstein frame, 
\bea && S_{IIb} =    {m _D ^8 \over 2 } \int  d ^{10} X \bigg  [
\sqrt {-g} \bigg ( R  - {\vert \dh _M \tau \vert ^2 \over 2 \tau _2 ^2 }
-  {1\over 2 \tau _2 } G_3 \cdot \bar G _3  - {1 \over 4 }
   \tilde   F_5  \cdot \tilde   F_5   \bigg ) 
 -{i  \over 4\tau _2} C_4 \wedge G_3 \wedge \bar G_3   \bigg ] 
\cr &&  = {2\pi \over \hat l_s^8}   \int  d ^{10} X \bigg  [
\sqrt {-g} \bigg ( R - \ud ( (\dh _M \phi )^2 + e ^{2\phi } (\dh _M
C_0 )^2 ) -\ud ( e ^{-\phi } \vert H_3   \vert ^2 + 
e ^{\phi } \vert  \tilde F_3   \vert ^2  ) -{1\over 4 } \vert \tilde F_5
\vert ^2 \bigg )     -{1\over 4 }   C_4\wedge  \tilde F_3 \wedge H_3 \bigg  ]
 ,\cr &&   
 [F_1= d C_0,\ F_3 = d C_2, \ F_5 = d C_5,  \ \tilde F_3 = F_3 - C_0 \wedge H_3 ,
 \cr && \tilde F_5 = \star _{10}  \tilde F_5  = F_5 - \ud C_2 \wedge H_3 + \ud B_2 \wedge F_3 ,\ \tau  \equiv \tau _1 +i  \tau _2 =  C_0 + i e ^{-\phi } ,\
G_3   = F_3 -\tau H_3 = \tilde F_3 -i e ^{-\phi } H_3
 ]      \label{sect1.eq1}   \eea 
where the action   is   derived from  the string frame action via the  metric tensor   rescaling,   $ g^s _{MN} \to e ^{\phi /2} g_{MN} $. The  gravitational   mass scale $  {2/ m_D   ^8}  =  2 \kappa ^2_{10} =
{(2\pi )^7 } \a ^{'4}   = {\hat l _s ^8  / (2\pi )},\ [\hat l _s   =
2\pi  l _s  ]   $   is  set   by the string  inverse tension  $\a
' = 1/ m_s^2$,   independently  of the   string coupling constant   
$ g_s = <e ^{\phi } > $.  Our  notational conventions and system  of units, 
$\hbar = c =1$, are same as  in our earlier work~\cite{chemtob16} which
    specialized, however,   to the    alternative  Einstein frame derived  from  the string  frame by the replacement   
$ ds ^2 _{10 }  \to  (e ^{\phi } / g_s )^{1/2}  ds ^2 _{10 }$,  changing the gravitational mass scale   $ \kappa ^2_{10} \to \kappa ^2  =  g_s^2 \kappa ^2_{10}.$    

The classical vacua  for  background   spacetimes  $ M_4 \times X_6$  
preserving  $\caln = 1$  supersymmetry involve  conic Calabi-Yau orientifolds $ X_6$   with  3-fluxes $\int _A F_3 = \hat l_s^2 M,\  \int _B H_3 = - \hat l_s^2 K $  across   dual 3-cycles $A,\ B $ 
sourcing    a warped  spacetime  region  near the conifold  singularity  of  $ X_6 $.   
The background   may also include spacetime filling $ O3/O7$-planes and
probe $ D3 / D7$-branes  that carry    effective  $D3$-brane  
charges   $ N_{O3}  ,\ N_{D3} $   satisfying   the $C_4$-tadpole
cancellation  condition,    $ MK +  N_{D3} - {1\over 4} N_{O3} =0$.  
We   consider  the     family of GKP classical  solutions~\cite{gkp01} 
involving an  imaginary self dual   3-form  field strength, 
$G_ - \equiv  (\star _6 -i ) G_3
=0 $, a constant  axio-dilaton,   $\tau (y) = {i /g_s }$,       metric  tensor  and  5-form  field strength of form,  
\bea && ds ^2 _{10} = d s ^2 _4  + ds ^2 _6  = 
e ^{2 A (y) }  d\tilde s ^2 _4 +e ^{-2 A (y) }
d\tilde s ^2 _6  ,\ F_5 = (1 +\star _{10} ) d \a (y) \wedge vol
(M_4) , \cr  && [ h ^{-1} (y) = e ^{4 A (y) }  = \a (y) ,\ 
d  \tilde  s ^2 _4 = \tilde g _{\mu \nu }  d x^\mu dx ^\nu  ,\ 
d  \tilde  s ^2 _6 =  \tilde g _{mn}  d y^m dy ^n ]   \label{sect1.eq2}  \eea 
along with  localized sources    whose     stress energy-momentum
tensor  $ T_{MN}^{loc}$  and  effective $D3$-brane  charge density 
$\rho _3 ^{loc}  $ satisfy the BPS-like inequality  
\bea &&  \hat T  ^{loc}(y) - \tau _3 \rho _3 ^{loc} (y)  \geq
0,\  [\hat T  ^{loc}(y) =  { (T^m _m - T _\mu ^\mu ) ^{loc}  \over
    4  } ,\ \tau _p ={ 2\pi  \over  \hat l _s ^{p+1} }    ] . 
\label{sect1.eq3} \eea  
(In the   dual  gauge theory  description,  the 3-fluxes  $M,\ K$ and the induced   5-flux   $N = MK$  dissolve into  regular and   fractional  $ N\ D3 + M\ D5$-brane stacks.) 
One can relate the  4-d  (Planck) gravitational  mass scale, $ \kappa ^{-1} _{4}  = M_\star  = M_P/ \sqrt  {8\pi } \simeq  2.43 \ 10^{18}  $ GeV, to the supergravity   mass scale, $\kappa ^2_{10} = (2\pi )^7 \a ^{'4} /2 $, 
by matching  the  10-d curvature  action   reduced on $ X_6$  to
the   standard  (Einstein-Hilbert)  4-d  curvature action,
\bea &&  {1\over 2 \kappa
  ^2_{10} } \int d ^{10} X \sqrt {-g_{10}} R ^{(10)}  \simeq {1\over 2 \kappa
  _4^2} \int d ^{4} x \sqrt {-\tilde g_4} g^{\mu \nu }  R ^{(4)} _{\mu \nu } 
\ \Longrightarrow \  \kappa ^2_{10} =\kappa _4^2 V_W \equiv  {V_W \over M_\star ^2}
, \ [V_{W} = \int d^6 y \sqrt {\tilde g} e^{-4 A (y)} ] . \label{sect1.eq7} \eea
The  resulting     relation   between gravitational scales,   involving 
the  6-d internal manifold  warped volume  $V_W = (2\pi L_W)^6$,  with $ L_W$  interpreted as the effective  compactification radius,
can be used  to trade   the string  scale  for  the reduced  Planck  scale, $  m_s= ( {\pi M_\star ^2 / L_W^ 6 }) ^{1/8} $.  Recall  that   warped  compactifications exhibit  a redundancy~\cite{gkp01,gidd05}
under  rescalings of the warp profile and   internal manifold $X_6$ unwarped metric, $e ^{2A} \to \l e ^{2A} ,\  \tilde g_{mn} \to \l \tilde g_{mn} ,\  [\l \in  R  _+]$    leaving  the    10-d   metric  unchanged, 
$ d s ^2  _{10} \to d s ^2  _{10} = \l  e ^{2 A(y) }\tilde g _{\mu \nu }  (x) d
x^\mu dx ^\nu +   e^{-2 A  (y) } \tilde g_{mn}  (y) d y^m dy
^n, $       up to the Weyl rescaling of the  non-compact spacetime   $M_4$  metric,  $g_{\mu \nu } \to \l  g_{\mu \nu } $.
For  a  given internal   manifold  $ X_6$  of  warped metric   $g_{mn } $,   the  classical    background  can then be described   by  a  one-parameter family  of  solutions   with  warp  profiles and    unwarped internal space metric rescaled   by the  parameter   $\l $.
This  establishes     an equivalence   between   descriptions (frames)  differing   in   the  4-d       gravitational   mass
scale  definition~\cite{gidd05,shiu08}.  The   frame  arbitrariness  is neatly 
delineated  by  defining   equivalence  classes of 
10-d  frames,  with respect to reference  (fiducial)  warp  profile  $A_0  (y) $, 6-d manifold of metric $\tilde g^0_{mn} $  and volume  $ V  ^0 = \int  d ^6 y \sqrt {\tilde g^0_6} $  (carrying suffix label $0$),
with parameter  dependent     metric and    4-d gravitational   scale,
\bea && d s ^2  _{10} = \l  e ^{2 A_0 (y) }\tilde g _{\mu \nu }  (x) d
x^\mu dx ^\nu +   e^{-2 A_0  (y) } \tilde g^0_{mn}  (y) d y^m dy
^n   ,\ { 1 \over \kappa _4 ^2 (\l ) } = {\l V_{W}^0 \over \kappa _{10}  ^2 },\ [  V^0_{W} =  \int  d ^6 y \sqrt {\tilde    g^0_6} e   ^{-4 A_0 }  ] . \eea
For the  10-d   Einstein frame  choice,   $\l =   V^0 / V^0 _W $,  one
finds $ { 1/  \kappa _4 ^2  } = {V^0 / \kappa _{10}  ^2 } $ and 
for the    4-d  Einstein frame   choice  $\l =1 $, the  result
  $ { 1 / \kappa _4 ^2 } = {V^0_W / \kappa _{10}  ^2 }$ reproduces the above matching relation,  $V^0_W = V_W $.  (Going from   the  4-d to 10-d Einstein frames   replaces the 4-d metric and   gravitational   scale as,
$g _{\mu \nu } \to (V^0 /  V_W ^0)  g _{\mu \nu } ,\ {1 / \kappa _{4}  ^2 } \to   (V^0 /  V_W ^0)  / \kappa _{4}  ^2  .$)
A  similar   situation holds  if one    incorporates  the  universal  volume modulus  $ c(x)$  through the $x$-dependent warp function $ A(x,y)$  and define   the  family  of 10-d metrics with respect   the fiducial warp  profile  and 6-d metric $ A _0(y)$ and $\tilde g^0_{mn} $ as~\cite{gidd05}  
\bea && ds ^2  _{10} = \l   e^{2 A(x,y)} \tilde g _{\mu \nu }  d
x^\mu dx ^\nu  + e^{-2 A(x,y)}  \tilde g  ^0_{mn}  d y^m dy ^n ,\ 
[e^{-4 A(x,y)} =  c(x) +   e^{-4 A_0(y)} ] . \eea   
The  equivalence under  the  metric  and gravitational   scale 
rescaling   is  then described by
\bea && e^{2 A(x,y)}  \to \l   e^{2 A(x,y)}   
,\ \tilde g^0_{mn} \to \l \tilde g^0_{mn} ,\ { 1\over \kappa _4 ^2 (\l ) } = {\l V_{W}^c  \over \kappa _{10}  ^2 } ,\ [V_{W}^c = \int d ^6 y \sqrt {\tilde g^0_{mn} }   e ^{-4 A  (x, y) } = c(x) V^0 + V_W ^0  ] \eea 
with the   10-d Einstein frame  defined  by $\l = V^0 / V_{W}^c  $
and $  { 1/\kappa _4 ^2 } = V^0 /  \kappa _{10}  ^2$   and the  4-d
  Einstein frame   by $\l =1 $    and  $ { 1/\kappa _4 ^2 } = V^c _W /  \kappa _{10}  ^2  $. (Going from  the 10-d to  4-d  Einstein frames  multiplies 
  $ g _{\mu \nu } $ and ${ 1/\kappa _4 ^2 } $ by $ (V^c_W  /  V^0)  .$) 
We note  incidentally    that  the non-trivial result~\cite{frey08} for  
  the  K\"ahler potential  of  the  chiral superfield $\rho (x) = i  c(x) +  a_0 (x)  $, comprising     the  universal  volume  modulus and   axion   field  from the 4-form potential,
  \bea &&  \kappa _4^2  K (\rho ,\   \bar \rho  ) = - 3 \ln  (2 c (x) +  2 V_W^0 / V^0 ) =-3  \ln  ( -i (\rho -  \bar \rho )  +  2 V_W^0 / V^0 ),   \eea
is reproduced  here by the  intuitive  construct, $ \kappa _4^2  K = - 3  \ln (2 V_W ^c/  V^0) .$
  In the large volume (dilute flux) limit,   $ c >>  V_W^0 / V^0 $,
the  additive  type volume  modulus  $ c(x)$   is related  as  $ c(x) = e ^{4 u (x) }  $   to  the multiplicative  type  modulus  $ u (x) $,   which is introduced via the 
metric and 4-form  fields rescaling,  $  g^0_{m n} \to e ^{2 u(x) } g^0_{m n} ,\ 
g_{\mu   \nu } \to e ^{-6 u(x) } g_{\mu \nu } ,\ \a (y) \to e ^{-12 u(x) } \a  (y)$.
The  (Einstein frame) K\"ahler potential is  then expressed  as $ \kappa ^2 _4 K (u)   = - 2 \ln \calv    ,\  [e^{6u}= \calv =  V_W^0 / \hat l_s ^6  ]  $
in terms of  the   internal manifold    volume in  string  units $\calv $,
related  to the string frame  volume  by $\calv = g_s ^{3/2} \calv_s  $. 

\subsection{Application to  Klebanov-Strassler   background} 
\label{sect1.sub2} 

The  computations are greatly  facilitated     if the  warped throat
region    is modeled by a   deformed
conifold  $\calc _6$ glued  to the   Calabi-Yau manifold.
As     reviewed   in Appendix~\ref{appwdc},  the  deformed
conifold is    a   non-compact K\"ahler manifold~\cite{candelas90}
of isometry group $ SO(4)$,  admitting 
a   single  complex structure   modulus $\e $  and   
a  metric  derived  from an  isotropic    K\"ahler  potential $
\calf (\tau ) $,   function of 
the   radial    coordinate  $\tau \geq 0 $.   The 
fixed-$\tau $ sections   of $\calc _6$    are  copies of the
compact manifold   $ V_{4,2} \simeq  T^{1,1} $.  
The   Klebanov-Strassler solution~\cite{klebstrass00} for type $ II\ b$
   supergravity   on  the    warped 
spacetime  $ M_4\times \calc _6$  is described   by   the non-singular 
metric tensor and   classical  profiles for   3- and 5-form  field
strengths, 
\bea && \bullet \ ds ^2 _{10} = h^{-\ud } (\tau ) d\tilde s ^2 _4 + 
 h^{\ud } (\tau ) d 
\tilde s ^2   (\calc _6) , \ \ [h(\tau ) =2 ^{2/3} (g_s M \a ' )^2 \e ^{-8 /3} I
(\tau )  , \cr && d\tilde s^2_6 (\calc _6) = {\e ^{4/3} K(\tau ) \over 2} \bigg (
{1\over 3 K ^3(\tau ) } (d \tau ^2 + (g^{(5)} )^2 ) + \cosh ^2 ({\tau \over
2})  ((g^{(3)})^2 + (g^{(4)} )^2 ) + \sinh ^2 ({\tau \over 2})
((g^{(1)})^2 + (g^{(2)})^2 ) 
 \bigg ) ,\cr &&  K(\tau )= { (\sinh (2\tau ) - 2 \tau) ^{1/3}  \over
  2^{1\over 3}\sinh 
\tau } ,\ I (\tau ) = \int _\tau ^\infty  dx {x \coth x -1\over \sinh
    ^2 x } (\sinh  (2 x) - 2 x) ^{1\over 3} ] .    \label{sect1.eq4}
\cr &&      \bullet \ F_3= {M \a '  \over 2} 
 \bigg ( (1-F(\tau ))  g^{(5)} \wedge g^{(3)} \wedge g^{(4)} + 
F(\tau )  g^{(5)} \wedge g^{(1)} \wedge g^{(2)}  + 
F '(\tau )  d\tau   \wedge (  g^{(1)} \wedge g^{(3)} +  g^{(2)} \wedge
g^{(4)} ) \bigg  )  ,\cr &&  
\ H_3= {g_s M \a ' \over 2}  \bigg (  d\tau   \wedge  
  ( f'  g^{(1)} \wedge g^{(2)} + k '  g^{(3)} \wedge
g^{(4)} ) + \ud (k -f )  g^{(5)} \wedge (  g^{(1)} \wedge g^{(3)} +
g^{(2)} \wedge g^{(4)} )  \bigg ) . \cr && 
\bullet \  F_5   =\calf _5  +  \star _{10}  \calf _5  ,  
\ \calf _5 = B_2 \wedge F_3 =  {g_s M^2 \a ^{'2} \over 4} l
(\tau ) g^{(1)} \wedge  \cdots  \wedge g^{(5)}, 
\ \star _{10}  \calf _5 = {4 g_s ^2 M^2 \e ^{-8/3} l (\tau ) \over K^2 h^2(\tau )
  \sinh ^2 \tau } \e  (M_4) \wedge d\tau     , \cr &&  
[F(\tau ) = {\sinh \tau  - \tau  \over 2 \sinh \tau  } ,\   
{f(\tau ) \choose  k(\tau ) }  =  
{ \tau  \coth \tau  - 1 \over 2 \sinh \tau } ( \cosh \tau  \mp 1 )  
,   \  {f'(\tau ) \choose  k '(\tau ) }  =   
{  (1- F(\tau ) ) \tanh ^2 (\tau /2 ) \choose  F(\tau )  / \tanh ^2 (\tau /2 ) } ,  \cr &&    l (\tau )= f (\tau )(1-F(\tau )) + k(\tau ) F (\tau  ) = 
{ \tau  \coth \tau  - 1 \over (2 \sinh \tau  )^2 } 
(\sinh  2 \tau  -  2 \tau )] .      \label{app2.eq1}   \eea  
(The  string frame  metric is $ g_s^{1/2}  d s^2
_{10} $.)    The  conifold deformation  parameter $\e $  is defined in
Eq.~(\ref{app1.eq0}) and   the basis of    left-invariant 1-forms
$ g ^{(a)} ,\ [a=1,\cdots   5] $  of fixed-$\tau $ sections
isomorphic to  $   T^{1,1}\sim S_1^3 \times S_2^3 / U(1) $~\cite{minasian99} is   parameterized   by    the    5   angle coordinates   $\T ^a ,\ [a=1,\cdots ,  5]$ of  $  T^{1,1} $ built  from  a   pair of  Euler  angles $[\t _i\in (0, \pi ),\ \phi _i \in (0, \pi ) ,\ \psi  \in (0, 4\pi ),\  i=1,2]$  as in Eq.~(\ref{eqKSmet1}).  
The  auxiliary functions $K(\tau ),\ I (\tau ) $ limits  at  $\tau
\to \infty ,\  \e \to 0$,  with fixed conic     radial  coordinate $ r^3
\sim \e ^2 e ^\tau   $, yield expressions    for the   warp  profile   and  classical unwarped metric,
\bea && K(\tau )  \simeq  2^{1/3} e
^{-\tau /3} ,\ I(\tau ) \simeq  2^{-1/3}  3 (\tau - {1\over 4})   
e ^{-4 \tau /3} ,   \   h^{1/2} (\tau ) \simeq   
2^{1/6}  3 ^{1/2} \e ^{-4/3}   g_s M \a ' 
(\tau - {1\over 4})^{1/2}  e ^{-2 \tau /3} ,  
\cr &&    
 d\tilde s ^2 _6 \simeq {\e ^{4/3}  \over 6 K^2 (\tau ) } d \tau  ^2 
+  2^{-5/3} \ 3 \ \e ^{4/3} e ^{2\tau  /3} d \tilde s ^2 (T^{1,1}) \simeq 
dr ^2 + r^2  d \tilde s ^2 (T^{1,1}) ,  \ \ \
[r = {3^{1/2} \over   2^{5/6} } \e ^{2/3} e ^{\tau /3} ,\ 
\cr &&   d \tilde s ^2 (T^{1,1}) =
  ({1\over 9 } g ^{(5) 2}  + {1 \over 6 } \sum _{a=1} ^4   g   ^{(a) 2}  ) 
  = {1\over 6} (d \t _1 ^2 + \sin ^2 \t _1 d \phi _1 ^2 +
  d \t _2 ^2 + \sin ^2 \t _2 d \phi _2 ^2 + {2\over 3} (d\psi +
  \cos \t _1 d \phi _1  + \cos \t _2 d \phi _2 )^2 ) ,   \label{sect1.eq6}  \eea 
  which   coincide    with  Klebanov-Tseytlin solution~\cite{klebse00}  for the  warped undeformed (singular)   conifold.     A similar   conclusion holds   for the  3- and 5-form   field strength limits. 
  The warp profile  exhibits    the  familiar      
power  law  behaviour  scaled by   the constant  curvature
radius  parameter $\calr $    common to the   $AdS_5$  and $ T^{1,1}$ submanifolds,  
\bea &&  h (\tau )  \simeq  ({\calr \over r })^4 (1 + c_2 ({1\over 4} +
\ln ({r \over   r_{uv} } ) )   =   {L_{eff}  ^4\over r^4}  \ln {r \over     r_{ir}} 
,\cr &&    
[c_2 =   {3 g_s M\over 2\pi K } ,\ \calr = ({27 \pi \over 4}  g_s N
  )^{1/4} l_s ,\ L_{eff}  ^4 =  c_2 \calr ^4 = {81\over  8}  (g_s M\a ')^2 ] 
\label{eq.wprof}   \eea  
where the logarithm  factor arises   from the  prefactor in $ I(\tau )  \propto  (\tau -1/4) \sim 3 \ln (r /
r_{ir} ) ,\  [ r_{ir} =  e ^{1/12} 2 ^{5/6} 3 ^{1/2} \e ^{2/3}] $ and the  ultraviolet and infrared  radius parameters   $ r_{uv} ,\ r_{ir}$
were introduced with hindsight  from holography. 
Recall that the   AdS/CFT gauge theory dual to the supergravity  model 
living  on   the  $ N\ D3 + M D5$-branes at the conifold apex
in the gravity decoupling limit $ N>> 1,\ K = N/M >> 1 $ is the
Klebanov-Witten gauge theory~\cite{klewit98}
of local  symmetry group  $\calg =  SU(N+M) \times SU(N) ,\ [N= MK]$ 
and     global symmetry group  $ G= SU(2) \times SU(2)\times U(1)_b \times U(1)_r $, with  two  pairs of  bifundamental  chiral superfields $ A_i,\ B_i $    coupled through  the
 quartic order superpotential  $ Tr (A_1B_1A_2B_2 - A_1B_2A_2B_1).$  The  renormalization group     flow     down the throat occurs through  a cascade     of   self-similar   Seiberg  dualities   typically ending   in  the
confining   gauge   theory   $ SU(2M)\times  SU(M)  $   with  
a  spontaneously  broken chiral  symmetry   $U(1)_r \to Z_2$. 
The radius parameters are related to the gauge theory  ultraviolet cutoff,     confinement  and gluino condensate mass scales   $\L _{uv},\ \L  _{ir},\  <\l \l > =  M  S  $  by the   formulas
$ r_{uv}  \simeq   \L  _{uv}  \a ' ,\ r _{ir}  \simeq  \a ' \L _{ir} \simeq  
\a ' S^{1/3}   \simeq \e ^{2/3}$.
The string-gauge  theory  duality    is  tested
through  the   following  order of magnitude relations    expressing 
the  gauge  theory confinement,  Kaluza-Klein   glueball and baryon,
$F1$-string and    domain wall  mass  or  tension  scales  in  terms of
the  supergravity parameters,
\bea && \L _{conf} \sim   \e  ^{1/3}  / ( g_s     M   \a  ^{'3})^{1/4}
,\ m_{K}  \sim {\e  ^{2/3}  / ( g_s     M   \a  ' ) } ,\
m_{b }  \sim M \e  ^{2/3} / \a  ' ,\ T_s \sim {\e ^{4/3} /(  g_s M \a     ^{'2} ) }  ,\ T_{dw}  \sim {\e ^{2} / (g_s \a  ^{'3} ) } . \eea   
 
The  limiting   formula for  the metric   at $ \tau \to 0 $, 
\bea && ds ^2 _{10} \simeq { \e ^{4/3} \over  \calc } 
d \tilde s ^2 _{4}  + { \calc \over  2^{2/3} \ 3^{1/3} } 
   ( \ud d \tau ^2 + {\tau ^2 \over 4}    d \O ^2 (S^2) + d \O ^2
(S^3) ),\cr &&    
[\calc =   2^{1/3} a_0 ^\ud  g_s M \a ' ,\ a_0 =  I(\tau =0 )  \simeq
  0.71805,\ K(\tau )    \simeq   (2/3)^{1/3}  ,\ 
h ^\ud (\tau ) \simeq   2^{1/3} \e ^{-4/3} a_0^\ud
  g_s M \a '  e ^{-2\tau /3 }  , \cr && d \O  ^2 (S^2)
  =( (g^{ (1) } )^2 + (g^{(2)} )^2 ) ,\  d \O  ^2 (S^3)   =  ( \ud (g^{(5) }
 )^2    + (g^{(3)} )^2 + (g^{(4)} )^2 ) ]   \label{sect1.eq5} \eea 
shows that  the  conifold   geometry      near the  apex  reduces to a real cone over a  base $S^3 $ of  constant  (unwarped)   square radius, 
$ \tilde r^2 (S^3) =  \ud ( {2\over 3 } )^{1/3} \e ^{4/3}  $,  times a collapsing 
$ S^2 $ fibre  (of     square  radius,  
$  \tilde r^2 (S^2) =  \tilde r^2(S^3)  \tau ^2 / 4
$)~\cite{herzog01}. 
(The   radii  referred to  the warped  metric  are 
$ r^2 (S^3) = \calc / 12 ^{1/3} = a_0^{1/2} g_s M \a' /6^{1/3} .$)
 The  term $ d\O ^2 ( S^3)$  above  gives the round metric  of 
the $ S^3$   manifold for the $SU(2)$ group element $
T= g_1 \s _1 g _2 ^\dagger \s _1 $,    in  the
notations of Eq.~(\ref{app1.eqKSp})~\cite{minasian99}. 
Changing   the radial variable from  $\tau \to r= 2 ^{-5/6} 3 ^{1/2}  \e ^{2/3} e^{\tau /3}$      gives  the       power law  for  the  warp profile,     
\bea && e ^{-2A} = 2 ^{-4/3} 3  a_0 ^{1/2} ({\calr  ' \over r })^2 ,\ 
[\calr '  \equiv  ({32\over 81}) ^{1/4}  \calr _- = (g_s M  \a ' ) ^{1/2}   
=  \calr ({27  \pi K \over  4  g_s M  })^{-1/4} = \calr    (-{81\over 8}  \ln
{\e ^{2/3} \over   l_s } ) ^{-1/4}   ]   \label{eq.radiusp}   \eea    
where  $ \calr ' \simeq \calr _-  $ characterizes   the curvature  radius
in the strongly   curved region  near  $ r_{ir}$~\cite{dougtorr08}.


The necessary patching  of  the throat  to the compactification  manifold
unavoidably introduces   additional     parameters~\cite{firouz06,dougba07}.
An  upper   cutoff  on the  radial   distances is clearly required 
to  mark  the  point where   the throat   merges into the bulk
region inside which the  solution  for the warp profile   makes no sense.
If one identifies the cutoff location $\tau _{uv}$,  consistently with holography,  as the  point   where the warp profile   reaches unity,  the
assigned  value is  related to the string and throat   parameters as, 
\bea && h^{-1} (\tau _{uv} ) \simeq 1 \ \Longrightarrow \ 
e ^{2 \tau _{uv} /3}(\tau _{uv} - {1 \over 4}) ^{-1/2} 
\approx 2 ^{1/6} 3 ^{1/2} \e ^{-4/3} g_s M \a '  \ \Longrightarrow
\ \tau _{uv} \simeq  3 \ln ( {(g_s M )^{1/2} l_s \over \e ^{2/3} } ) =
3 \ln ({\calr ' \over  \e ^{2/3} } ) .    \label{eq.ktoffuv}   \eea
Combining the    corresponding  cutoff value    for the  conic   radial variable,    
$r _{uv} = 3 ^{1/2} 2 ^{-5/6} \a ' \vert S \vert ^{1/3}  e^{\tau _{uv} /3 }   
\simeq     (3/2)^{3/4} (g_s M )^{1/2}l_s \simeq (g_s M / (2\pi  K) ) ^{1/4} \calr $,     with  the  relation, $ r_{uv} =\a ' \L_{uv} $,
links the   supergravity  and gauge theory   ultraviolet cutoffs
via the  (shape) complex structure modulus $S$ as,  
 $\tau _{uv}= \ln (2^{5/2} 3 ^{-3/2} \L _{uv} ^3 / \vert S \vert ) $.

Another important auxiliary    parameter  is  the   mass hierarchy in the warped throat, defined in the notations of Eq.~(\ref{eq.wprof})  by 
the ratio  $ w_s = m_{eff} / m_s = r _{ir}/ r_{uv}  = \min  (e ^{-A}) =  e ^{-1/4 - 1/c_2}  $.  Since the  solution for  the  6-d   warped metric is  independent of $\e $, it is natural   to identify  the  throat shape  parameter     to
the bulk    manifold   shape     modulus, $ \e  ^{2/3}  \simeq  \a ' S ^{1/3}   $.  This is consistent  with the   fact that  the  dependence on     $\e $
cancels out    in  the internal  part  of  the
warped metric   solution   ($ d s^2 _6 $). 
Using then the effective  field theory description  for the modulus $S$ 
in the special-K\"ahler geometry  limit allows relating the warp factor to the  string and flux parameters~\cite{gkp01},  
$ \ln w _s \simeq  \ln ( \e ^{2/3} / l_s )  =-2\pi K /  (3 g_sM)   $.
Although  the deformed  conifold  has no K\"ahler   modulus   in proper,
the internal  manifold  should typically     admit a universal volume modulus
which is introduced    through the  already  mentioned  (large volume)     metric ansatz,   $ d s ^2 _{10}  = e ^{-6u } h^{-1/2} (\tau ) d \tilde  s ^2 _4  +     e ^{2u} h^{1/2} (\tau )  d \tilde s ^2 _6 $. 
Identifying  the warp factor   $  w_s $  to the  ratio of the  warp  profile  
along $ M_4$ at  the  throat  horizon   and   boundary
$ h^{-1/4} (\tau _{ir} ) /  h^{-1/4} (\tau _{uv} ) $, regardless of the metric along $X_6$,   would  give a result   independent of the volume  modulus $u (x)$. Since  the region  near the horizon  is strongly  warped,
it is  more satisfactory to   modify the 10-d  metric  near $\tau =0$
so that  the    internal manifold part   becomes   independent   of   $u (x)$,
while keeping the metric at the boundary unchanged.  
Replacing  the war profile near the apex as, 
$ h^{\pm 1 /2} (\tau ) \to  h^{\pm 1/2} (\tau )  e ^{\mp 2u  (x) }  $,
yields a  10-d metric $ ds ^{2}  _{10} \simeq  e ^{-4 u} h ^{-1/2} (\tau ) d \tilde s ^2 _4 + h ^{1/2} (\tau ) d \tilde s ^2 _{6} $ with the 6-d  warped metric   $ d s^2 _6 $   independent   of  $ u $. (The   Riemann curvature tensor components  near the apex~\cite{douglasroba08}  become  likewise  independent   of both $ u $ and $\e $, as illustrated  by   the scalar curvature,  $ R  ^{(10)}   \propto   1 /   (g_s   M \a ' ) $.) 
Retaining  the initial   form  of the  metric solution elsewhere
and defining    the warp factor  by the ratio
$ w_s = e ^{-2  u} h^{-1/4} (\tau _{ir})  /  ( e ^{-3 u} h^{-1/4}  (\tau _{uv}   ) )  \simeq  e ^ {u }  h^{-1/4}  (0 )  $   with the warp profile  value at the
boundary  set at $h^{-1/4}  (\tau _{uv}   ) = 1$,   as in Eq.~(\ref{eq.ktoffuv}),  yields the   formula  for   the warp factor,   previously  proposed in~\cite{browndW09}, depending   on both    string and compactification  parameters
\bea  w_s = {e^u \e ^{2/3} m_s \over 2 ^{1/6} I ^{1/4} (0) (g_s M ) ^{1/2} }
\simeq  {L_W \e ^{2/3} \over l_s ^2 (g_s M)^{1/2}  } , \ [
  e ^{4u} = \rho _2 = \Im (\rho ) = 2 \bar \rho  
  \simeq \calv ^{2/3} ]   \label{sect3.eq5} \eea
where we assumed  the identification  $\calv = V_W / \hat l_s ^6 = (L_W / l_s ) ^6$.    The volume dependence in the above relationship between $\e $ and $ w_s$,   $\e = 2^{1/4} a_0^{3/8} (g_s M \a ' )^{3/4} w_s ^{3/2}  \calv ^{-1/4}  $ 
should modify  the familiar  parametric    relations  for the string and Kaluza-Klein masses,  ${m_s / M_\star }  \sim  \calv ^{-1/2}, \  {m_{K} / M_\star }  \sim \calv ^{-2/3}  $.  If one trades   $\e $     for the 
warp  profile minimum value,   $ e^{A_0} \sim \e ^{2/3} /( l_s(g_s M)^{1/2} )$, then    $ w_s \sim e ^{A_0} \calv ^{1/6} $ and   the effective  string  and Kaluza-Klein  mass scales   $w_s m _s $ and  $m_{K} \simeq \e ^{2/3} / (g_s M \a ') $  can be expressed in units of Planck mass  by the  parametric    relations,  ${ w_s m _s}/ M_\star  \sim \calv ^{-1/3} e ^{A_0} ,  \   {m_{K}/ M_\star }  \sim  \calv ^{-1/2} e ^{A_0} / (g_s M)^{1/2}= \calv ^{-1/2} e ^{A_0} {l_s / \calr '} .$ 

It is useful to examine   how  the ultraviolet cutoff 
impacts   the  throat       properties. 
Consider first the  contribution  to the warped throat volume, 
\bea && V_{T} = \int d ^6 y  \sqrt {\tilde g_6 } h(\tau ) 
=  {9  V_{X_5} \over 2 ^{7/3} }  \e ^{4/3}    (g_s M \a ' )^2 J
(\tau _{uv} ) \approx 1.2\ 10^{2}
w_s^2 {(g_s M \a ')^3   ({ \bar \rho  })^{-1/2}  } e ^{0.718 \tau _{uv} },
 \cr &&  [J (\tau _{uv} ) =\int _0 ^{\tau _{uv} } d
   \tau \sinh ^2 \tau   I(\tau ),\
V_{X_5} =  \int _{T^{1,1} } d^5 \T \sqrt {\tilde \g   _{5} } ={1\over 108}
\int \prod _{a=1}^5  g^{(a)}   = {16\pi ^3 \over    27}]  \label{eq.vol5} \eea
where the  above approximate formula for $ V_{T} $  was deduced from 
a  rough  fit of the   integral $J (\tau _{uv} ) $  for $ \tau _{uv}\in (10, 40) $.  We  see that the        exponential  growth of   $ V_{T} $ 
 with  $\tau _{uv}$   is   mitigated  by  the   warp and overall volume suppression factors,  $ w^2 / \calv ^{1/3} $. 
 For comparison, we recall that the   volume   modulus    is set at  $ \calv ^{1/6} = V_W ^{1/6}  / (2\pi l_s)  \simeq 5 $
 in the    brane  inflation  scenario~\cite{kklt03}
 and ranges from   $ \calv \sim  5  $  
 to  $ \calv \sim  ( 10^5 - 10^7)$
in  the  studies~\cite{choi05} and~\cite{bala05,blumeLVS07,blumeLVS09}   of supergravity  mediation    of soft  supersymmetry breaking.
For a  rough  orientation on  the relative 
contributions from the bulk and throat regions, we  tentatively
consider   splitting up the   internal manifold    volume  into bulk and throat parts $ V_W= V_{B} + V_{T}  $      over  the    respective    intervals  
$\tau \in [ \tau _{uv} , \tau _{max}  ] $ and $\tau \in [0, \tau
  _{uv}] $. The  bulk  contribution to the volume
is   estimated  by the  integral 
with a constant warp  profile $ h (r ) \to  h (r _{uv}) $,
\bea && V_{B} \simeq h (r _{uv})  \int ^{r_{max} }_{r_{uv} }
dr r^5   \int d ^5 \T \sqrt {\tilde \g _5}  = { V_{X_5}  \over 6} (
{\calr  \over  r_{uv} } )^4 (r _{max}^6 - r _{uv} ^6 ).     \eea 
Comparison of the throat  volume in Eq.~(\ref{eq.vol5})  with  the    bulk volume,  using  $r _{max} \simeq L _W= ( V_{bulk}  / 2\pi ) ^{1/6} $,     shows  that one   can   satisfy  the   inequality $ V_{B} >   V_{T}$
provided   $ w_s  < 10^{-3} \calv ^{2/3} K ^{1/2} / (g_s M)^2 $.
It is also instructive  to  consider  the  throat   proper radial length,    
\bea && l_W = \int _0 ^{\tau _{uv} } d \tau
\sqrt {g _{\tau \tau } } = 2 ^{-1/3} 3 ^{-1/2} (g_s M \a ' )^{1/2}
\int _0 ^{\tau _{uv} } d \tau  { I^{1/4} (\tau ) / K (\tau ) }
\approx  0.37 \sqrt{g_s M \a '} \tau _{uv} ^ {5/4} , 
\label{sect1.eq8}  \eea
where the   result  in the   last  step was obtained
using a rough   fit of  the integral for  $  \tau _{uv} <  10$. 
The intuitive expectation that stronger warping    (smaller  $ w_s$)
is consonant with    longer  throats (larger  $  l_W $) 
is indeed verified  if one recalls the relation for  the cutoff  parameter  from Eq.~(\ref{eq.ktoffuv}), $\tau _{uv}   \simeq 
3  \ln ( \calr ' \e ^{-2/3} )  \propto -\ln (w_s)$. One could also consider   the proper length  referred to  the unwarped  metric, 
$ \tilde l_W   \simeq 3 ^{1/2} 2 ^{-5/6} \e ^{2/3} ( e
^{\tau _{uv}  /3 } - 1)  \sim w_s e^{\tau _{uv}  /3 } l_s   $.
We conclude  this discussion  with  the  following table displaying the three radial  regions
from the  conifold apex  $ r_{ir}$ to the bulk  manifold in which
warping evolves  from    strong to intermediate   to weak regimes.
\vskip 0.2 cm 
 \begin{center}  \begin{tabular}{|c|c|c|c|}  \hline 
Warping  &  Strong $(e ^{-4A} >> c)  $   &  Intermediate     $(e ^{-4A}
\sim c  )$  &  Weak $(e ^{-4A} << c )$  \\ \hline 
Radial intervals   &   $ r_{ir} \to  \calr ' \sim (g_sM )^{1/2} l_s    $&
$ \calr  \sim  (g_sN )^{1/4} l_s  \sim \calr ' (K/g_sM)^{1/4} l_s$ 
& $l_W  \sim (g_sM )^{1/2} \tau _{uv}  ^{5/4} l_s  $  \\  \hline 
\end{tabular}\end{center} \vskip 0.2 cm 

\section{Type $II\ b $ supergravity action reduction on  the  conifold}
\label{sect2} 
The non-separability  of  the   deformed  conifold  radial and angular coordinates    has so far hampered  the progress in applying Kaluza-Klein theory  to supergravity  in Klebanov-Strassler background.
The harmonic  analysis  remains an arduous task in spite of  the existence of    analytic~\cite{krishtein08} and     group theory~\cite{pufu10}  methods 
motivated by the mathematical literature~\cite{levine69,gelbart74}.
For illustration,  we  remark  that  the harmonic   decomposition  of  10-d   fields~\cite{pufu10},
\bea &&    \phi  _q ( x, y ) =  \sum _{m }   \phi _{q} ^{(m)} (x)   
\Psi _{q }^{m}    (\tau, \T ) ,\ [m = (jlr),\ \Psi _{q} ^{m}  (\tau, \T )
  = \sum _{r} f _{q,r}  (\tau  ) Y^{j l r  (M) } _q   (\T )]  \label{app2.eqHARM}    \eea
introduces     field  modes $\phi _q ^{(m)} (x)  $ in $M_4$
with square normalizable  wave functions  $\Psi ^{j l r} _{q}  (\tau, \T )$  given by linear  combinations  of   harmonic  functions  $  Y^{j l r , M } _q   (\T )  $  of the  undeformed   
conifold  base   manifold $ T^{1,1}\sim   SU(2) _L  \times SU(2) _R / U(1)_H$ with   coefficient  functions $f _{r,q}  (\tau  )$    obeying  coupled  linear   differential equations of second order~\cite{pufu10}. 
The    suffix   $ q$  labels the fields tensor  type, 
$j,\ l $  and  $  M = (m_j, \  m_l ) $  label   the conserved     angular  momenta and magnetic  quantum numbers   of  the   isometry  group
$ G=  SU(2)_L \times  SU(2) _R \times Z^r_{2} $  irreducible representations,
and the charge  $r\in (-\tilde j , -\tilde j +1,  \cdots , \tilde j ) ,\ [\tilde j = \min (j,l)]$  for  $ U(1)_r  \supset   Z_2 ^r $
labels the $ 2 \tilde j +1 $   representations
$  (j, l)$   part of the  harmonic    basis.
For scalar  fields, the charges $r  $ and $q$    are related  to the magnetic  quantum numbers  $(m_j, \  m_l )$   as $ r = (m_j-m_l )/2 , \ q=
(m_j+m_l)/2 $.
More  details   on this   construction are    provided  in Appendix~\ref{appwdcsubII1}.

We  shall  make use in this work     of  an
approximate  version  of Klebanov-Strassler metric  in which 
the   5-d  compact   base  metric  is replaced 
by  a  large  $\tau  $   limit setting
$ \cosh ^2 {\tau \over 2} \simeq \sinh ^2 {\tau \over 2} \simeq  e ^{\tau } / 4
\simeq 1/ (2 K^3 (\tau ) ) $.  In the
internal space part of the 10-d metric, $ds ^2 _{10} =   h^{-1/2} (\tau ) d\tilde s ^2 _4 +   h^{+1/2} (\tau ) d\tilde s ^2 _6,$  the
 radial and angular variables   then separate     as,     
\bea && d \tilde s ^2 _6 \simeq {\e ^{4/3} \over 6 K^2
  (\tau ) } d\tau ^2 + {3  \e ^{4/3} \over 2  K^2 (\tau ) }  ds ^ 2 ({ T^{1,1}}) 
= d r^2 + r^2  ds ^ 2 ({ T^{1,1}})  , \
r = \sqrt {3\over 2}  {\e ^{2/3} \over   K (\tau ) } 
\simeq  3^{1/2} 2^{-5/6} \e ^{2/3} e ^{\tau /3} ] \label{sect2.apmet}  \eea
 and the     geometry      
 is asymptotic  to the spacetime  $ AdS_5 \times T^{1,1} $.
The Kaluza-Klein ansatz for scalar  fields, 
 \bea &&  \d \phi (X) =  \sum _{\nu }  \phi ^{(\nu )} (x, \tau )  Y^{\nu   } ( \T )  ,  \ [x ^{\mu } \in     M_4, \  \tau  \in  R _+ ,\ \nu = (j,l, r) ]    \label{eq.approx1}    \eea 
introduces    field   modes   $ \phi ^{(\nu )} (x, \tau ) $   in $ AdS_5$    
in   unitary irreducible representations of   the  superconformal  group   $ SU(2,2|1)$  and  the conifold     isometry group
$G=  SO(4) \sim SU(2)\times   SU(2)  \subset SO(5) $ of wave functions 
$ Y^{\nu   } ( \T ) $  along  angle  directions  of the radial sections
$T^{1,1} \subset \calc _6 $. 
Further       decomposition  on   plane waves of   four-momentum     $  k_m $  in $ M_4 = \dh (AdS_5 ) $,  
\bea && \phi ^{(\nu )} (x, \tau )  = \sum _{m}  \phi ^{(m)}  
(x )  R_m (\tau )  ,\ [\phi ^{(m)}  (x ) = 
\phi ^{(m)} _{k_m}  e ^{i k _m \cdot x } ]    \eea  
introduces   4-d  mode  fields  whose labels    $ m = [ \nu = (j, l
  , r) ,\  n ]$   include the integer  radial  quantum number $n$. 
The  radial  wave functions  $ R_m (\tau ) $  belonging to
 the vector space  of  normalizable   solutions  of 
a  Sturm-Liouville   type    equation can be organized
into   orthonormal  bases   labelled   by    5-d and 4-d  masses set as, 
$ \tilde \nabla ^2  _{T^{1,1}} \to - M_5^2,\  \tilde \nabla _4 ^2 (M_4)
\to  - k^2 =  E_m ^2 $.  The discussion for other   components 
of the  supergravity  multiplet is similar   to that  developed
in~\cite{kiroman84}, modulo  modifications    discussed
in~\cite{ceresole99,ceresoII99} and  partly summarized in~\cite{chemtob16}.  
In the next  subsections   we  derive     the wave
equations   for  the  metric tensor components along $ M_4$  
and   the  scalar    modes in $ M_4$ descending  from  
the  axio-dilaton field and the  4-form potential   $C_4$ along $
T^{1,1}$.     For completeness, we also   briefly  consider  in
Appendix~\ref{appwdcsub2} the  modified  geometry 
near the apex region   of~\cite{firouz06}   
corresponding   to   a  real  cone over  an   $ S^2 \times S^3 $ base.  


\subsection{Graviton  modes  from  metric tensor field  reduction} 
\label{sect2sub1}

The  metric  tensor field  fluctuations $\d  g_{MN } $ in
the  10-d   gravitational (curvature)  action are governed  by  the     linearized    wave equation, 
\bea &&  \d  (G_{MN }  -\kappa ^2 _{10}  T_{MN } ) =0 ,\  
[G_{MN }  = R_{MN } - \ud g_{MN } R ,\ T_{MN} = -
{2  \over \sqrt {-g} } {\dh L_m \over \dh g_{MN }} ] \eea  
linking variations of  the  Einstein tensor   $  G_{MN } $  
to those   of the stress  energy-momentum  tensor   $ T_{MN}  $,
representing contributions  from  other fields   and $Op$-plane   or  $Dp$-brane  sources  described  by  the  matter  Lagrangian  $L_m $. 
We shall restrict  consideration to the metric  tensor components along $ M_4$,  $\d g_{\mu \nu } (X) = e ^{2 A(y) }  h _{\mu \nu } (X)  $, and specialize to the transverse-traceless gauge.
The       field equation for Ricci tensor  variation takes then the
form,
\bea && \kappa  _{10} ^2 \d (T_{\mu \nu } -{1\over 8} g_{\mu \nu }  g^{MN} T_{MN} )
=  \d R _{\mu \nu } \equiv - \ud (e ^{2A} \nabla ^2 _4  + \nabla _{6} ^2 ( e ^{2A} h _{\mu \nu } ) + e ^{-2 A} (\nabla _m e ^{2A}) (\nabla ^m e ^{2A} ) h  _{\mu \nu }) ,  \label{sub2eq13p} \eea 
where we have  set $ \d  \tilde R ^{(4)}  _{\mu \nu }= - \ud e ^{2A}\nabla ^2 _4 h
_{\mu \nu } $.   We   choose to treat  all  other fields   and
sources as  non-dynamical degrees of freedom
and  also  impose the important  condition~\cite{firouz06}   $ \dh L_{m} /\dh  g^{\mu   \nu } =0 $  which   relates  the stress  energy-momentum   and metric tensors  as,  $T_{\mu \nu } \equiv    g_{\mu \nu } L_m - 2 (\dh  L_m  /\dh g_{\mu \nu } )\to  g_{\mu \nu } L_m $,
and implies  in turn  the proportionality relation between  Ricci
and  metric tensors,    $R_{\mu \nu }
=  - {\kappa  _{10} ^2 \over 4} (L_m   - g^{mn} {\dh L_m \over \dh g^{mn} })  g_{\mu \nu }  $.  Combining Eq.~(\ref{sub2eq13p}) with 
the  equation   for  the  Ricci tensor variation   deduced from  this   constraint,
\bea &&  \d  R_{\mu \nu } =   
({1\over 4}  R ^{(4)}  -\ud (\nabla _{6} ^2 (e ^{2A}) +
  e ^{-2A} (\nabla _m e ^{2A}) (\nabla ^m e ^{2A}) )  ) h  _{\mu \nu } , \label{sub2eq13px}   \eea
  simplifies the    linearized wave  equation
for the metric tensor $ h _{\mu \nu }$   to a  form where the wave  operator reduces   to the   scalar   Laplacian~\cite{firouz06}, 
  \bea && 0= \nabla _{10} ^2  h_{\mu \nu } (X)=
 (e ^{- 2A (\tau ) }\tilde  \nabla _{4} ^2    + e ^{2A (\tau ) } \tilde \nabla _{6} ^2 )  h_{\mu \nu } (X) . \label{sect2.eqLap} \eea
  The generalized     equation   for spin $2$
  string  like   excitations of  squared  mass  parameter $\mu ^2$  is given    by
 $0= ( \nabla _{10} ^2 -\mu ^2 ) h_{\mu \nu } (X) .$     
Substituting   the  decomposition  on 4-d  graviton   mode  fields,
$h _{\mu \nu } (X) = \sum _{m }  h ^{(m)} _{\mu \nu } (x )   \Psi  _m
(\tau, \T ) $,    gives the wave equations for the Hilbert vector space of wave   functions  in     $ \calc _6 $, 
equipped with   an  Hermitian  scalar  product,
\bea &&    (e ^{- 2A (\tau ) }\tilde  \nabla _{4} ^2    + e ^{2A (\tau
  ) } \tilde \nabla _{6} ^2  - \mu ^2 )\Psi  _m (\tau, \T ) = 0   
,\  [\int d ^6 y \sqrt {\tilde g_6}h (y) 
  \Psi ^\dagger _m (\tau ,\T ) \Psi _{m'}  (\tau ,\T )= \d _{mm'} ] \label{eq.norsq}    \eea
where we  have  chosen  a    normalization  condition consistent   with the matching  relation $ {M_\star   ^2 /(m_D ^8V_W) }=1 $    in Eq.~(\ref{sect1.eq7}). 
For     a  6-d unwarped metric (distinguished   from the warped
metric  by the  tilde  symbol) of  general    form, 
\bea && d \tilde s ^2  (\calc _6) = \tilde  g_{\tau \tau } (\tau )
d \tau ^2 + d \tilde s ^2 _5  ,\   [d \tilde s ^2 _5 =  \tilde g _{ab}
  (\tau , \T )d \T ^a d \T ^b   ,\  \tilde  g _{ab} (\tau , \T )  =
  K_{ab} (\tau )\tilde  \g _{ab} (\T ) ,\  \tilde g ^{ab} (\tau , \T )
  =  K^{ab} (\tau )  \tilde \g ^{ab} (\T )] \label{eq.kkbkmet}  \eea  
involving   the  symmetric matrices 
$ K _{ab}  (\tau ) ,\  \tilde \g _{ab}   ( \T ) $,
of  inverses $ K ^{ab}  (\tau ) ,\  \tilde \g ^{ab}   ( \T )
$,  the 6-d Laplacian in Eq.~(\ref{sect2.eqLap})  splits   into
radial and     (mixed) radial-angular   parts     satisfying the  identities, 
\bea &&  \nabla _{6} ^2=  {1\over \sqrt {g_{10} } } \dh _m g^{mn} 
\sqrt {g_{10} } \dh _n = {  1 \over G (\tau ) g _{\tau       \tau } }  
\dh _\tau G (\tau ) \dh _\tau + \nabla _{5} ^2  , \   
    [g _{\tau \tau }= h ^{1/2} \tilde  g _{\tau \tau }  , \cr &&
      \nabla ^2 _{5}=h ^{-1/2}   \tilde  \nabla _{5} ^2  = {1 \over \sqrt {g  _{10} } } \dh _a  g ^{ab} \sqrt {g  _{10} }  \dh _b
,\  \tilde  \nabla _{5} ^2 = K^{ab} (\tau )O^{ab} (\T )   ,\ 
O^{ab}  = {1 \over \sqrt {\tilde \g _5 } }\dh _a \tilde \g ^{ab} \sqrt {\tilde \g _5 } \dh _b   ,  \cr &&   \sqrt {\tilde g_6}   = \sqrt {\tilde
  g_{\tau \tau }  (\tau )   \tilde \g _5 (\T ) }
= {\e ^4 \over 96} \sinh ^2 (\tau ) \sqrt {\hat \g _5}
={ 9  \over 8}\e ^4 \sinh ^2 (\tau ) \sqrt {\tilde \g _5} ,  \cr &&  
G (\tau ) \equiv \sqrt {g_{10} }  g ^{\tau \tau }
={ 27 \over 4} e ^{8/3}   K ^2 (\tau )   \sinh ^2 (\tau ) \sqrt {\tilde  \g  _{5} } , \ \sqrt { \tilde g_6   } {h  (\tau ) \over G(\tau ) } = {\e
    ^{-4/3} \over 2 ^{1/3 }   3} (g_s M\a '  )^2
{I(\tau )\over K^2 (\tau ) } \sqrt { \hat \g_5 } ]   \label{sect2.eq4}   \eea
    where  $\dh _a = {\dh  / \dh \T ^a   } $  and
the base manifold volume 5-form is given by
$vol (T^{1,1} )= (\prod _a g^{(a)} )/ 108,\   [V (T^{1,1} )=\int  \sqrt {\hat  \g _5 }  / 108 =   \int  \sqrt {\tilde \g  _{5} }]$.
The wave equations are then given by  
\bea &&  0= (\nabla _{10} ^2 -\mu ^2 ) \Psi   _m (\tau ,\T ) 
=  (h^{1/2} \tilde \nabla _{4} ^2 + h^{-1/2}
( \tilde g ^{\tau \tau } { 1   \over  G  } \dh _\tau G \dh _\tau +
\tilde \nabla _{5} ^2 )   -\mu ^2 ) \Psi   _m (\tau ,\T ) 
\cr && = g ^{\tau \tau } [ g _{\tau \tau } (h^{1/2}\tilde \nabla _{4} ^2
-\mu ^2  + h^{-1/2} K^{ab}   O^{ab} )   + { 1   \over  G  } \dh _\tau G \dh _\tau  ] \Psi   _m (\tau ,\T )   . \label{eq.SLeq}  \eea

Specializing  now to  the  approximation in Eq.~(\ref{eq.approx1}) involving the  diagonal metric,    $ K _{ab} (\tau ) \propto \d _{ab} K(\tau 
)  $, one can  consider  the product  ansatz for the
radial and  angular wave functions,   
$ \Psi  _m (\tau, \T ) = R_m  (\tau) \Phi _m (\T ) ,\ [\Phi _m (\T ) =
  Y^{\nu } (\T ) ] $   where  we  used the approximate relations,
\bea && K^{ab}  O^{ab}\simeq    {2 \over  3 \e ^{4/3} } K^2(\tau ) 
 \tilde  \nabla   _{T^{1,1} } ^2 ,\  V_5 = 
- \tilde g_{\tau \tau } \tilde  \nabla _5 ^2 =- \tilde g_{\tau \tau }  K^{ab}  O^{ab}   \simeq  - {1\over 9}  \tilde  \nabla _{T^{1,1} }  ^2 .\eea
with  $ Y ^{\nu }(\T )$  denoting   eigenfunctions of
the  base manifold scalar Laplacian,
$ (\tilde \nabla ^2 _{T^{1,1  } }  +  H_0 ^\nu  )    Y ^{\nu } ( \T ) =
0,\ H_0^\nu  \equiv H_0 (j,l, r) = 6 ( j(j+1) +  l(l+1) - {r^2 / 8 } ) ,$
For  modes  of fixed      5-d and 4-d   squared masses  $M_5 ^2  $ and $ E_m^2 $,    the scalar Laplacians are set  as
$\tilde \nabla  _{T^{1,1} }^2  \to  - M_5^2 ,\
\tilde \nabla _{4} ^2   \to  E_m   ^2   $ and
the resulting   diagonal  radial  wave equation is given by
\bea 0= [g_{\tau \tau }  (E_m^2  h^{1/2} -\mu ^2
-{1 \over 9}   h^{-1/2} M_5 ^2  ) +   {1\over  G(\tau  )} 
\dh _\tau    G (\tau   ) \dh _\tau  ]  R _m (\tau ) . \label{eq.SLeq1} \eea 
  The wave function  redefinition $R _m =   {B_m (\tau )  / G^{1/2} (\tau ) } $,    removing    the first order derivative  term,  transforms the  wave  equation  to   the   Schr\"odinger  type  equation,  
\bea && (\dh _\tau ^2 - V_{eff} (E_m ^2 ,\tau )  ) B _m  (\tau ) =0 ,
\ [R _m =   {B_m (\tau )  \over   G^{1/2} (\tau ) } , \cr &&  
V_{eff} (E_m ^2 ,\tau ) = - \tilde g_{\tau \tau }  (
E_m ^2 h (\tau ) -  \mu ^2   h^{1/2}(\tau ) )  + V_5  + G_1 
 ,  \ G_1 (\tau ) = {(G^{1/2} (\tau ) )'' \over  G^{1/2} (\tau )} ]   \label{sect1.schrod1}  \eea
where  $ f'  = d f  ( \tau  ) / d \tau $.   The dimensionless  effective potential can be expressed in terms   of the  dimensionless string and glueball   mass parameters   $ \hat   \mu  $ and $ \hat  E_m$ as 
\bea &&   V_{eff} (\hat E_m ^2 ,\tau )
= - {2 ^{-1/3} \hat E_m ^2  I(\tau ) \over 3 K^2 (\tau ) }  
+   \hat \mu ^2  {I ^{1/2} (\tau ) \over K (\tau ) }  + {H^\nu _0\over
  9}   + G_1    , \   [\hat  E_m =  E_m \e ^{-2/3 }(g_s M \a ' ),\ \hat
  \mu ^2 =   {2 ^{1/3} \over 6}  (g_s M \a ')  \mu  ^2 ] .  \eea 
The   wave  function and    normalization integral are given by  
the explicit  expressions, 
\bea && \Psi _m  = {B _m  (\tau ) \Phi _m  (\T ) \over  A J_{(m)}
  \sqrt { \tilde G (\tau )} }  = \caln _m   {B _m  (\tau )\Phi _m  (\T
  ) \over  \sqrt { \tilde G (\tau )}}   ,  \ 
[G(\tau )   =   {\e ^ {8/3}    \tilde G (\tau ) \over    16} ,\ 
 A= {g_s M \a '  \over     2 ^{1/6} 3 ^{1/2} \e ^{2/3} } 
,  \ \caln _m = {4 \e  ^{-4/3 }\over  A J_{(m)} }  = 
{2^{13/6} \sqrt 3 \e ^{-2/3}  \over g_s M \a ' J _{(m)}} ,\cr &&    
{1\over A ^{2}} \int d ^6 y \sqrt {\tilde g_6} 
{h (\tau )  \over G(\tau ) } B^\star _m B_{m'}     \Phi ^\star _m
\Phi _{m'}  =  \int d \tau  {I(\tau ) \over K^2 (\tau )}  B_m ^\dagger (\tau ) 
B_{m'} (\tau ) \int d ^5 \T  \sqrt {\tilde \g _5} \Phi ^\dagger _m (\T  )
\Phi _{m'} (\T  ) =  J_{(m)} ^2  \d _{mm'} ]         \label{sect2.eq7}\eea
where  we have extracted out the  dependence on the  dimensional parameters $ \a '  $ and $\e $    and  included  it in    the  constant  factors  $A$  and $ \caln _m $.
The  limiting  behaviour for the measure factor
$\sqrt {\tilde  g_6 } $ in the  normalization  integral in Eq.~(\ref{eq.norsq}),  evaluated using Eq.~(\ref{sect2.eq4}),  and
that for the  rescaling    factor, 
\bea  &&  \lim _{\tau \to 0}   G ^{1/2} (\tau ) =  2 ^{-5/3}
\e ^{4/3} \tau    ,\ \lim _{\tau \to \infty } G^{1/2}(\tau ) =  2 ^{-8/3}
\e ^{4/3} e ^{2 \tau /3} ,\eea
show   that   one can select normalizable    wave
functions   by requiring  $B_m (\tau ) $  to
be   finite  near the origin  and  at  infinity.
The singlet massless graviton mode,    
$M_5 ^2=0,\ \mu ^2 =0  ,\   E_0 =0$, is  assigned  the 
radial wave function  $B_0 (\tau )  \propto   G^{1/2}(\tau )$,
yielding in accordance with  Eq.~(\ref{eq.norsq})   
the constant normalized    wave  function,
$ \Psi _0 (\tau )  \equiv  {B_0 (\tau )  /  G^{1/2} (\tau ) 
} =    {1 /  \sqrt {V_W}} $, which is then orthogonal to the   wave  functions of  all  massive graviton modes. 

The     modes radial  wave    equation,  
$( \dh _\tau ^2 - V_{eff} (\tau ) )  B_m(\tau ) =0$,     
looks   formally    as   a  time-independent  (zero   energy)  Schr\"odinger equation    over the radial  variable half-axis  $\tau  \in [0, \infty ] $.
The potential depends    in a non-trivial way
on the  modes  mass  $ \hat E_m $  which are derived along 
with   the wave functions as solutions of   a Sturm-Liouville  boundary  eigenvalue  problem.
The  radial dependence   of  $ V_{eff} (\tau )$ typically features   
an  attractive (negative sign) well    of depth set by $ \hat  E_m$, followed  
beyond   the  turning   point   at $ V_{eff} (\tau _0(\hat E_m) ) =0$  by   a  plateau     of  height  set by   the  mass independent 
term $ G_1 (\tau )  $.   Only  non-tachyon (massless or massive) modes are allowed  consistently with the  correspondence to  the graviton and  
the  confining  dual gauge theory glueballs.

The  normalizable solutions are given by    linear combinations of   regular
and irregular  solutions of the  second order  linear differential   equations. 
 In the absence of   analytic  solutions,  the    eigenvalue  problem is  commonly  solved  by means of the  shooting   technique.  One
considers   suitable  wave  function ansatz
at small and large $\tau $ (near the horizon   and  boundary),
evolve these   via the  wave equation  in steps  of  increasing and 
decreasing  $\tau $ and determines   the   mass
 parameter    by matching the   solutions at some intermediate  radial
 distance. The integration  is   performed numerically~\cite{berg06},   by means of  series  expansions~\cite{demello98,zyskin98} or by  adapting  the  WKB approach~\cite{csaking99,minahan99,russo99}.
For  a  warped conifold throat  glued to a compact    manifold, the  background   can be crudely described  by truncating  the  radial semi-axis
to  a  finite interval   $0 \leq \tau \leq  \tau _{uv} $ ending   at the  throat-bulk  interface. For a hard  wall type boundary at $ \tau _{uv}$,
the modes masses and wave  functions are then determined  by solving  the wave equations  subject to     (Neumann  or Dirichlet)   conditions
at the origin and  boundary. 
We    specialize hereafter to  the  so-called throat
domination case  (large  $ \tau _{uv} $)  where the   
bulk  is far   smaller than  the throat.  This  selects  the   unique normalizable
solutions  regular at $ \tau \to 0  $  and  $\tau \to \tau _{uv} $.  In the   alternative throat  domination case,
for which an   interesting  realization is  proposed  in~\cite{firouz06},  
one must supply  the information  on     how the metric 
extrapolates inside the    bulk. 

In the (semi-classical) WKB   approach that we use hereafter,   
the  radial  wave functions are  evaluated at leading order
by means of   the familiar textbook formulas  (see Chapter $VII$  of~\cite{LLqm} or Chapter 7  of~\cite{merzbacher}), 
\bea && B_m (\tau ) = {C_m \over \sqrt {p(\tau ) } } \sin
({\pi \over 4}  + \int _{\tau }  ^{\tau _0} d\tau  ' \ p(\tau ')) ,\  
[\tau \leq \tau _0 ,\ p(\tau ) = (-V _{eff} (\tau ) ) ^{1/2} ] \cr &&
B_m (\tau ) = { C'_m \over \sqrt {p(\tau ) } } e ^{- \int _{\tau
    _0 } ^ \tau d\tau ' p (\tau ') } , \ 
[\tau \geq \tau _0 ,\ p (\tau ) = (V_{eff}(\tau ) )^{1/2}]
  \label{sect2.eq9} \eea 
where the  effective  potential zeros,  $ V_{eff} (\tau _0 ) =0$,
separate    the  classically allowed and forbidden regions  
on the left and right  hand  sides  of the 
(mass dependent) classical turning points  $\tau _0(E_m)  $
and continuity at the
turning point is  approximately  fulfilled  by setting $C'_m  \simeq  C_m $.
(The    normalization   condition 
$ \int d \tau  \vert B_m  \vert ^2  = 1 $ approximately relates  
the constant coefficient   to the classical
period  $T$  inside the potential well  region, 
$ C_m \simeq  (2 /\int _0 ^{\tau _0}  d \tau  /  p (\tau ) )^{1/2} = 
(2/T )^{1/2} $ while continuity at the  turning point $\tau _0$  is usually ensured   by  using adjustable  linear combinations of Airy   functions~\cite{berry72,bender99}.)
It is safe to ignore    the  external  region at  $\tau  > \tau _0 $ where wave functions   decrease exponentially.      For  charged  modes of finite 
angular  momenta, the centrifugal  force  typically contributes a   repulsive potential  producing    an inner turning point  $\tau '_0 >  0$ near the origin.
Each  member  of   the towers of 5-d modes 
develops   a  sequence  of 4-d  radial
excitations  whose masses  are  determined   by 
means    of the quantization   rule    for 
the    resulting phase  integral  over   the  well region    between the  pair of
turning points~\cite{krasnitz00,caceres00},
\bea && \int  _{\tau  '_0 }  ^{\tau _0 } d \tau   (- V_{eff}   (\tau )
) ^{1/2}   = (n -\ud + {\d _{W}  \over  4 } )\pi  , \ [V _{eff} (\tau
_0 ) =0,  \  V _{eff} (\tau '_0 ) =0,  \ 
n=1, 2, \cdots ]     \label{sect2.eq10} \eea  
where the   integer radial  quantum  number 
$ (n -1) =0, 1,\cdots $ counts the  number of zeros
in   radial wave functions  and the barrier parameter $\d _W$  
is set to   $1 $ if  the potential  near  the origin  is    finite 
and  to  $0 $   if it is sloping,   as  happens in the presence  of a repulsive centrifugal   energy  barrier term~\cite{benna07,pufu10}.  
The  phase integral  $ (n - {1   \over  4 } )\pi  $   in  the  former  case     amounts to   imposing a  hard wall that forces   the wave function to vanish   at the origin,
\bea && \lim _{\tau \to 0 } B_m (\tau )
= {C_m \over (-V _{eff} (\tau ) )^{1/4} } \sin  ({\pi  \over  4}  + \int _\tau ^{\tau _0}  d \tau ' (-V _{eff} (\tau ') )^{1/2} ) = {C_m \over (-V _{eff} (0 ) )^{1/4} } \sin  (n  \pi  )  =0.\eea 
In the large radial  distance   approximation that we use, 
the centrifugal barrier  in the  potential
(for both singlet and charged  modes)  is
smoothed   out, so   one must   set $\d _{W}  =1 $  in the  quantization rule.
The  mass eigenvalues  can  be    conveniently evaluated  
by means  of   the   procedure initially devised in~\cite{krasnitz00}.  
For each 5-d mode   of fixed   angular  and  radial  quantum   numbers 
$ m = (\nu ,\ n )$,  one    searches   for the  
constant  parameters $\hat  E_m  $ and $ \tau _0 $  
solving  the pair of  equations in Eq.~(\ref{sect2.eq10})  
with the phase integral  set  to $(n -  {1/ 4 } ) \pi $.  

\subsection{Scalar  modes   from   4-form   and axio-dilaton  fields}
\label{sect2sub2}

We   continue  our discussion  of the reduced  supergravity action  with     a study  of the  scalar   modes descending  from   the  4-form    and metric tensor trace field
components      $\d  C_{abcd} (X)  $ and $\d g ^a_a  $  along
$ T^{1,1}$  of type $ II\ b $  supergravity.  With   the
approximate   formula for the metric in Eq.~(\ref{sect2.apmet}), one can use
the  Kaluza-Klein  ansatz with factorized   radial and         angular   wave  functions, 
\bea && \d C_{abcd} (X) =\sum _m b ^{(m)} _{k_m }  e ^{i k_m\cdot x } b_m (\tau ) 
 \e _{abcd e } \cald ^e  Y^{\nu  } (\T ) ,\ 
\d h ^a_a  (X) = \sum _m \pi ^{(m)}_{k_m }  
 e ^{i k_m\cdot x } \pi _m ( \tau ) Y^{\nu  }  (\T  ) , \ [m= (\nu , k_m)] \label{sect2.eq11}  \eea
where the  coupled  mode  fields  $(b ^{(m)}  (x) ,\  \pi
^{(m)} (x)  )  $ in   $ M_4 =\dh ( AdS_5)$  are assigned  the wave functions $(b_m (\tau ) , \pi _m ( \tau ) )$. Instead of   the  usual   procedure~\cite{kiroman84}
combining   the first order variations  of  Einstein equation   $ R_{MN} =
-{1\over 6} F_{MPQRS}  F_{N} ^{\ \ PQRS} $   with 
the    self-duality  constraint equation $ F_5 = \star _{10}  F _5
$,   we  shall  adopt an approximate  derivation 
using the  second  order variation  
  of    the    4-form potential kinetic   action,
\bea && \d ^{2}  (\sqrt {\g _5 } \vert F_5 \vert ^2) =  
\sqrt {\g _5 }   [2 \d ( F_5 )  \d   \cdot F _5 + \g _5 ^{-1}    
  \d \g _5 \d ( F_5 )   \cdot F ^{cl} _5 ] \ 
\sim [ H^\nu _0   ( b ^{(\nu )}  \cdot b^{(\nu )}  )   + {4 \over 5}  (b ^{(\nu )}
  \cdot \pi ^{(\nu )}   )   ]  (Y^\nu    \cald ^2   Y^\nu) , 
\label{sect2.act2}  \eea
where $F^{cl} _5= 4 \calr ^4 vol (T^{1,1}  ) $.   Consider  first
the  terms    depending on the 4-form only, 
\bea && \d S ^{(2)} =  +{ g_s ^2   M_\star ^2  \over 8  V_W} 
\int d^{10}  X \sqrt {\vert \tilde  g_{10} \vert }b ^{(m) \dagger }  (x)  Y^{\nu \star } (\T   )   \bar b _m (\tau ) \cr &&
 \times [e ^{4 A}  Q(\tau ) \tilde \nabla _4 ^2   
 + {1\over \sqrt {\tilde  g_{10}  } } \dh _\tau e ^{8 A} Q(\tau )  
\sqrt {\tilde  g_{10}  }  \tilde g ^{\tau \tau }  \dh _\tau   
+ e ^{8 A}   P(\tau )  \tilde \nabla  _5^2 ]   b  _m (\tau )
\cald ^2 Y^{\nu } (\T ) b ^{(m)}  (x)  ,  \cr &&  
[Q(\tau ) =  (\tilde K^{aa} \cdots \tilde  K^{dd}) 
= {2 ^4 3 ^{-4} \e   ^{-16/3} \over  K^4(\tau ) \sinh ^4 \tau  }  ,\  
P(\tau ) =   ( \tilde  K^{aa} \cdots   \tilde  K^{ee} )     =   
{2 ^{5} \ 3^{-5}  \e ^{-20/3}   \over   K^5  (\tau ) \sinh ^5\tau   } ]   \eea 
where the auxiliary function  $ Q(\tau ) $, in contrast to  $P (\tau )$, 
is unambiguously defined only in  the   large $\tau $ 
limit  of the metric in  Eq.~(\ref{sect2.apmet}).
Since  the  covariant  derivative   square $  \cald ^2 
$ identifies to the Laplacian  of $ T^{1,1}$,  one can simply  replace for
modes    of fixed  5-d  mass eigenvalues $M_5^2 $ 
the    wave functions  $ \Psi _m (\tau, \T ) =  b_m (\tau ) (\cald ^2
)^{1/2} Y ^\nu (\T )  \to  \Psi _m (\tau, \T ) =   b_m (\tau )   Y
^\nu (\T )   $,   since     this   amounts    to     a  constant rescaling.    
The  resulting wave equation
\bea &&  
 ( e ^{4A } Q( \tau ) \tilde \nabla  _4^2  + e ^{8A } 
\tilde g ^{\tau \tau }  Q ( \tau ){1\over \calg ( \tau ) }  
\dh _\tau \calg ( \tau )\dh _\tau    +e ^{8A }
P( \tau )  \tilde \nabla _{5  } ^2  )    \Psi _m ( \tau  ,\T )  =0  ,
\cr && [\calg ( \tau ) =\sqrt {\tilde  g_{6}  }   \tilde g ^{\tau \tau } e
  ^{8A }    Q ( \tau ) = \sqrt {\tilde \g _5 \over \tilde g _{\tau \tau } } 
  e ^{8A } Q ( \tau ) = {2 ^{2/3} \e ^{8/3} \over 3 ^{-1}  (g_s M \a ' ) ^4 }  
{\sqrt {\tilde   \g _5 }
  \over \sinh ^2 \tau   K^2 ( \tau )    I^2 ( \tau ) }   ,\
 \tilde g _{\tau \tau } = {\e ^{4/3}\over 6 K^2 (\tau )}  ] 
\eea
admits the  normalization condition for wave functions, 
\bea  &&  
\d _{mm'} = \int d ^6 y \sqrt {\tilde g _{6} } e ^{4A} Q (\tau )    \Psi _m  ^\dagger (\tau ,\T ) \Psi _{m'}   (\tau ,\T )  
  = {2^{1/3} \e ^{4/3}    \over 9 (g_s M \a ')^2  }
  \int d \tau   {b_m^\dagger  b_{m'} (\tau )
\over K^4 ( \tau ) I( \tau )\sinh ^2 \tau }    
\int d ^5\T \sqrt {\tilde \g _5} \Phi ^\dagger _{m}   \Phi _{m'} (\T
)   . \label{sect2.eq13}  \eea
The wave function    rescaling,  
$b_m (\tau )    \to  \tilde    b_m (\tau )  /\calg ^{1/2}    (\tau ) $,   
transforms the  wave  equation  to  the    Schr\"odinger  type equation, 
\bea &&  (\dh _\tau ^2  - V_{eff}  )  \tilde    b_m (\tau )  
=0, \  [V_{eff}(\tau   )  =  -  \tilde g_{\tau \tau }  h (\tau ) E_m ^2 + 
V_5 +  \calg _1 =  - {2 ^{-1/3} \hat E_m ^2 I (\tau ) \over 3 K^2
  (\tau ) }  +V_5  +  \calg _1  ,\cr &&  \Psi _m  (\tau ,\T ) = {\tilde  b_m (\tau )  \over     \calg ^{1/2}   (\tau ) } \Phi _m (\T )  ,\
  \  V_5  = - \tilde g_{\tau \tau } {P(\tau   )  \over Q(\tau   ) }
\tilde  \nabla _5 ^2  =  {2  M_5 ^2 \over 9 K^3 e ^\tau } \simeq
 {M_5 ^2   \over 9}, \ \calg _1 = {  (\calg ^{1/2}  (\tau )) '' \over 
  \calg ^{1/2}  (\tau ) } ] \label{sect2S.eq13}  \eea 
where we  observe that the   term     $ \calg _1 (\tau )  $ contributes a repulsive wall    $  O( 1/ \tau ^2 )  $    in the potential near the  origin. 
Using   the   large $\tau $   limits   of the auxiliary  functions,   
$P(\tau ) \to r ^{-10} ,\ Q(\tau ) \to r ^{-8}  ,\ \sqrt {\tilde g
  _{6} } \to r ^5  \sqrt { \tilde g_5 } ,\  V_5   \simeq  {P\over Q}
\tilde \nabla _{T^{1,1} } ^2 = {1\over 9} M_{5 } ^{\nu 2},$
one can verify  that the  known  wave equation  in   the  undeformed      conifold  case~\cite{chemtob16} is reproduced.   
The limiting   behavior of the rescaling
factor    $ \calg ^{1/2} (\tau ) $
at the origin  and  boundary,    
\bea && \lim _{\tau \to 0} \calg  ^{1/2} (\tau ) =  {2  \ 3 ^{1/3} \e ^{4/3}   
  \over a_0 (g_s M \a ' )^2  \tau },\ 
\lim _{\tau \to \infty } \calg  ^{1/2} (\tau )  =   {2 ^{7/3} \e
  ^{4/3}  e ^{2\tau / 3} \over   3 (g_s M \a ' )^2  (\tau - 1/4 )    }
,  \label{sect2S.eqP13}  \eea  
 shows  that  the normalizable  modes  must  be assigned
 radial wave functions $ b _m (\tau )$  
that  are     finite   at small  and large  $\tau $. 

The    mixing  with  modes descending  from  the    metric trace  fields  in Eq.(\ref{sect2.act2})  can be taken   approximately into account 
 by  restricting to the angle dependent contribution  
contained in the  $2\times 2 $ mass
matrix $ M_{5\mp }  ^{\nu 2} $  in the   vector space 
$ (\pi  ^{( \nu   )} ,\ b  ^{(\nu )} )  $.   Noting that     the  
 diagonalization  of the 5-d mass matrix admits      the   pair of eigenvectors  and eigenvalues~\cite{chemtob16},  
\bea && S ^{(\nu )}  _{\mp } (x,\tau    ) = 10 ( (2 \mp \sqrt {H_0^\nu  +4} ) \pi
^{(\nu )}  (x,\tau    )  + b ^{(\nu )}    (x,\tau    ) ) ,\  M^{\nu 2}
_{5 \mp }  = H_0^\nu  +16 \mp 8 \sqrt {H_0^\nu  +4},    \eea 
one finds that the radial wave  functions  $\tilde  b_{\mp, m }  (\tau )$
for the eigenmodes  of 5-d and 4-d   squared masses $M_{5 \mp } ^{\nu  2} $
and $ E_{m, \mp } ^2 =  \hat E_{m, \mp } ^2 \e ^{4/3 } / (g_s M \a ' )^2 $   obey the Schr\"odinger  type  diagonal  wave   equations
\bea && (\dh _\tau ^2  - V_{eff, \mp }  ) \tilde  b _{\mp, m }  (\tau ) =0, \
[V_{eff, \mp } =  \tilde g _{\tau \tau }  (-E_{m, \mp }  ^2 h
+{ P \over Q} \tilde \nabla _5 ^2 ) + \calg _1 = -{I (\tau ) 
\over  2 ^{1/3} \ 3 K^2 }\hat E_{m, \mp } ^2   + {1 \over 9}  M_{5 \mp } ^{\nu 2} + \calg
_1 ]. \eea  
 
\begin{figure} [t] 
\includegraphics[width=0.5\textwidth]{   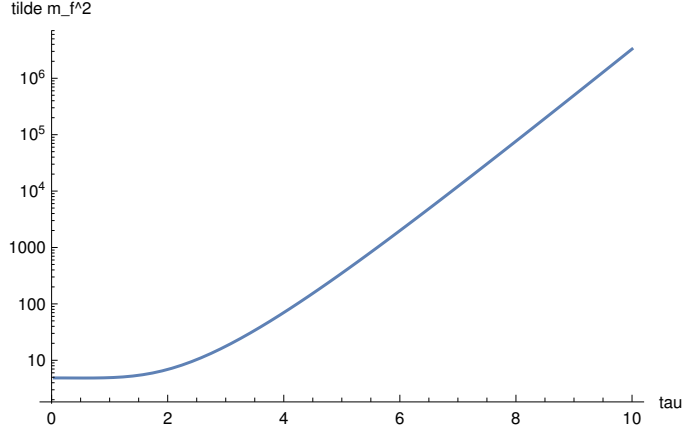}
\caption{\it \label{wcofb2}   
Plot  versus $\tau  $ of the   mass  parameter for  axio-dilaton modes  from  
3-fluxes,    $m_f ^2(\tau ) ,\  [\xi \simeq  {5.25 \ 10^{-2}  / g_s}
]$  in    Eqs~.(\ref{sect2.eq17})   and~(\ref{sect2.eq18}) evaluated
for $ g_s =1.$} \end{figure}

We  discuss next the   axio-dilaton field  
fluctuations  in   the simplified   case ignoring        the couplings to 
the metric and 2-form    fields   components~\cite{frey06,chemtob16}. 
The reduced    action,   
\bea &&  \d S ^{(2)}   = {1\over 2\kappa ^2  _{10} } \int d ^4 x \sqrt {-\tilde g _4 } \int  d ^6 y   \sqrt {\tilde g _6 } \d {\bar \tau } [{1\over 4\tau _2 ^2} 
(e ^{-4 A} \tilde \nabla _4 ^2 + \tilde \nabla _6 ^2)  
- {e ^{4 A} G_3 \tilde \cdot \bar G_3   \over 12 \tau _2 }  ] 
\d \tau , \label{sect2.eq14}  \eea 
is   evaluated using
the  6-d Laplacian in the large radial  distance limit.
Substituting the Kaluza-Klein   decomposition on harmonic modes $t ^{(m)}(x ) $, labelled by $ m = (\nu , k_m)$,   yields   the   radial  wave equations   for the wave functions $t _m (\tau ) $, 
\bea &&  0= 
[\tilde g _{\tau \tau }  (   h  E_m^2 +  \tilde   \nabla _5 ^2-
h^{1/2} M_f ^2 (\tau ) )   +{1 \over G(\tau ) }    \dh _\tau    
G  \dh _\tau  ] t _m (\tau ), \  
[\d \tau (X) = \sum _m   t ^{(m) } _{k_m}      e ^{i k_m\cdot x
  }  t _m (\tau ) Y ^{\nu } (\T)  ,  \cr &&     
G (\tau ) =\sqrt {g _{10} } g ^{\tau \tau }   ,\    M_f ^2 (\tau ) = 
{h^{-3/2} \over 12 \tau _2} G  \tilde \cdot  \bar G 
=   {e ^{6 A} \over 12 } ( g_s F_3\tilde \cdot F_3  + {1\over g_s}
H_3\tilde   \cdot H_3 ) ]    \label{sect2.eq15}   \eea 
where $ M_f ^2 (\tau ) $   denotes the  effective mass term  contributed
by  3-fluxes.   (The  10-d  mass term $\mu $ can be  included    
by  replacing $ E_m ^2 e ^{-4A} \to E_m ^2 e ^{-4A} - 
\mu ^2e ^{-2A} $)
The  wave  function rescaling  $t_m (\tau
  )  = \tilde  t_m   (\tau )  / G  ^{1/2} (\tau ) $     yields  
the Schr\"odinger   equation 
$ (\dh _\tau ^2 - V_{eff} (\tau ) ) \tilde   t_m (\tau ) =0, $ with  the
 effective potential     \bea && V_{eff} (\tau ) 
= - \tilde g _{\tau \tau }  (E_m ^2   h  - h^{1/2}
M_f ^2   )   +  V_5  + G_1 (\tau )   ,\ [G_1 (\tau ) = {(G^{1/2} ) ''
\over   G^{1/2} }   ,\  V_5 = -  {\tilde g _{\tau \tau }} \tilde
  \nabla _5 ^2 \simeq   { M_5 ^2  \over 9}    = { H_0^{\nu } \over 9} ].  \label{sect2.eq17} \eea   
The  results  are   same as those for graviton modes  except 
for the   additional mass term  contributed by the  classical  3-forms  in
Eq.(\ref{app2.eq1}), 
\bea &&   M _f ^2(\tau )    = 
{e ^{6A} g_s \over 12} ( F_{mnp} F^{\tilde m\tilde n\tilde p} 
+ {1\over g_s ^2} H_{mnp} H^{\tilde   m\tilde n\tilde p} ) =
{e ^{6A} g_s \over 2} ({ M \a ' \over  \e ^2 } )^2   \varphi (\tau ),\cr &&   
[\varphi (\tau ) = {(1- F)^2 + k^{'2} \over \cosh ^4\tau /2 } + { F^2
    + f^{'2} \over     \sinh ^4  \tau /2 } +{ 2 (F^{'2} + (k - f )^2
    /4 )\over \cosh ^2 \tau /2       \sinh ^2  \tau /2 } ],   \label{sect2.eq18}   \eea 
which  adds the extra term   to    the effective  potential,  
\bea && \d V_{eff}   =  m_f^2(\tau  ) = 
\tilde g _{\tau \tau }(\tau  )  h^{1/2} (\tau  )   M_f^2 (\tau  ) =
\xi   \tilde  m_f^2(\tau  ) ,\ [ \xi = {1 \over  2 ^{8/3} 3 g_s} ,\ 
\tilde  m_f^2(\tau  ) = {\varphi (\tau )\over K^2 (\tau )     I(\tau )
}]. \eea   
The  mass profile    $\tilde   m_f ^2(\tau )  =  m_f ^2(\tau ) / \xi  $   from 3-fluxes is displayed in Fig.~\ref{wcofb2}.  We see that this is  small
and nearly  constant inside the throat  for $\tau < 2  $,
where warped modes wave functions are  mostly  concentrated,
but  that it grows exponentially at larger $\tau  .$

\subsection{Predictions  for  warped modes   masses and wave functions} 
\label{sect3sub2} 

The   free parameters  at our disposal 
consist of   the   string coupling constant  
and mass scale $ g_s $ and $m_s$, the internal manifold    
 warped  volume  $ V_W = (2\pi L_W)^6$, the  3-fluxes $
 M, \ K  ,\  [N=MK] $ and the   warp  factor for  the mass
 hierarchy  relative   to the    Planck mass scale,  
$ w  = m_{eff} /   M_\star  $. 
We  choose  $M_\star $   as the reference  energy scale  
and   express      predictions 
in terms of the  dimensionless       geometric  and flux
parameters   given by   the ratio of  bulk to throat  radii  $ \eta
= L_W/ \calr $   and Planck  mass times   and     $ AdS$
 curvature,  $z=M_\star \calr =M_\star / k$.   
The      string     coupling constant  is   set at the value,  
$ g_s = g_{D3} ^2  / (4\pi  )  \simeq  0.1 $, appropriate to a $D3$-brane  setup  realizing   a grand  unified theory.   The ratio parameters $\eta  ,\ z $   
are    naturally  of  $ O(1)$,  although  we expect values  well above  unity 
upon   matching    predictions to data.   For instance, the  Chen-Tye  study~\cite{chen06}  assigns values    $\eta = 10 \  - \ 100$.   The  phenomenological analyses for  the standard  Randall-Sundrum   models,  using   flat branes embedded in $ AdS_5  $  or   $ AdS_5
\times S^\d $ spacetimes,      select   values
$ z  \in  (10 -\ 100) $~\cite{davoudheriz02,grzadgunion06},  with   larger  values $z =  O(10 ^{4} )$  required     in the non-standard  type 
gauge unification     invoking the Weyl anomaly~\cite{dienes99}. 
Large uncertainties also affect the  Calabi-Yau volume parameter.    For a fixed   string  compactification,  it seems natural  to   identify the  
total  warped volume $ V_W   $  (in Einstein frame)
to the   volume    in the   low energy effective action, setting   $\calv = V_W /\hat l_s^6= (L_W/l_s)^6$.  
The  analyses of $D3$-brane  inflation~\cite{kklmmt03}  (after inserting
the proper   $(2\pi )$ factors)  and  those of   electroweak supersymmetry
breaking  effects~\cite{choi05}  both   favour  the value 
$ V_W  ^{1/6} / (2\pi l_s ) = L_W/l_s  \simeq 5 $. 
Larger values    covering  the wide range   $\calv \simeq (10^5 -
10^{30})$ are   quoted in applications   of the   large volume
scenario~\cite{bala05,burgessSB06,spdealwis12}.
The  parameters  satisfy  the     useful  relations in Einstein   frame,
\bea &&  {w_s \over  w } =
{M_\star \over    m_s}=  {(L_W m_s )^3 \over  \pi ^{1/2}  } = 
\sqrt {\calv \over  \pi  }  = {(\eta z)^{3/4}  \over  \pi ^{1/8}  }  ,
\ \rho _2 \equiv  2  \bar \rho  = e ^{4u} =
\calv  ^{2/3}  =     ({V_W  \over  \hat l_s^6 } )^{2/3}  = ({L_W \over
  l_s} )^{4} \cr &&  
{\eta  ^3\over   z}  =  {\sqrt {\pi } \over \l _N^{4}   g_s } ,\ 
\eta = { m_s L_W \over \l _N  g_s^{1/4} } ,\  z= {M_\star \over    m_s}
\l _N  g_s ^{1/4} = m_s \calr \sqrt { \calv \over \pi } ,   
[ \calr = \l _{N }  g_s ^{1/4}   l _s  ,\  \l _{N } = ({27 \pi   \over
    4 } N ) ^{1/4}] .   \label{eqparams}   \eea 

The WKB    approach that we use should   hopefully be trustable  
in identifying  the  lightest  charged  Kaluza-Klein
particles    that   sets the mass   gap between Kaluza-Klein  and moduli  modes.   The   masses   of  radially   excited   modes  (of  fixed   charges) grow  linearly   with the   radial quantum number,  $ \hat   E_{m_n}
\sim n $,  as is   inferred  from  the  limiting  formula  
$ n \simeq  \int _0 ^{\tau _0 }  d \tau (- V_{eff}  )^{1/2} 
\sim E_m^{(n) }\int _0 ^{\infty }  d \tau ( I (\tau ) / K^2 (\tau ))^{1/2} 
$, assuming   that  $ \tau _0 (\hat  E_m) $ recedes to   infinity
for  large $\hat   E_m$.   The  use of  a truncated  radial interval 
$ 0\leq  \tau  < \tau  _{uv}  $       with    $\tau _{uv} \simeq 10 $ 
has  little  effect on  the accuracy of 
predictions   since    the  region   where the  effective
potential is sizeable   does not   extend far   beyond  the   classical  turning  points, 
typically    located at  $\tau _0   \leq   5  - 7 $. 
Since  the massive modes wave functions are  concentrated near the  throat apex,  the  estimates of  masses and couplings are  insensitive to the ultraviolet  radial cutoff and  justify using the throat domination
case. 

We evaluate  the   two unknowns  $ \hat  E_m $  and  $ \tau _0 $ 
by  solving simultaneously  the   Bohr-Sommerfeld    quantization
condition,  $ \int  _0 ^{\tau _0}  d   \tau  (- V_{eff}(\tau ) )^{1/2} = (n- {1\over 4} ) \pi $,  and   the  turning point equation,  
$ V_{eff} (\hat  E_m ^2 ,\tau _0) =0 $ in  Eq.~(\ref{sect2.eq10}). 
In   the  presence    of  a repulsive wall   producing  
an inner  turning point $\tau '_0 >0 $, one can  extend   the   search procedure    to   the  three unknowns $ E_m ^2 ,\ \tau _0 , \ \tau '_0 $,   by including   the  additional condition fixing the  location of the   inner  turning point,   $ V_{eff} (\hat  E_m ^2 ,\tau '_0) =0 $,   and  matching the    phase integral  to   $ (n -  {1/ 2 } ) \pi $.  

\begin{figure} 
\includegraphics[width=0.6\textwidth]{   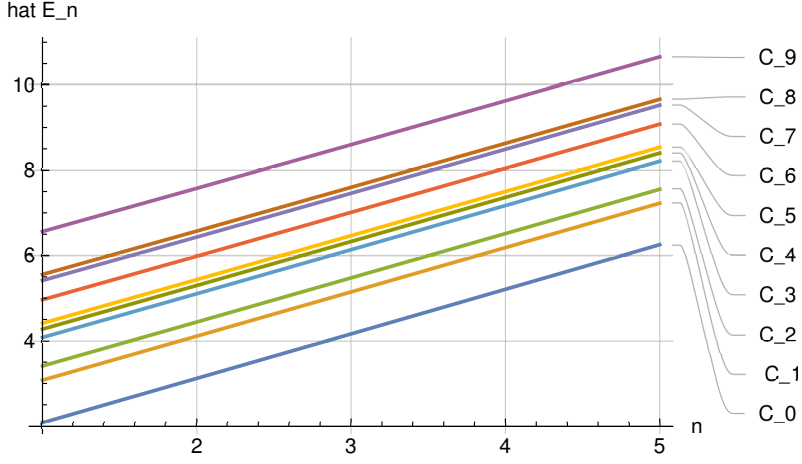}
\caption{\it \label{GRwf0} 
The reduced  masses $   \hat E_{m} $ of the ground and first four radial 
excitations of  the 10  singlet  and charged modes $ C_i $  of the 
graviton  field  (Fig.~\ref{GRtabxp1}) ordered   according
to increasing masses     are plotted  versus the radial 
excitation number  $m +1 = n= 1, \cdots  , 4  .$}\end{figure}

The  numerical applications  were all performed with the help  of  'Mathematica'  programming tools.  
For each of the graviton, 4-form scalar and axio-dilaton fields, 
we have selected  10 modes $ C_0 ,\cdots ,\ C_9 $ in low-lying
representations of the isometry group. 
The  mass spectra   for the modes  $ C_i$, identified  by their  conserved  charges  $(j ,\ l, \ r ) $,  are   listed  in   Table~\ref{GRtabxp1}
where we present    results for the  4-d  masses   of  ground states and  first  few  radial excitations.   The  predicted masses are seen to    increase with the   angular momenta  $j,\ l $,   just like the 5-d masses  $ M_5^2$ but 
much less  rapidly. The   favoured candidate  for the LCKP is the
mode   $ S_- (j l r)=  S_- (1 0 0) $  which saturates the unitary
bound  on $ M_5^2$.      The  mass splittings are independent 
of the    field types  or the  modes charges and  
grow  linearly    with  the radial quantum number $
\hat E_{m_n} \propto  a n $, as appears clearly  on the plot  of  gravitons masses in    Fig.~\ref{GRwf0} where $a \sim 1 $.
The scalar and axio-dilaton fields $ S_m ,\  \tau _m $ feature a   
 faster    slope $a\simeq  1.5 $.  

It is useful to compare our   predictions for gravitons  
to those using  the  full-fledged   solutions  in the deformed conifold throat~\cite{pufu10}. Note that     each $(j,l,r)$   mode in our case splits up       into $ 2 \tilde j + 1 $ sub-modes $(j, l)$ in the exact case labelled by $ r  \in (- \tilde j ,\cdots ,\tilde j ),\ [\tilde j = \min (j,l)] $.   The  comparison of our  results for the   sample  of ground states  masses,  $\hat E_m \  [C_{0, 1,2,3, 4}]  =  [2.1,\  3.1,\  3.4,\  5.0,\  6.6] $   
with  the    numerical results (averaged  over $r$ values)  from~\cite{pufu10},  $   ( 2 (3/2)^{1/6} / (1.139)^2 ) \times 
\hat m  \ [C_{0, 1,2,3, 4}]   \approx  [1.7,\ 2.6, \ 3.0,\  4.6,\  5.9] $
show agreement to   within    $(20 - 10 )   \% $. 
To  assess the impact  of the deformed   conifold  geometry,  we  compare  the  reduced  masses  $\hat  x_m$,  defined    by   the formula $
E_m  =  \hat x_m w /\calr =\hat x_m M_\star w / z    $,   to the
corresponding  quantities    $x_m$   in the undeformed   conifold  case,   $ E_m  =  x_m w /\calr $~\cite{chemtob16}. Using   Eq.~(\ref{eqparams}),  one   finds the expression for  the effective dimensionless masses, 
\bea && {\hat x_m  }  = {2^{1/6} a_0 ^{1/4}  z  \hat E_m\over  
g_s ^{1/4}  (g_s M )^{1/2} \calv ^{1/6}   } \sim  ({K \over g_s M
})^{1/4}   \calv ^{1/3}  \hat E_m \sim \vert \ln w_s \vert ^{1/4} \calv ^{1/3}   \hat E_m ,\eea
indicating an   enhancement   effect
from  stronger    warping   or  larger    compactification volume,
relative to the (parameter independent) hard wall   model $ x_m $.     
We note for  orientation that  our  prediction  for the graviton   ground state mass,   $ \hat x_0 $,   agrees with   the  value~\cite{chemtob16}    $x_0 = 3.83 $  for $\calv  ^{1/6}  =5    ,\ g_s M = 1 $  and  $z= O(10)$.

\begin{figure}[t]
\begin{subfigure}{0.45\textwidth} 
\includegraphics[width=0.99\textwidth]{   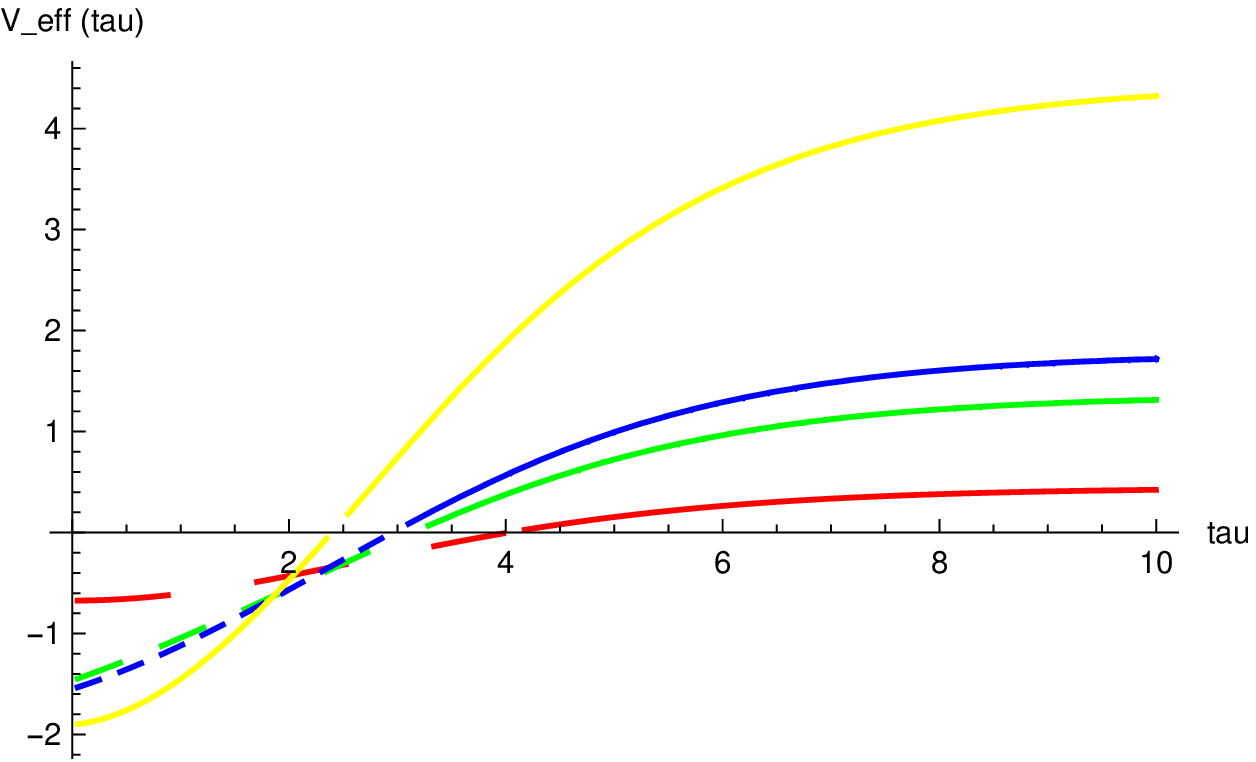}
\caption{Potential  profiles $ V_{eff}  (\tau  ) $.}
\end{subfigure} 
\begin{subfigure}{0.45\textwidth} 
\includegraphics[width=0.99\textwidth]{   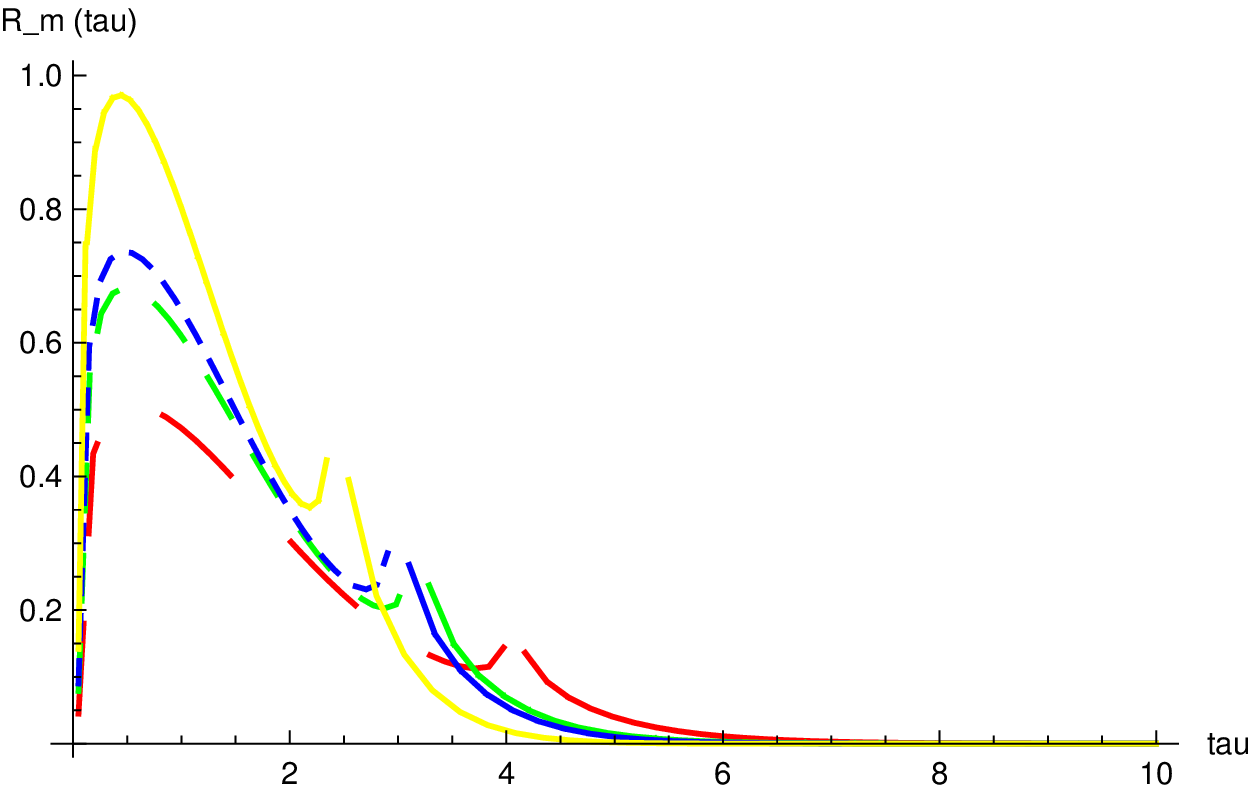}  
\caption{Wave functions profiles $ R_m (\tau   )  = B_m / \sqrt { G} $.} 
\end{subfigure}
\caption{\it  \label{GRwcofbp1}
The   radial effective potentials and  (normalized) WKB    wave functions   for   ground states  of   massive gravitons  are   plotted versus $\tau $.
The  red-green-blue-yellow coloured  curves  
with dashes of  decreasing   sizes  refer to the  singlet and  charged states 
$ h (000),\ C_1 (\ud \ud  1 ),\   C_2 (100)  ,\ C_3 (200)  $ of 
masses $ \hat E_m = (2.07,\ 3.07,\ 3.41,\ 4.96) $.} \end{figure}

The effective potentials  and  wave  functions   for  graviton modes are
displayed in   Fig.\ref{GRwcofbp1} for  illustrative  cases. 
The attractive  well regions in the potentials  stem    from 
the compensating  contributions of    the   negative warping term  
$ -\tilde g _{\tau   \tau } h (\tau )  E_m ^2  $   
and  the   positive curvature  term  $\calg _1 (\tau ) $,  which  dominate 
at small and  large $\tau $,   respectively.  
With  increasing  angular momentum, the  turning points   
move  to lower  values     $\tau _0 \in   (2-4 ) $. 
Although  the calculations    extend  over  the  complete 
interval $  \tau \in [0,  \tau _{uv}] ,$  the masses and 
wave functions   are     chiefly  sensitive to the 
inside      well regions    $ 0< \tau < \tau _0$. 
Deeper  and  narrower potential wells (with smaller 
turning   points  $\tau_0$)   and
more peaked wave  functions   occur  for modes
of larger  masses $ \hat E_m$.  The    flat potentials  
at     $\tau  > \tau _0$  causes the normalizable  wave functions to  decay 
   exponentially  at large  distances.     Note that the   small  discontinuities 
in the curves  of  wave functions   is an  artefact of  our
approximate matching prescription at turning points.   

\begin{figure}[b]
\begin{subfigure}{0.45\textwidth} 
\includegraphics[width=0.99\textwidth]{   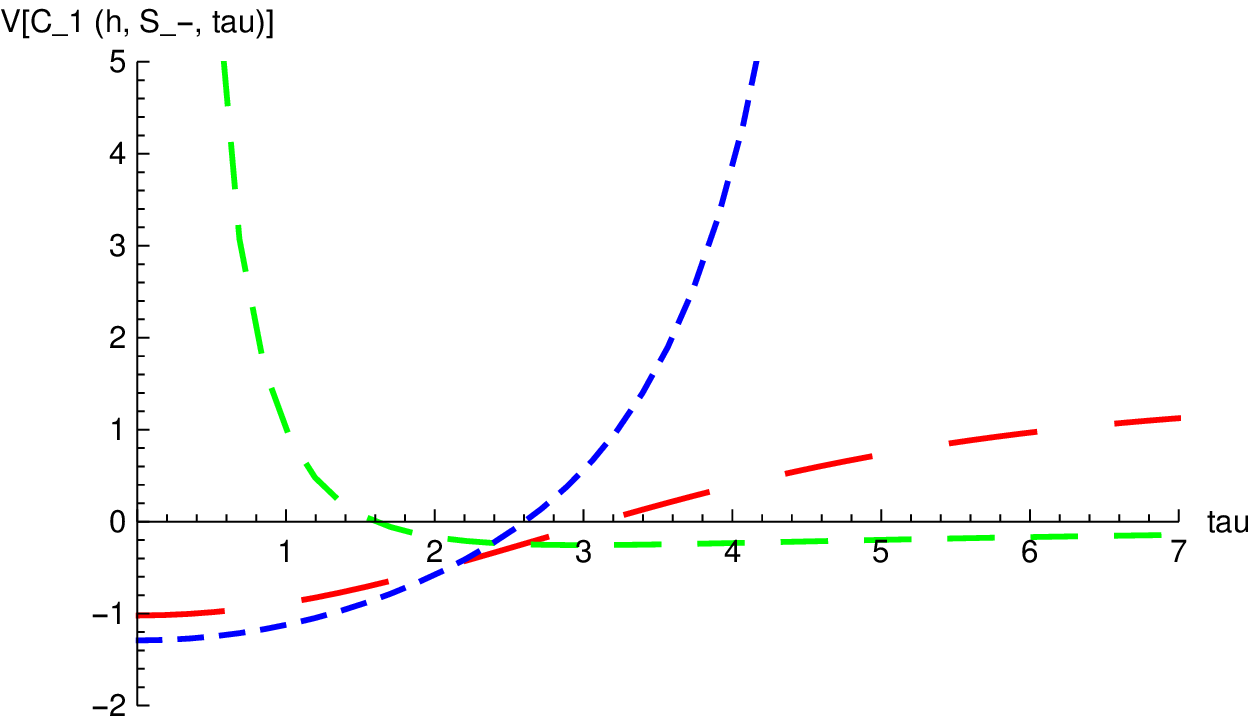}
\caption{$ C_1 (\ud \ud 1) $ mode.}
\end{subfigure} 
\begin{subfigure}{0.45\textwidth} 
\includegraphics[width=0.99\textwidth]{   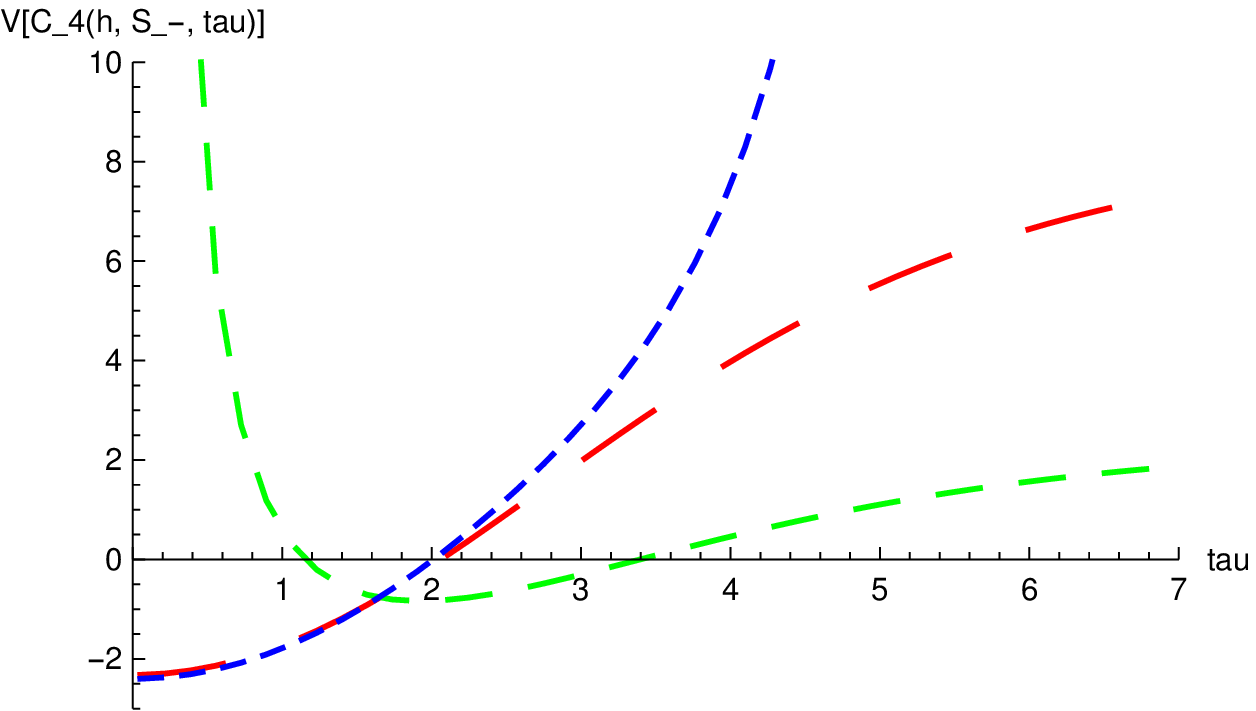}
\caption{$ C_4 (300) $ mode.}
\end{subfigure}
\caption{\it \label{GRwpC14tgds}
Plots versus $\tau   $ of the effective  potentials  $ V_{eff}   (\tau   ) $
for  charged  modes of the   graviton, scalar  $S_-$ 
and axio-dilaton   fields   given by the   red-green-blue  coloured   curves with dashes  of decreasing   sizes.}  
\end{figure}

The comparison   in    Fig.~\ref{GRwpC14tgds} of the  effective  potentials in  different modes  reveals  two  novel  features. Firstly,   the  inner well region for   axio-dilaton    modes is  similar  to that of graviton modes 
but  the  outer  region    receives    additional    contributions from     
3-fluxes.  The turning point location $\tau _0 $  is   significantly larger
for   lighter modes   and it grows   slowly with the radial excitation.  For the     $ C_1  $   graviton modes    of  radial charges   $m= [0,1,2] $,  
$ \tau _0  = [3.13, 4.59 , 5.57]$  and  for the axio-dilaton modes
$ \tau _0  =[2.62,    3.42, 3.84]$.  Larger values occur for 
scalar modes.  Secondly,   the scalar  modes $ S_- $  feel a    sloping potential  near the origin stemming   from  the divergent   term $\calg _1  \sim 1/\tau ^2$.  (This forces  the   choice $\d _{W}=0$
in Eq.~(\ref{sect2.eq10}).)   
 The  repulsive  wall    overwhelms  the attractive  contribution  from the
warping  term   and     the repulsive (or attractive) contributions  from
the 5-d   mass term   for $ M_5^2 >0$ (or   $ M_5^2 <0$). 
 The  resulting     inner  turning point     lies  typically at $ \tau '_0 \simeq   (0.5 \ -  \ 1.  ) $   for  $ M_5^2 >0$  and  at      $
\tau '_0 \simeq (1.  \ -  \ 1.5) $   for $ M_5^2 <0$,   while      
the  outer turning point    is   pushed   to larger values.
The  wide mass  gap    between  the  modes $ S_-( C_1) $ and $  S_- (C_4 ) $  is   explained by  the  difference  between 5-d masses ($ M_5^2 (C_1)  = - 15/4   ,\  M_5^2 (C_4) =  73/4 $).

\begin{table}   \begin{tabular}{|c |c| c| c|}\hline 
&  GRAVITON $ (h) $   &  SCALAR  $ (S_- ) $ &  DILATON  $(\tau )  $  \\
    Mode  $ (jlr) $  &   $  \hat E_{n=0,1,2,3, 4} $  &
 $ \hat  E _{n=0,1,2} $  & $  \hat  E_{n=0,1,2}  $ \\   
\hline  $  C_0   (0  0  0) $ &$ 2.08 \ \ 3.11\ \ 4.16 \ \ 5.20 \ \ 6.25 $  
&  ${1.90\ \ 3.06\ \ 4.17} $ &
$ {2.67\ \ 4.24\ \ 5.80}$  \\ \  $ C_1  (\ud  \ud 1) $  &$  3.07\ \ 4.10\ \ 5.14\ \ 6.18\ \ 7.22 $
&  ${0.66\ \ 1.56\ \ 3.04} $ &
$  {3.41\ \ 4.79\ \    6.24}$ \\ \  
$ C_2  (1 0  0 ) $ &$  3.41 \ \ 4.44\ \ 5.47\ \ 6.51\ \ 7.55 $ 
&  ${  1.58 \ \ 1.41\ \ 2.96} $ &
$    {3.69\ \ 5.02\ \ 6.43}$  \\
$ C_3 ( 2  0 0 ) $ &$ 4.96\ \ 5.98\ \ 7.01\ \ 8.04\ \ 9.07$  
&  ${2.20\ \ 3.34\ \ 4.44}  $ &
$ {5.11\ \ 6.27\ \ 7.51}$ \\
$ C_4( 3  0  0) $ &$ {6.56\ \ 7.57\ \ 8.59\ \ 9.62\ \ 10.6} $  
 &  ${4.17\ \ 5.22\ \ 6.26} $ &
$  {6.66\ \ 7.73\ \ 8.87}$  \\
$ C_5  (2  1 2) $  &$ {5.41\ \ 6.43\ \ 7.45\ \ 8.48\ \ 9.52} $ 
 &  ${2.79\ \ 3.89\ \ \ 4.96} $ &
 $    {5.54\ \ 6.67\ \ 7.88}   $  \\
 $ C_6  (1  1  2) $  &$ {4.07\ \ 5.10\ \ 6.13\ \ 7.17\ \ 8.20}$  
 &  $ {0.79\ \ 2.06\ \ 3.35}$ &
${4.28\ \ 5.52\ \ 6.86}  $      \\
 $ C_7  ( 1  1  0) $  &$ {4.27 \ \ 5.29\ \ 6.32\ \ 7.36\ \ 8.39} $ 
 &  ${1.00\ \  2.35\ \ 3.55} $ &
 $  {4.46\ \ 5.68\ \ 6.99}   $  \\
 $ C_8  ({3\over 2}  \ud  1 ) $ 
&${4.41\ \ 5.43\ \ 6.46\ \ 7.49\ \ 8.53}$
&  ${1.34\ \ 2.62\ \ 3.73} $ &
$   {4.59\ \ 5.80\ \ 7.09} $      \\
$ C_9 ( 2 1  0) $ &$  {5.55\ \ 6.57 \ \ 7.59\ \ 8.62\ \ 9.66} $ 
&  $  {2.96\ \ 4.06\ \ \ 5.13} $ &
$  {5.68\ \ 6.80\ \ 8.00}$  \\  \hline 
\end{tabular}
\caption{\it \label{GRtabxp1} 
Reduced masses  $\hat E _n $ for  warped modes  of the graviton  $h $
real scalar $S_- $   and axio-dilaton   $\tau $   fields\ \ appearing in
column entries,  for a selection of ten  singlet and    charged  states
modes  $  C_i , \  [i=0, \cdots , 9] $  appearing in
line entries. The    sequences of masses for each  mode refer to the  
ground state   and  first radial  excitations  labelled by $n$.
The  5-d squared masses of the listed modes  $ C_i (jlr),\ [i=0,\cdots , 9]$   are  given  for   gravitons and axio-dilatons  by  $M_5^2  (h ,\tau ) = H_0= (0,\ 8.25,\ 12,\ 36, \ 72,\  45, \ 21,\  24,
\ 26.25 ,\  48)      $  and     for    $ S_- $   eigenmodes  by      $M_5^2(S_-)  = H_0 + 16 - 8 \sqrt{ H_0+4}=(0 ,\  -3.75, \ -4 , \ 1.40, \ 18.25, \ 5, \ -3, \ -2.33, \ -1.75, \ 6.31)  $.}
\end{table}

\section{Interactions of  warped  modes}
\label{sect2p}

The    information on   warped modes interactions is  nicely encoded within  the supergravity  action in  Eq.~(\ref{sect1.eq1}). 
The coupling constants at  tree level are given by    the overlap integrals over $\calc _6$ of wave functions products.  We   start the discussion with the   cubic and higher order local  couplings of graviton  modes,   discuss  next  how deformations  of the classical   warp profile   
 $ e^{-4 A}$  caused by  the conifold  embedding in a  
compact  manifold~\cite{gandhi11}     affect  the  cubic  
couplings  and finally     consider   the      graviton
modes couplings   to pairs of  bulk 4-form scalar, axio-dilaton or
geometric  moduli  modes   and  of $D3$-branes  modes. 

\subsection{Couplings  of  massive gravitons}
\label{sect2psub1}

The graviton  modes    couplings   can be inferred  from the  perturbed  10-d  curvature   action,   
\bea &&  S  ^{(2)}  _{grav}     = { m_D ^8 \over 2} \int  d^4 x 
\sqrt {- (\tilde g_4 +h ) } \int d^6 y \sqrt {\tilde g_6 }  
e ^{-4 A(\tau )}  \tilde  R ^{(4)} (\tilde g_{\mu \nu }  +  h_{\mu \nu }  ), \eea
by substituting  the  decomposition for the
metric   tensor  fluctuations   on   the ground (real)
and  radially excited  (complex)   singlet and  charged states 
of   wave functions $\Psi _m (\tau , \T )$
\bea && h_{\mu \nu }  (X) = \d \tilde g_{\mu \nu }  =  \sum _m   
(h  ^{(m)} _{\mu \nu } (x) \Psi _m (\tau , \T )  + \bar h  ^{(m)} _{\mu
    \nu } (x)  \Psi  ^\dagger _m (\tau , \T )  ) . \label{eq.effag1} \eea   
The    expansion  in powers of the  4-d  mode fields  $ h_{\mu \nu } ^{(m)} (x) $
produces,   in addition to   the  classical term $( M_\star ^2 / 2) \int d  ^4 x
\sqrt {- \tilde g_4 } \tilde   R  ^{(4)} $ and the  field equation   term of linear order,   the modes kinetic  energy and  mass terms
along with   their  cubic and higher order mutual   couplings
which are expressed in the transverse-traceless   gauge  by
the  schematic  formula,  ignoring    the  spacetime   structure of
couplings,   \bea && \d  S _{grav}  = \int  d^4 x  \sqrt {- \tilde g_4 } 
 \int _x  ( \sum _{m } \l _2 \bar  h ^{(m) \mu \nu } 
(\tilde \nabla _4 ^2 -    E_m^2 )  h ^{(m)}_{\mu \nu }  +  \sum _{(m _i )} \l
'_3 h ^{(m_1)} \dh h ^{(m_2)}  \dh h ^{(m_3)} 
+ \l '_4 h ^{(m_1)} h ^{(m_2)}  \dh  h ^{(m_3)}   \dh h ^{(m_4)}  +
\cdots  ), \cr &&    
[\l _2=  {1 \over 4\kappa _{10} ^2 }= {m_D^8\over 4} = {M^2 _\star  \over 4 V_W }   ,\    \l '_n = {m_D^8 \calr ^3 J_n\over 8}  = {M^2 _\star  \calr ^3 \over 8 V_W }  J_n =\ud \l _2 J_n, \ 
J_n  = \int d^6 y  \sqrt {\tilde g_6 }  e ^{-4 A(\tau )} 
\prod _{i=1} ^n \Psi _{m_i } (\tau ,\T ) ].         \label{sect3.eq9}    \eea   
The normalization  condition   in Eq.~(\ref{sect2.eq7})  sets the  constant
$\l _2 $     and  ensures that kinetic terms are
diagonal  while the    numerical factor $1/2 $   is due to   the   
convention   of   summing  over pairs  of complex
conjugate  modes~\cite{chemtob16}.  The
$n$-point coupling  constants $\l '_n $    are  given by
overlap integrals  over the   conifold  volume  of  products  
of the  participating modes     wave functions. 
  The transformation to canonically normalized  fields, 
$ h ^{(m)} _{\mu \nu }   \to   h ^{(m)} _{\mu \nu }  / \sqrt {\l _2} ,
  $   replaces  the  $\l '_n $  by the rescaled   coupling  constants,  
  \bea &&   \l _n =  
\l '_n / \l _2^{n\over 2} = \ud \l _2 ^{1-n/2} J_n ,\ [1/\sqrt {\l _2 }  = 2 \sqrt {V_W} /M_\star  = 2 (2\pi )^3 L_W ^3  /M_\star =  2 (2\pi )^3 \eta ^3 \calr ^3 /M_\star ]  \eea    
which   are assigned   the energy    dimensions    $[\l _n ]= E^{2-n} $. 
We consider the expression for  the normalized    modes   wave  functions,  $\Psi _m ={B _m (\tau ) \Phi  _m /(A  \tilde G ^{1/2}   J )  } ,\ 
[G (\tau ) =\e ^{8/3}  \ 2 ^{-4} \ \tilde G (\tau ) ]$
explicitly exhibiting    the   normalization integral $ J  $.
It is then useful   to factor  out  the dependence on parameters  in the overlap integrals $  J_n$  and $ J $, hence  expressing 
the coupling  constants  for canonically normalized  fields
in  terms of   new    overlap integrals   $ \hat J_n$
and    $J_{(m_i)}$    as, 
\bea &&  \l _n = {1\over 2} \l _2 ^{1- n/2} J_{n}   =
(A^2 \e ^{8/3} 2 ^{-4})^{1- n/2}  {\hat J_{n}\over  \prod _{i=1}^nJ_{(m_i)} }  
=    ({2^{13/6} 3 ^{1/2} \over \e ^{2/3} g_s M \a ' \sqrt {\l _2 } }  )
  ^{n-2}   { \hat J_n  \over  2  \prod _{i=1}^nJ_{(m_i)} }  ,\cr &&
[\hat J_{n} =\int d  \tau d ^5 \T \sqrt {\tilde \g_5}   {I (\tau ) \tilde   G(\tau )   \over  K^2 (\tau ) } \prod _{i=1}^n
{B  _{m_i}   \Phi _{m_i} \over   \sqrt {  \tilde   G(\tau ) }
} ,\  J _{(m)}= ( \int d\tau {I(\tau ) \over K^2 (\tau ) }  
\vert B_m \vert ^2 \int d^5\T \sqrt {\tilde \g _5} \vert \Phi
_m \vert ^2   ) ^{1/2} ] .     \label{sect3.eq11} \eea  

We   recall that the  angular wave  functions  $\Phi
_m $  are     expressed in terms of trigonometric  type  polynomials  related to   Hypergeometric   functions~\cite{chemtob16}.
Only  the  $ \t _{1},\ \t _{2}$   polar angles
 integration  are non-trivial,  while those     over 
 azimuth  angles   $\phi _1,\ \phi _2 ,\ \psi  $    implement the selection rules on   magnetic  quantum numbers imposed  by the throat isometry,  
$\sum _{(i)}  m_j^{(i)} =0 ,\  \sum
_{(i)}  m_l^{(i)} =0  ,\  \sum _{(i)}   r^{(i)} =0 $ modulo $ Z $. 
This  suggests   factoring out the contribution
from  the three   azimuth angles in   $ d^5 \T $  and retaining only the
angular integrals,  $ \int  d^5 \T \to  \int  _0 ^\pi d  \t _1 \sin \t _1   \int _0 ^\pi  d \t _2 \sin \t _2  $,   in both    the    overlap and normalization integrals   which are   then denoted  with primes,  $ \hat J _n '  $ and $J '_{(m_i)} $. The resulting     formula  for  the  $n$-point coupling  constants reads 
\bea && \l _n = ( {2 ^{13/6} 3 ^{1/2} \over \e ^{2/3} g_s
  M \a ' \sqrt { \l _2 } } \sqrt { 4 \over V_{X_5} } ) ^{n-2}  {\hat J'_n  \over 2
  \prod _i^n  J'_{(m_i)} } ,\ [{\hat J_n  \over \prod _i J_{(m_i) }} 
=  ({V_{X_5}  \over 4})^{1-n/2} {\hat J _n '  \over \prod _i J '_{(m_i) } }  ,\ V_{X_5} = { 16  \pi     ^3 \over   27 } ] .    
   \label{sect3.eq16} \eea
   It    is finally convenient   to   trade the $\l _n$   for   reduced  dimensionless coupling constants   $\hat \l _n =  O(1) $,
   extracting out the  dependence on parameters by using the definition 
   \bea &&  \l _n =  ({\xi \eta ^3   \over M_\star w })^{n-2} \hat  \l _n , \ 
[\xi = {2^{13/6} 3^{1/2} \sqrt { 4 \over V_{X_5} } 
 \over  \e ^{2/3} g_s   M \a ' \sqrt    {\l _2 } } {M_\star w \over \eta ^3 }
  =  {2^{19/6} 3^{1/2} (2\pi )^3 ({\calr \over \sqrt {\a '}})^3  \sqrt { 4 \over V_{X_5} } }      ,\  \hat \l _n = { w ^{n-2}  \hat J'_n \over 2 \prod _i^n J'_{(m_i)} } ]  .  \label{sect3.eq16p} \eea     
The  approximate relation  for  the   ratio of 3-fluxes,  $  g_s M/ K \simeq {2\pi \over  3 }  c_2  =  
- 2\pi / (3  \ln  (w_s  \hat \calc  ^{1/2} / \calv ^{1/6} ) ), $
yields the  useful  formula  for the auxiliary parameter $ \xi $, 
 \bea &&  \xi =  5.1 \ 10^{3} \bar \rho ^{1/4} ({\calr \over \sqrt {\a '}})^3 ({ K \over  M g_s }) ^{3/4}
= {5.1 \ 10^{3} \calv ^{1/6} \over 2 ^{1/4} } (\l _N g_s ^{1/4} )^3  
(- {3 \over 2\pi }  \ln ( w_s  \hat \calc ^{1/2} / \calv
^{1/6} )  ) ^{3/4}  ,\     [\hat   \calc = 2 ^{1/3} a_0 ^{1/2} g_s M ]  \label{sect3.eq16q} \eea 
which is seen to depend  weakly on  the   flux and  compactification volume parameters  and  to have  a   logarithmic dependence  on the   warp factor. For   $ \calv ^{1/6} = 5 ,\ g_s M= 1 , \ w_s = 10^{-4} $ and 
$  \l _N g_s ^{1/4} = 1$, one finds  the   numerical  value, $\xi \simeq 7.3\ 10^{4} $, which is  significantly larger than the  natural   estimate   $ \xi =   O(1)$ assigned in~\cite{chen06}  and  lies   well above the
value $ \xi = 6 \sqrt {12}  \pi ^{3/2} \simeq  1.15\ 10^2 $
found in  the undeformed  conifold case~\cite{chemtob16}.
The    resulting  enhanced  couplings   for warped modes   
is a  manifestation of  the  softer infrared geometry of  the
deformed conifold, as we  discuss  at the end of Subsection~\ref{subsecthD3}.

The  couplings of   massless and massive  gravitons (denoted  below by $ g$ and $h$)   differ  significantly in size     due
to the  orthogonality   conditions on the   wave functions  and the  fact that 
massless gravitons have constant   wave functions. An examination of the overlap integrals shows  that the massless gravitons   couplings
  $ g^{m-2} \dh g \dh g$     are      independent of  $ w$,    
the  massive   and  mixed massless-massive gravitons
couplings behave  as $ h^{m-2} \dh h \dh h \sim 1/ w^{m -2}$
and    $ h^{m-2} \dh g \dh g \sim 1/ w^{m -4}$,  while  
the    single   massive graviton  couplings    $  h  g^{m-3}   \dh g \dh  g$  vanish.   We display in the table  below   order of magnitude  (dimensional analysis)  estimates   for  the  amplitudes   $  \l _{M, N}  $   and    reaction   cross sections $  \s
_{M, N}  $   of  the processes $ g^M,\  h^N,\  h^N g^M  ,\ [M, N \geq 2]$.
 In the pair  annihilation  cross sections, 
$ \s  (h^2\to h ^2)  \sim  ({ \xi / (M_\star w) })^4 
E_h ^2 ,\   \s  (h^2\to h g)  \sim  ({ \xi / (M_\star w )})^2  ({E_g /
  M_\star  })^2  \hat \l _3^2 ,$  the   energy  scale  factors
are set  at  the reaction energy $E_h^2  \sim  \max (E^2,
m_K^2 ) $    for    massive  initial state
modes         and at $E_h^2  \sim m_K^2$   for    massless  modes.
\begin{center}  \begin{tabular}{|c||c|c|cccc|} \hline 
Configurations   & $      g^M   $ & $   h^N $ & $   h g^M$ &  
$  h^2 g  $ & $  h^3g  $ & $     h^2 g^2 $   \\ \hline 
Coupling Constant  $\l _{M,N} $  & $ {2 ^{M-3}\over M_\star ^{M-2} }
$   & $    ({\xi \over M_\star w   }) ^{N-2}   $ 
& $    0  $ & $    {1\over M_\star}  $ & $ {\xi \over M^2_\star w } $ &
$   {2 \over     M^2_\star}$  \\  \hline 
Cross Section  $\s _{M,N}  $ &$ {E _g ^2 \over M_\star ^{2(M-2)} }   $& $ ({ \xi \hat  \l  \over M_\star w })^{2(N-2)}   E_h^2   $&$ 0  $
&  ${1 \over M^2_\star}  $ &  $ ({ \xi \hat \l _3^2 E_g
  \over M^2 _\star w })^2  $   &$  { 4  E_g^2 \over 
M_\star^4}    $   \\   \hline 
\end{tabular}\end{center} \vskip 0.2 cm

The approximate  matching   of    WKB   wave functions at turning  points
confronts us   with    a  technical  difficulty in the  numerical  evaluation   of  overlap  radial integrals  $\hat J'_n$. 
Since   the turning  points lie at    mode dependent locations,
a piece wise  decomposition of the radial interval $\tau $ is required for  modes  of different masses. In practice,  we  limit  the integration   intervals    for $ J'_{(m_i)} $   to the inner  region
$\tau \leq \tau _0 (E_{m_i } )$ and  account   for the
mode dependent   locations of turning  points  in $\hat J'_n$ 
by  limiting the interval of integration
to  $ \tau \in [0, \min  (\tau _0   (E_{m_i } ))$.
  The     reduced coupling constants  for the  3- up to 6-point   self interactions  in  allowed configurations $ \hat \l _3  (C_0 ) ^q  C_i \bar C_i  , \ [ C_0 = h, \ i= 1,\cdots , 9   ,\  q=1,\cdots , 4 ] $ are   listed in   Table~\ref{tabx2}.   We see that predictions  are   not   very sensitive  to   the   participating  modes charges  and that the 
  typical values for cubic  couplings $ \hat \l _3 = O(10^{-1})$ decrease
by  a factor $ 2\ - \ 3$ at  each unit incremental   step in
the  number of coupled modes,   $\d q = 1 $. 

\begin{table}  \caption{\it    \label{tabx2} 
Reduced coupling constants $  \hat \l _{2+q} $ 
for     3-, 4-, 5- and 6-order local   
couplings of massive  gravitons    in    configurations   
$(h )^{q}  C_i \bar C_i      ,\ [i=1, \cdots , 9]$  for  $q=1,2,3,4$
singlet  modes $ h$    coupled to    pairs of conjugate  
charged  modes  $ [h=  C_0 , C_1,\cdots , C_9 ] $  listed
in Fig.~\ref{GRtabxp1}.} 
\begin{tabular}{|c|ccc ccc ccc c|} \hline 
Couplings  & $ (h)^q h h $ & $   (h)^q C_1 \bar C_1  $    & $     (h)^q
  C_2 \bar C_2 $  &   
$    (h)^q C_3 \bar C_3 $ & $(h)^q  C_4 \bar C_4  $ & $ (h)^q C_5 \bar C_5$& 
$ (h)^q C_6 \bar C_6  $ & $(h)^q C_7 \bar C_7 $ & $    (h)^q C_8 \bar C_8$ & 
$   (h)^q C_9 \bar C_9 $   \\  \hline  
 $q=1$ &$ 0.128$&$ 0.158$&$ 0.165$&$
  0.191$&$ 0.209$&$ 0.197$&$ 0.17$&$ 0.182$&$
  0.184$&$ 0.198 $ \\ \hline
 $q=2$&  $0.0464  $&$ 0.0717  $&$ 0.0782  $&$ 0.108  $&$ 
     0.136  $&$ 0.116  $&$ 0.0917  $&$ 0.0972  $&$ 0.0988  $&$ 
    0.118   $ \\  \hline
$q=3$&  $ 0.00950   $&$ 0.0150   $&$ 0.0165   $&$ 0.0236   $&$ 
   0.0306   $&$ 0.025   $&$ 0.0197   $&$ 0.0210   $&$ 
   0.0214   $&$ 0.0261   $  \\  \hline
 $q=4$&$0.00405   $&$ 0.00653  $&$ 0.00723   $&$ 0.0106   $&$ 
    0.0140   $&$ 0.0116   $&$ 0.00871   $&$ 0.00932  $&$
    0.00951  $&$  0.0117 $ \\  \hline 
\end{tabular}    \end{table}    \vskip 0.3 cm

\subsection{Throat deformation by compactification effects} 
\label{sect2psub3} 

The modification of the  classical  vacuum  solution   ($\Phi _- =0,\ G_- =0$)
resulting from  embedding  the conifold in a   compact  Calabi-Yau  manifold  can  affect the modes couplings.   These  effects     are  amenable to a   perturbation theory description provided one restricts  to radial    distances   in the    throat   region   $ 0 < \tau  < \tau _{uv}$   intermediate between the horizon  and boundary where   deformations     are small.      Within the  AdS/CFT  duality approach of~\cite{gandhi11},   the  $x$-independent  fluctuations   of  the  various supergravity fields, $\varphi  (y) = [\d \Phi   _\pm (y)  =\d ( e ^{4 A} \pm \a (y) ) ,\ \d   G_\pm    (y), \d g   _{ab} (y), \   \d \tau (y)] $, are  split up  into  homogeneous     and inhomogeneous   parts. The  homogeneous parts    correspond to  zero modes of the  Laplace-Beltrami  wave operators  that one can  then    decompose   on   harmonic  functions    $ Y^{\nu } (\T ) $   of   the    conifold   base $ T^{1,1}$  times     radial  scaling functions  $ f_\nu (\tau ) $
obeying same  radial   wave equations as the  massless  warped modes.
Only the non-normalizable  (NN) solutions for $ f_\nu (\tau ) $,   which dominate in the  ultraviolet, need be retained. 
The  calculations    in   the large   $\tau $ region, far     from  the   conifold apex,   can be  conveniently  carried out in terms of  the  conic   radial  variable,  $r  =3^{1/2} 2 ^{-5/6} \e ^{2/3}   e   ^{\tau /3 } $. 
The   decompositions  of homogeneous parts on radial scaling functions,
\bea && \varphi (y) = \sum _\nu c^{NN} _\nu ( \tau _\star , \varphi )   f_\nu (\tau , \varphi )  Y^{\nu } (\T ) \simeq \sum _\nu  c^{NN} _\nu (r _\star , \varphi ) (r / r_\star )^{\D _\nu (\varphi ) -4 }  Y^{\nu } (\T ) ,\eea
introduces  the dimensions $\D _\nu (\varphi ) $  for    operators $ O_\nu  $  of  the  dual  conformal gauge theory
and the  (floating) coefficients   $ c_\nu (r _\star , \varphi)  $,  both    
depending  on  the field  $\varphi  $ type.  The operators $ O_\nu  $  are selected among  the class of    (gauge  invariant)  composite operators of   quantum number $\nu $  and the coefficients  are  expressed  in terms of their unknown  values  at the ultraviolet scale $  c_\nu (r _{uv} , \varphi) ) $,   using the  radial scaling  laws,  $ c_\nu (r _\star , \varphi)/ c_\nu (r _{uv} , \varphi )  = (r _\star  / r _{uv} ) ^{\D _\nu (\varphi ) -4 } $.
The  coupled field    equations   obeyed by  the  inhomogeneous source and mixing type fields fluctuations     are solved iteratively by  expanding the   $\varphi  (y) $   in powers of  the small  ratio  $ \tau _\star / \tau  _{uv}$ (matching radius  $ \tau _\star $  over ultraviolet  cutoff   radius)  and the  warp factor $ w $. Matching the solutions to small field deformations 
at the ultraviolet boundary  $r _{uv}$  yields  linear equations for the  constant   coefficients  in these expansions  that can be solved algebraically in terms of the $O(1)$    coefficients $  c_\nu (r _{uv} ) $  describing the homogeneous parts.  In practice, the  coupled  system   of  differential equations  for inhomogeneous   parts is of  small   dimensionality because the  leading  operators of  lowest   dimensions   are  few in numbers. 

Since the  coupling  constants  $\l _n $ of gravitons  local interactions  are   evaluated  from   overlap integrals of the  participating modes wave functions   weighted  by the warp profile,   as  in   Eq.~(\ref{sect3.eq9}),   the  leading   corrections  should  arise     from deformations 
of the warp  profile,  $ \d  e ^{-4 A }  = 2 \d  (\Phi _ - + \Phi _+ ) ^{-1}   $. 
Based on  the    available classification~\cite{ceresole99,ceresoII99}
of operators  of  the  dual superconformal gauge   theory  
in terms of  the dimensions    and   supersymmetry
character of deformations,  the    dominant contributions are  those  induced  by  fluctuations  of  the field     $ \d \Phi _-$ which corresponds to the
scalar modes  $ S_-$. The perturbed  warp profile 
is  then   expressed at  large radial  distances by a sum over
radial scaling terms,  
\bea &&  \d (e ^{-4 A(\tau , \T ) } ) \simeq e^{-4 A_0  (\tau ) } \sum _\nu  c^{-} _\nu (r_{uv} )  w ^{ Q ^- _\nu }  ({r \over \calr }) ^{\D  ^- _\nu
  -4 }   Y^{\nu  } (\T ) ,   \  [\D  ^- _\nu = -2 + (4 + H_0  ^\nu )^{1/2} ] \label{eq.defprf}  \eea  
where  the ultraviolet  coefficients   $ c^{-} _\nu (r_{uv} ) $  of $O(1)$  are    corrected   by  warp  factor powers $ w^{Q ^- _\nu } $  with   index parameters  $Q ^- _\nu  \geq  0$  set by the  supersymmetry  breaking character of the  initial  background   produced by  the gauge theory operators of  dimensions
$\D  ^- _\nu  $.   The  $F$-  or $D$-type operators in   the  highest superspace components ($\t ^2,\ \t^2 \bar \t ^2$) of   chiral or vector   supermultiplets  are unaffected  ($ Q ^- _\nu =0$) while those in  lower superspace   components    are assigned the  power index $ Q ^- _\nu =  2 $ or 4.  Details on the    notations     are provided in~\cite{chemtob16}.   The  deformation effects on
$ e^{-4 A (y)} $    can be taken into account  by  inserting   inside the
overlap integrals, denoted $J_n $   in   Eq.~(\ref{sect3.eq9}),
the spurion  modes  wave functions $ f_\nu  (\tau ) Y^\nu (\T )  ,\  [f_\nu  (\tau ) \simeq   r ^{\D _\nu ^- -4 } ]   $.   These effects   modify  the  selection rules      imposed  by     the
throat  isometry   in    an easily identified way  by allowing  otherwise     forbidden couplings. The   reduced  dimensionless coupling   constants   for the  deformed   couplings  $\hat \l _n $ can  be  defined
in a similar     fashion as   Eq.~(\ref{sect3.eq16}), 
\bea &&  \l _n = c _\nu w ^{-4 + \D  ^- _\nu   +  Q ^- _\nu }   
\l  '_n  ,   \ \ 
\l  ' _n   \simeq   ( 5.1 \ 10^3    { \eta ^3 (N /M^2 )^{3/4}  \bar \rho ^{1/4}  
  \over w M_\star g_s^{3/4}   } )^{n-2} \hat \l _n  .\eea 
The   leading  contributions   to $\d \Phi _-$  
arise  from supersymmetry  breaking operators  in lowest components 
of  vector supermultiplets,   $  Q ^- _\nu = 4$,  whose    
dimensions $\D ^- _\nu $  increase with the  modes charges. 
The coupling  constants of  the cubic interactions 
$(h h C_i ), \ [i=1,\cdots , 9]  $  which  vanish
in   the undeformed  background case,  take  the finite values  displayed 
in Table~\ref{tabDEF}.   

\begin{table} \caption{\it \label{tabDEF} Coupling constants
of  the    cubic  local  couplings $  h  h C_i ,\ [i=0,
  1, \cdots , 9] $   induced through   the  warp profile perturbation $ \d
(e ^{-4 A   (r,\T ) } )$   due to compactification     effects.
The deformed coupling constants  $ \hat  \l
_3  $  are  given by   overlap integrals involving  products   of    wave
functions for  the  participating
modes     and  for the  spurion    harmonic modes   
$\d \Phi _- ^ {\nu  _i} ( \bar   C_i) $    needed  to neutralize the
total charge.  The warp  factor  dependent factors 
$ w ^{-4 +\D ^-_\nu  + Q ^-_\nu } $   are  determined  
by the  supersymmetry breaking parameter,  which is
set at $Q ^-_{\nu _i}  =4 $ in all
cases,    and   the   dual  gauge theory operator dimension
$\D  ^-_{\nu  _i} (S_-)$   which is mode dependent.  
The  unknown   constant coefficients   $\ud \hat c ^-_\nu =    O(1)$
were  factored out. The  power  index of   $w $
is determined  by the dimensions,    by $\D  ^-_{\nu  _i} (S_-) = -2 + (H_0 +4)^{1/2} =  (0, \  1.5, \  2, \  4.32, \  6.71, \  5, \  3, \  3.29, \  3.5,  \  5.21)  $.} 
\begin{tabular}{|c|c|ccc ccc ccc|}  \hline 
 Couplings & $ C_0 C_0 C_0 $ &  $ C_0 C_0 C_1 $ & $ C_0C_0 C_2 $ & $ C_0C_0
  C_3 $ & $ C_0C_0 C_4   $ & $ C_0C_0 C_5 $ & $ C_0C_0 C_6 $ & 
$ C_0C_0 C_7 $ & $ C_0C_0 C_8   $ & $ C_0C_0 C_9 $  \\ \hline
$ \hat  \l _3  / ( -\ud \hat c ^-_\nu  ) $ &  $ 0.00761  $& 
$  0.0180 w ^{3/2}   $  & $ 0.0235 w^2   $ & $ 0.0747 w^{4.3}   $ & $
0.208  w^{6.71 }  $ & $    0.101  w^{5 } $ & $  0.0394  w^{ 3 }
$ & $0.0458  w^{3.29 }  $ & $0.0505  w^{3.5 } $ & 
$ 0.111  w^{ 5.21 } $  \\ \hline
\end{tabular}   \end{table}    \vskip 0.3 cm

\subsection{Gravitons  couplings   to  bulk  scalar  modes}
\label{sect2psub2} 
 
The interactions    between  different types  of warped  modes  
can be   computed    at   tree level  by applying   the familiar  
perturbation theory rules  to the action in    Eq.~(\ref{sect1.eq1}).
The    metric  tensor   field couplings  are  encoded
within   the universal  type   operator,     $ h^{(m) \mu \nu }  T_{ \mu \nu }  $,    proportional to  the  energy-momentum stress tensor.
For  a  single graviton  mode $ h^{(m)} _{\mu \nu } $
coupled to pairs of  scalar  modes  $b ^{(m)} (x)$  from the 4-form  field in Eqs.~(\ref{sect2.eq11}) and~(\ref{sect2S.eq13}), the effective Lagrangian for 
canonically normalized  fields  is given by, 
\bea &&   L_{EFF} = \l  _{m_1m_2 n}  ^{b}  h^{(n) \mu \nu } \dh _\mu b   ^{(m_1)  \dagger }  \dh _\nu b ^{(m_2)} ,\cr &&
     [\l  _{m_1m_2 n}^{b} = {1\over  2 \sqrt {\l _2 } }
       {\int d\tau  d ^5 \T {\sqrt {\tilde g_6 } e ^{4A}  Q(\tau ) \over  \calg (\tau ) G^{1/2} (\tau ) }     \tilde b _{m_1} \tilde b _{m_2}
B _n (\tau ) \ \Phi _{m_1}  ^\dagger   \Phi _{m_2}    \Phi _{n}
    \over  ( \prod _{i=1,2}  \int d\tau  \int d ^5 \T
  {\sqrt {\tilde g_6 } e ^{4A}  Q(\tau ) \over \calg  (\tau ) } \vert \tilde b _{m_i}  \Phi _{m_i} \vert ^2 )^{1/2}  (\int d\tau d ^5 \T
  {\sqrt {\tilde g_6 } e ^{-4A}   
    \over G   (\tau ) }  \vert  B _n \Phi _n \vert ^2   )^{1/2} } ].  \label{sect4.eqGS1}    \eea  
     Although the wave functions   of scalar modes  $ S^-_m  $  differ widely  from  those  of gravitons,  we  anticipate that  this is compensated by  the   different measure  factor in  overlap  integrals. The  expectation that  the strengths are comparable to those  of gravitons   self couplings,  $\l _3 ^{b, t}   \simeq \l _3 $,   can be   verified   by  analyzing  the overlap integrals at   small $\tau $  and is  also  borne out from   the hard wall  model case~\cite{chemtob16}. 

The couplings  of graviton modes to  pairs of
axio-dilaton  mode fields  $   t ^{(m)}  (x)$,   of wave equations given by Eq.~(\ref{sect2.eq15}),   have  same   overlap integrals as those 
for   the  graviton modes   self couplings in Eq.~(\ref{sect3.eq11}).
The  effective Lagrangian  for    trilinear couplings of
canonically normalized  modes   is    given by 
\bea && L_{EFF} = \l  _{m_1m_2 n}  ^{t}  h^{(n) \mu \nu } \dh _\mu t ^{(m_1)  \dagger }  \dh _\nu t  ^{(m_2)},\cr &&
 [\l  _{m_1m_2 n}  ^{t}    =  {1 \over  2 \sqrt    
 \l _2} {\int d  \tau  {I(\tau )\over K ^2 (\tau )} 
 \tilde t_{m_1}    \tilde t _{m_2}   B_n (\tau )   \int   d^5 \T   \sqrt {\tilde \g _5}  \Phi _{m_1}  ^\dagger  \Phi _{m_2} \Phi _{n}
\over   \prod _{i=1,2}  (\int d\tau d ^5 \T {\sqrt {\tilde g_6 } e ^{-4A}   
  \over G   (\tau ) }\vert \tilde t _{m_i}  \Phi _{m_i}  \vert ^2  )^{1/2}
(\int d\tau d ^5 \T   {\sqrt {\tilde g_6 } e ^{-4A}  
    \over G   (\tau ) } \vert   B_{n}  \Phi _{n}  \vert ^2 )^{1/2} }  ]  \label{sect4.eqGS2}  \eea  
 where $\l ^t $  are expected to  be  comparable    to  those  of
 graviton  modes  self couplings.
 The   interactions of   the geometric (complex  structure
 and K\"ahler) moduli can be  inferred     in a similar    way starting from  the    reduced kinetic  action, as   discussed in~\cite{chemtob16}.
 Nevertheless,  the information on the moduli wave functions is still  uncertain  in spite of   the insights   provided  in  the initial studies~\cite{dougba07,dougtorr08,freyroberts13}.  For instance, the  K\"ahler metric    for the  complex structure modulus  $S$
 is found to   acquire  a    divergent  contribution
 $ G_{S\bar S} \sim 1/ S $  near the point  $S=0$  of the moduli space where the deformed  conifold   3-cycle   collapses~\cite{dougba07,dougtorr08}.
 The predictions for  radial profiles    of the universal  volume  modulus   mixing  with the metric 
 tensor, $ c(x) - g _{MN} (x) $,  derived  in  the deformed conifold
 within the  gauge compensator approach~\cite{frey08}, $ \d  _c g_{\mu \nu  }\sim r^2 / N^{1/2} ,\ \d _c g_{rr } \sim N^{1/2} /r^2 $,
differ   from the  naive classical  estimates  inferred
 from  expanding   the  warp profile  ansatz    at large  radial  distances,   $\d _c g ^{(0)} _{\mu \nu }  \sim r^6 / N^{3/2}  ,\ \d _c g^{(0)} _{rr } \sim r^2 / N^{1/2}  $.

\subsection{Couplings  of gravitons to  $D3$-branes}    
\label{subsecthD3}

We consider  next the   interactions   between  bulk  and brane fields
mirrors the  string theory couplings between  closed  and open strings.
The  scalar  and  Majorana-Weyl  spinor massless fields, 
$X^M(\xi ) $ and $\T (\xi )  $,      describing  the   $Dp$-brane
world volume  embedding  in   super-spacetime $ M_{10}$, couple  to  the  pull-back    transforms  of the    bulk supergravity  multiplet fields.
The   action principle    for   superbranes  is
formulated using  the invariance  under  diffeomorphisms   of  the  world volume  $ M_{p+1} $  intrinsic  coordinates $\xi ^\a ,\  [0,   1,\cdots ,   p]$   and local  Lorentz-Poincar\'e,  supersymmetry  and   fermionic  $\kappa $-symmetry   groups.  We   shall    consider    $ D3$-branes    located at points of  $ X_6$ using the  general       formalism   developed in~\cite{marino99,gauntlett03,martucci05,lustmartsim08}  and  applied in~\cite{marche08,marche10,chemtob16}.
The  couplings     of massive  graviton  field modes  $ h ^{(m)}  _{\mu    \nu } (x)  $  to bosonic and fermionic massless field  modes $ \varphi ^m = X^m / (2\pi \a ' ),\   \psi ^m  $  of  $D3$-branes  
are set by the graviton wave function  value
at the brane location, $ y  = y_\star  = (r_\star ,\ \T  _\star )$. 
The    dimension $ 5$ effective Lagrangian 
contributed by the Born-Infeld   action~\cite{chemtob16}  for  fields   of canonical kinetic energy   is   given  by
\bea && \d S (D3) =  -  {\l _3 ^B \over \sqrt 2 }  
\int d ^4 x \sqrt {\vert \tilde  g_4 \vert } \
h _{\mu \nu } ^{(m)}  (x) \tilde g _{np}   
( \dh ^\mu \varphi ^n \dh ^\nu \varphi ^p + i \bar \psi ^n \dh _\mu \g
_\nu  \psi ^p ) ,\  \l _3 ^B = \sqrt 2 {2 ^{13/6} 3 ^{1/2} \over
  g_s M \a ' \e ^{2/3} \sqrt { \l _2 } }   \Psi _m (y_\star ) 
,\cr && [{1 \over \sqrt { \l _2 }} = {2 \sqrt {V_W} \over M_\star } = {2(2\pi  L _W )^3 \over    M_\star } ,\ \Psi _m (y_\star )  ={B _m (\tau_\star   ) \Phi  _m (\T _\star )\over \tilde G ^{1/2}   J_{(m)}   }  ] .\label{eq.reducc1} \eea
One  can use the  familiar definition  of the  effective
action for a  graviton  field $h_{\mu \nu }$  coupled to a  scalar matter field $\phi $ to  evaluate the two-body  decay rates   of gravitons, 
using the identification, $1/ M_\star    \to   \l _3 ^B  $,
\bea &&  \d S (D3)= {1\over M_\star \sqrt 2 } \int d^4 x  h^{\mu \nu }
T_{\mu \nu } (\phi )  \ \Longrightarrow \  \D \G  (h \to \phi + \phi )
\simeq  {m_h^3  \over 960 \pi M_\star ^2 } (1 - {4 m_\phi ^2 \over m_{h}^2 })^{1/2}.  \label{eq.DEC}  \eea 
It is    convenient  to  define a reduced  dimensionless  coupling   constant  $ \hat \l _3 ^B  = O(1) $,  similar  to the prescription  used previously for bulk   modes in Eq.~(\ref{sect3.eq16p}),  
\bea &&  \l _3 ^B= {\xi ^B \eta ^3 \over M_\star  w } \hat \l _3 ^B ,
\ \   [\hat \l _3 ^B= {w B _m (\tau _\star   ) \Phi  _m (\T _\star ) \over \tilde G ^{1/2}   (\tau _\star   ) J'_{(m)}   } , \
\xi ^B =  \sqrt 2 {2^{13/6} 3^{1/2} 2 (2 \pi )^3 \calr ^3    \over  \e ^{2/3} g_s   M \a '  }  \sqrt { 4 \over V_{X_5} }    = \sqrt 2 \xi  ] \eea
with  $\xi ^B = \sqrt 2 \xi  \approx 10^5.$
For $ D3$-branes  near the conifold apex $\tau  _\star =0$,
the  numerical values   for $\hat \l ^B_3 $   are  listed  in  Table~\ref{tabD3}  for  four  cases  associated to different locations
in the  base  manifold. The   variations   between Cases $I, II, III $ reflect on the    dependence of  charged graviton wave functions $ C_i$ on  the base manifold angles. Note that the harmonic  wave functions  for modes $ C_{1, 5, 6, 8}  $  of  charge  $  r _i\ne 0$    vanish at the p\^oles $\t _{1,2} =0$
and those of    singlet modes are  angle independent. 
In the smeared distribution  Case $IV$, the
couplings   have the anticipated     smooth   dependence on   modes charges. 

The effective 4-d  gravitational  mass scale for   $D3$-branes  localized at
$\tau _\star  $,  usually defined by the ratio $\L _{KK} = M_\star /\Phi _m (\tau _\star )$,  can   also  be evaluated  from the formula,   $\L _{KK} \simeq 1/\l _3 ^B  = M_\star w /(\xi ^B \hat \l _3 ^B \eta ^3) $.
For   $ \xi ^B= 10^{5} ,\   \hat \l _3 ^B = 10^{-2} ,\ w= 10^{-14} ,$
one finds   the   rather small  value  $\L _{KK} \sim 10^{-2} / \eta ^3 $ TeV.
The  further   strong   suppression from $\eta  >> 1 $  might  be
compensated  if  the brane   were  located  at a finite distance from the tip.   The predicted mass  scale  lies  well  below  that  found  in  studies
using Randall-Sundrum  model, $ \L _{KK} \approx 51 $ TeV,  
the undeformed  conifold background~\cite{chemtob16}, $ \L _{KK} \approx  19$ TeV, and  the softened warp profile model~\cite{shiuetal07,guirkshiuzur07}, $ \L _{KK} \approx  2 $ TeV,  for the same value  of $ w $.

\begin{table}  \caption{\it    \label{tabD3}  
Reduced coupling constants for    bulk graviton modes $ h,\ C_i
,\ [i=1,\cdots , 9]  $   coupled  to pairs of $D3$-brane
modes in   four distinct  choices  for  the      
brane    location  in the  base  manifold,   all using  
vanishing azimuthal angles    $ \phi _1 =
\phi _2 = \psi =0$.    Case  $I$ refers
to polar angles $  \t _1 =0,\ \t _2 =0$,   Case   $II $ to $ \t
_1 =2 \pi /5 ,\ \t _2 =0 ,$ Case  $ III $ to
$\t _1 = \t _2 =  \pi / 5 , $ and Case $ IV $
to  smeared distributions    averaged over $\t _1 $ and $\t _2 $.}   
\begin{tabular}{|c||ccc ccc ccc c|} \hline 
& $h $ & $C_1 $ & $C_2 $ & $C_3 $ & $C_4 $ & $C_5 $ & $C_6 $ &
$C_7 $ & $C_8 $ & $C_9 $ \\ \hline 
  $\hat \l _3 ^B (I) $  & 0.022  & $0.39\ 10^{-4}$  &0.077 
  &$0.15 $  &$ 0.25 $   & 0.     &$ 0.$   & $0.18 $    & 
  $8.  \ 10^{-4} $ & 0.30  \\  \hline
  $\hat \l _3 ^B (II) $  & 0.022 & $0.32 \ 10^{-4} $    &  0.023 &
 0.056   &  0.10    &0. 
 & $0. $   &$ 0.055  $& $0. $& 0.11  \\  \hline
  $\hat \l _3 ^B (III) $  & 0.022 &  0.053    & 0.062
 & 0.075  &  0.027   & 0.08    & 0.075     & 0.    & 0.083    &   0.   \\  \hline
 $\hat \l _3 ^B (IV) $ &  0.022   &  0.035   &   0.038   &
0.060   &   0.084    &   0.059    & 
0.041   &   0.044     &   0.052     & 0.060   \\  \hline
\end{tabular}  \end{table}    \vskip 0.3 cm

\section{R\^ole of  warped modes in early universe  cosmology}  
\label{sect4}

The    discussion of   $D3$-branes moving in    Klebanov-Strassler type  background  deformed  by  the presence  of  $\bar D3$-branes near the deformed conifold apex has  provided useful insights on  the slow roll  inflation scenario~\cite{kklt03,baumann06,baumann08}. 
The  energy released  through   $D3 -\bar D3$-branes annihilation
is assumed to  produce    at inflation exit
massive  closed strings   fastly  decaying
to   massless    closed strings~\cite{lambert03}  that
produce  massless particles  and massive Kaluza-Klein  modes~\cite{kofman05,chialva05,chen06,kofman08}.  Based on the
information     collected  so far,   we  wish to examine    whether   the  gas of warped   modes  present   in the throat  could   provide an  attractive  mechanism  for the   post-inflation universe reheating in this   context.
The  resulting   system   of  multiple    species of  metastable  particle   coupled by gravitational interactions at   effective  scales  lying  well   below the  Planck mass scale  should hopefully  have a predictable thermal  evolution.

\subsection{Preliminary considerations}  
\label{secT0} 

We assume  that the exit from inflation  leads  to 
a   Friedman-Robertson-Walker (FRW)  universe  filled    by a   gas of relativistic  Kaluza-Klein warped modes  localized in  the inflationary throat,
called hereafter $A$-throat. The radiation dominated  regime is  characterized  by  the scaling laws for the temperature,    Hubble  rate and energy density as a function of cosmic time,  $T \propto    a^{-1} \propto     t^{-2}  ,\  H\equiv  \dot a (t) / a = 1/(2 t   )  ,\ \rho = 3 M_\star ^2 /(4 t ^2) $. We  shall use the simplified
formulas   for the   particles  masses    $ m_K = x_m  w /\calr $  and      couplings $  \l _n  h^{n-2}  \dh h \dh h ,\  [\l _n \propto 1/ (M_\star w )^{n-2} ] $  depending   on the $A$-throat warp factor and  curvature radius parameters,
$ w $  and $\calr $. 
The  statistical   number  of   degrees of freedom,   counting
the  number of  degenerate       harmonic modes in $ T^{1,1}$,  is described  by
the temperature dependent Chen-Tye   ansatz~\cite{chen06}, 
  $g^b_\star (T) = \t (T -  m_K ) (T/ (w /\calr ))^
\g  ,\ [\g =5 ]$.  In case   the  $A$-throat    also hosts  Standard Model $D3$-branes,     this is combined with  (or replaced  by)
 the  (constant)    number of     massless   brane modes, $ g ^{SM} _\star (T) \sim O(100) $.

The   initial  temperature   and  cosmic time $ T_I,\ t_I$  can be determined  
by  matching the   energy density   
$\rho _K (t)  = (\pi ^2 g_\star / 30)  T   ^4 (t)  $     of the relativistic   gas to  that  deduced from the     Hubble  expansion rate at
inflation exit,    $ \rho _I   (t)  = 3 M_\star ^2H^2(t) ,\ [H(t) = 1/(2t) ] $.
Assuming that the energy  released through  either $ D\bar D$-brane  annihilation at time  $ t_I  $, or massive closed strings   decays  at time $t_I+ \D t_{S} $,   is efficiently  transferred to    warped modes,
one obtains the  balance      equation, 
\bea &&\rho _I (t) \equiv  {3 M_\star ^2 \over 4 t^2 }  =  
\rho _K (t)\equiv {\pi ^2 g_\star
  \over 30} T  ^4  (t),\ [ t= t_I +\D t_{S} ].  \label{RHtemp1}\eea 
The   temperature at  time $ t_I$  can be explicitly determined in    two  limits   which we   examine   in turn    for  a constant $ g_\star (T) $.  
If the energy     from    by $ D3-\bar D3$ annihilation
 is transferred  instantaneously,  
 $\D t _{S} =  1/\D \G _{S}  <<  t_I  $, then    equating $ \rho _K (t_I) $  to  
$ \rho _I(t_I) \simeq  2 N_A \tau_3 w ^4  
= {4\pi  N_A w  ^4 / \hat l_s ^4  } $,  where $ N_A $ is the   5-form
flux  supported by the  $A$-throat,  can be used to    determine the initial temperature  and time,  
\bea && T_{I}  =  ({90 \over 4\pi ^2 g_\star })^{1/4}
\sqrt { M_\star  \over   t _I }   \simeq  ({ 2 N_A  \tau _3 w ^4 \over 3
M_\star ^2  } )^{1/4}   \ \Longrightarrow  \ 
{T_I \over M_\star } \simeq  0.51 \  w  ({N_A
  \over  g_\star (\eta z )^3})^{1/4}
\simeq    1.57  \ 10^{-7}  N_A^{1/4}  {w \over 10^{-4}  }
({10^4    \over  \eta z } )^{3/4}   . \label{eq.teI}  \eea
If the  decay  lifetime  is larger  that the Hubble time,
$ t_I  <<  \D t _{S }   $,        then  using   the estimate for the
total   decay rate of   massive
closed strings~\cite{chialva05},  $ \D \G _s (S\to K+K)
=  c '_{KK} g_s  w  m_s ,   $ one can evaluate the reheat temperature,
\bea && T_{RH} \simeq   ({90  \over  4 \pi ^2 g_\star })^{1/4} 
\sqrt { M_\star \D \G _S }  \ \Longrightarrow  \ 
{T_{RH} \over M_\star }  =    ({90 \pi ^{1/4} \over 4 \pi ^2  }) ^{1/4}  
{ ( c '_{KK} g_s  w  ^{1/2} \over g_\star ^{1/4} (\eta z   )^{3/8} }
\simeq   0.31  \ 10^{-6}  ( {w \over 10^{-4}  }   {c ' _{KK}\over 10^{-4}  } )^{1/2}   ({ 10^4 \over  \eta z })^{3/8} .\label{eq.tRH}  \eea 
The numerical results  in Eqs.~(\ref{eq.teI}) and~(\ref{eq.tRH})     were obtained   for $ g_s =0.1,\ g_\star =100 $  setting $ c' _{KK} = 10^{-4} $~\cite{chialva05}  which is numerically close  to the estimate $c '_{KK} \simeq g^3_s$~\cite{kofman05}. 
The two  predictions  for the initial  temperature
are sensitive  to  the warp  factor,  have acceptable magnitudes and
satisfy the relationship,  $\sqrt {T_I} /T_{RH} \approx  1.3 \ 10^{3}  N_A ^{1/8}  $.   Had one    used  instead the  statistical factor ansatz,   $ g_\star (T) = (T / \a  )^{\g} ,\  [\a = w  /\calr ] $,   the  resulting 
reheat temperature,   expressed in terms of  that  for   $g_\star = 100$  as, $({T_{RH} \over M_\star} )   \simeq (\sqrt {10}({w   \over z  }) ^{\g \over 4  ( \g +4)}  )   ({T_{RH} \over   M_\star})^{1 \over \g +4}$,
reaches  the   excessively   large   value,  $ ({T_{RH} \over M_\star} )  \approx  10^{-5/2} $  for $z= 10^3, \ \g =5  $. 

We wish  to   study  in this Section   the  thermal evolution of   warped modes 
in the   twofold goal   of   determining  their  ability
to  thermalize   and   leave     
a cold thermal  relic  component. The answer clearly    depends  on
whether the Standard Model branes setup is located 
in the  $A$-throat  or, if  a fraction of    modes   can  stream out   of  the  $A$-throat,  in the  Calabi-Yau  bulk    or   in  another  distant 
throat.  We  consider two  main   cases along same lines as~\cite{chen06}.   The   single throat case   in
Subsection~\ref{subT1} deals  with  an  $A$-throat  accommodating
both inflation  and   the  Standard Model and the  double throat case   in
Subsection~\ref{subT2}   assumes the existence of   an additional $S$-throat hosting  the Standard Model  branes  near  its apex.

A few  preliminary  remarks   are in order before  moving on to the main discussion. In  the context of 10-d   supergravity, the   cosmic  bath
consists  of  infinite towers    of massive particle species     
differing   by  the  spin  $\leq
2$,  the charge  under   the  throat isometry group and the    radial
excitation.   The tree level      action includes
cubic  and higher order  couplings   that
can  induce   decay  channels for    most     modes. 
The fate    of   massive modes  depends  on   how their   
decay rates $\G  _{1\to 2} $ compare to  the  Hubble  expansion   rate
$H( t )$.     For instance, the  decay  rates   for  massive
gravitons   with  $\calf $  open channels of conjugate  pairs of modes $\phi $, inferred from  Eq.~(\ref{eq.DEC}),  
\bea && {\D \G   (h\to \phi +\bar \phi ) \over   (1    - 4  m_{\phi }^2 / m_{h}^2
  ) ^{1/2} }  \approx  {\l _3  ^2    m_{h } ^3  \calf   \over 960 \pi }  
\simeq   {x_m^3 \xi ^2   M_\star \over 960 \pi }  
{\hat \l _3 ^2  \eta  ^6  w  \calf  \over z^3 } , \label{eq.DECq}   \eea  
yield   the estimates $\D \G \sim  O(10^{27} ) \ GeV $, for $ x_m = 3.83  ,\ \eta = 100 ,\ z=10 , \ \calf = 100, \   w= w_A = 10^{-4}$.
The resulting  lifetime  $ \D t _{1\to 2} = O(10^{-52} ) \ s  $
is comparable to  the estimate   $ O(10^{-49} ) \ s  $ of~\cite{kofman05}  
but considerably shorter  than  the lifetime  of  massive closed strings
$ O(10^{-33} ) \ s  $  or the inflation exit   time, 
$ t_I= \sqrt 3   \pi^{5/4} (\eta z )^{3/2} / ( w ^2  M_\star ^{-1})   \simeq  O(10^{-30})  \ s$.
The gravitons   decays  to  pairs of  brane modes
are of same size   up to  large   uncertainties   due to the dependence on
the brane location.

Similar  conclusions    hold   for    2-body decays
between  neighbour   radially  excited   graviton modes, $  \G _{1\to 2} (h_{m_n} \to h_{m_{n-1}}+ \bar h_{m_{n-1} } ) $, given  the comparable ratios of coupling constants~\cite{chemtob16},   $  \l _3 (h_1   hh )    / \l _3 ( hhh )  \simeq 0.2/0.5 $ and  the smaller  ratios  $  \l _3 (h_2  hh ) / \l _3 (hhh ) \simeq  0.04/0.5  $ for non-sequential   decays. 
We also note that the     axio-dilatons  have similar couplings as   
gravitons  while  the   scalars      $ S^-$ have suppressed
couplings.    The contributions 
from  deformation  effects to disallowed couplings    
are  suppressed  by    powers of the warp  factor.
If    massless modes  were  present in the  mass spectrum in addition to the
graviton,  one expects their  couplings to be $ O( 1/ M_\star) $
suppressed,  hence   causing  an  early     decoupling  from  the thermal bath  with   a    tiny   primordial  abundance.    Their   delayed   
production    through pair   annihilations   of massive   
gravitons, $  h +  h \to  g +g, \ 
   h +   g\to  g +g, $ is expected~\cite{chen06} to  contribute a 
small  radiation component   today of order $(g_\star / g_s )^{1/4} (m_s  w  / M_\star )$.    We  conclude from the  above   discussion  that the  excited 
warped modes  should   fastly  decay   to  a population 
of  weakly interacting  ground state modes  formed   mostly  from 
graviton  and   scalar   singlet modes.  This  should justify 
focusing  on a simplified  treatment of the  thermal     evolution
restricted  to a   single   species  of  graviton modes.  

\subsection{Single throat case}
\label{subT1} 

\begin{figure}[t]
\begin{subfigure}{0.45\textwidth} 
\includegraphics[width=0.99\textwidth]{   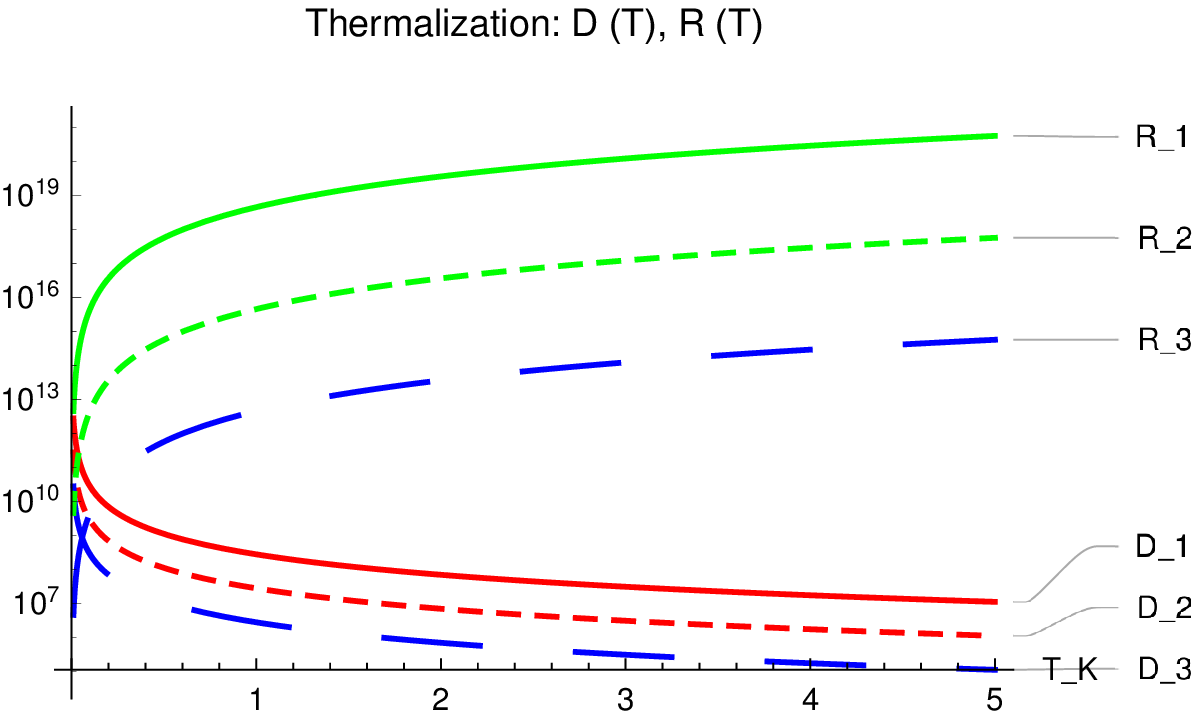}
\caption{\it  Plots of the  bulk and brane thermalization  ratios
  $ R (T) ,\ \D ^B (T) $   versus  temperature  $ T_K = T / m_K$
using  $\hat \l_3=   0.128 ,\  \hat \l ^B _3=  0.0225$ 
  at fixed   $\eta = 1$   for  three   values 
  $(1, 10 ,100) $ of     $ g_s N \simeq 0.0836 \times  (z / \eta ^3 ) = (1, 10 ,100)$
  associated  to $R _{1,2,3} (T) $ and $   \D ^B_{1,2,3}(T) $.} 
\end{subfigure}
\begin{subfigure}{0.45\textwidth} 
\includegraphics[width=0.99\textwidth]{   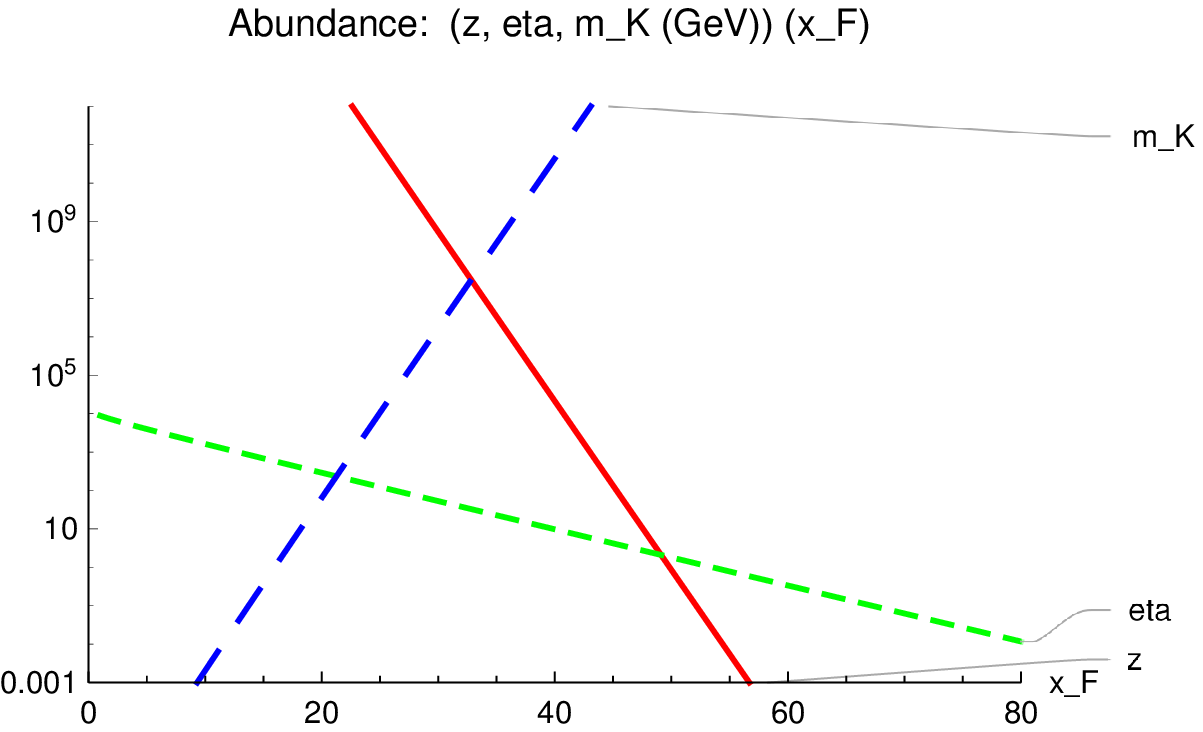}
\caption{\it  Plots versus  the
freeze out  variable  $ x_F = m_K / T_K$  of the solutions for  $\eta,\ z $  and   $ m_K (GeV) $   reproducing the  observed   relic  abundance $\O _A \simeq 0.11$.}
\end{subfigure}
\caption{\it \label{therm.inf} 
Thermal  equilibrium  and  cold thermal  relic  abundance  
constraints on the $A$-throat parameters $z,\ \eta $
at the  warp factor value $w= w_A = 10^{-4}$. The left panel $(a)$
displays  the   ratios  to Hubble rate
of the pair  annihilation  of bulk  modes $ R (T)  $  and 
of the decay rates of bulk  modes  to  pairs of brane modes  $\D ^B (T)$.
The right (panel $(b)$) displays the  admissible  range of values
for    the parameters $z,\ \eta $ and  the relic particle mass $ m_K$ in GeV units.}
\end{figure}

We begin  with the case  of a single deformed conifold throat 
hosting   both    $ D3-\bar D3$-brane inflation~\cite{kklmmt03}  and the Standard Model, using the  warp factor  value   $  w=  w _A  = 10^{-4}  $  that reproduces   the expected  value of the Hubble rate at  inflation  exit,
$ H_I \sim 10^{-4} M_\star   $. 
For  Standard Model   $ D3$-branes  
located  near the conifold apex, the  ultraviolet  cutoff mass scale  lies near       the   grand unified theory (GUT) value,
$ M_{GUT} \simeq  w  M_\star $. Recall that  the throat  thermalization is      realized as long as the  elastic  and inelastic  scattering  rates 
exceed the  Hubble  expansion rate $H(T)$. We examine this possibility
by considering pair  annihilation  among  bulk   modes and   decay  of  bulk   modes to pairs of brane  modes  controlled    by   $2\to 2$
 and  $ 1 \to 2$   processes, respectively, described by   the  dimensional    analysis   estimates of the rates,    
\bea && \D \G _ {2\to 2} =   {g_K \calf  \zeta
  (3) \over \pi ^2 } \l _3 ^4   T^5 = 1.73  \  10 ^{20} \hat \l _3 ^4
     {T^5   \eta ^{12} \over M_\star ^4   w ^4 }  , \  
\D \G  ^B _{1\to 2 }  = {\calf '\l   _3 ^2 m_K^3  \over 960 \pi } =
1.19  \  10 ^{9} \hat \l _3 ^2  {M_\star w \eta ^6 \over 
z^3 } , \label{WTeq1} \eea 
where $ \calf ,\  \calf ' $  count the    numbers of open
channels and the  numerical values   were obtained  from 
the  input data, $  x_m = 3.83, \ 
\calf =10  ,\ \calf ' =10  ,\   \zeta (3) =1.202 , \  \xi =
7.3 \ 10^{4} ,\     \xi ^B = \sqrt 2 \xi \approx  10^5$.     
The Hubble rate  $ H(T) = (\rho (T) /(3M_\star ^2 ) )^{1/2} 
= (\pi ^2 g _\star / 90 )^{1/2} T^2 / M_\star  \simeq
0.33  g_\star ^{1/2} T^2/ M_\star $   is  
evaluated  with $ g_\star (T) \simeq 100$.
The resulting conditions  for   bulk  or brane  thermalization 
as a function of the   temperature scaled  by  the modes mass   
$ T_K = T/ m_K$  are  then given by
\bea && 1 <   R(T) = {\D \G _ {2\to 2} \over H (T) } = 
C _R {\eta ^{12}  T_K^3 \over w z^3 } ,\   
[C _R   = { (\xi \hat  \l _3 ) ^4  x_m^3 g_K \zeta (3) \calf 
\over \pi ^2 ( 0.33 g_\star ^{1/2} ) }  \simeq 2.91 \ 10^{21} \hat  \l _3 ^4 ]
 \cr &&    1 < \D ^B (T) = {\D \G ^B_{1\to 2 } \over H(T)} =
C _{B}  {\eta ^{6} \over w z T_K^2} ,\    
[C _B = {(\xi _B  \hat \l _3 ^{B} )^2   x_m \calf ' \over  960 \pi (0.33
g_\star ^{1/2} ) }\simeq   3.8\  10^7 \hat  \l _3 ^{B 2}].  \eea   
It  makes sense   to examine  the variations of  these    ratios
in  the interval  $ T_K  \in [0.1 ,\ 10]$.
Writing    the   parameter  dependence      as,   
$ R (T)  \propto  (\eta ^3 / z) ^3  \eta ^3, \ \   \D ^B  (T) \propto  
(\eta ^3 / z)   \eta ^3,$   we    realize  that 
  both ratios are  proportional to powers of the parametrically small ratio,
$\eta ^3 / z = \sqrt \pi / (\l _N ^4 g_s)  \sim 1/ (g_s N ) $,  times the
same   factor    $\eta ^3$.  
The   plots of  $ R(T) ,\ \D ^B(T) $ in   Fig.~\ref{therm.inf}   at a discrete set  of values of $\eta ^3 / z$ with $\eta = 1$  
should give  lower bounds for these    quantities. We see that  both ratios  lie
comfortably  above unity,  with  the ratio   $ R (T)/ \D ^B  (T) > > 1  $   meaning that  the     throat    thermalization  well precedes  brane thermalization.   

We  now wish  to  check  whether  some   fraction of
metastable ground state  modes  could survive  as a   cold thermal relic.
For  this purpose, we examine the conclusions implied  by the   freeze out mechanism
based on  the boundary-layer  approach~\cite{bender12}. The non-relativistic   modes       decoupling     at  $ T/ m_K >  1 $    
is    described by  the familiar   kinetic   equation 
\bea && {d Y (x) \over d x} = -\l x ^{-n-2} (Y^2(x) -Y^2_{eq} (x) ),\  
[x= {m_K \over T},\ Y(x) = {n_K(t)  \over  T^3 },\  
Y_{eq} (x) = a \int _0^\infty d s {s^2  \over  e ^{(s^2
    + x^2)^{1/2} } \pm   1  }  ,\cr &&   
a  \simeq   0.145 g_K   / g_\star  ,\ g_K =5  ,\ 
\l = ({90   \over  g_\star \pi ^2 } )^{1/2}  M_\star m_K  <\s _{2\to
  2}  v >  = ({90   \over  g_\star \pi ^2 } )^{1/2}  {\xi ^4  \hat
  \l _3 ^4  \over x_m^3}     { \eta ^{12}  \over z^3 w }  ]\eea 
where  the annihilation rate    has been set to  
$ <\s _{2\to 2}  v >  = \s _0 x ^{-n} , $  with $
n \geq 0 $    denoting the orbital momentum  for the leading partial
wave amplitude.    For  $s$-wave annihilation, 
  $  <\s _{2\to 2}  v > = \l _3 ^4  m_K^2 = 
\xi ^2 (\eta ^3 / (M_\star w )) ^4  \hat \l _3 ^4 m_K^2$.   
The    boundary-layer solution~\cite{bender12}  is derived 
by solving the equation  analytically    in two distinguished  limits:
 one near the  thermal equilibrium  regime, $ Y \simeq  Y^{eq}$,   
in terms of the  asymptotic series  in  the rate  parameter $1/\l $, and  the
other in the post freeze out  regime  at  large  $x $ and small  $Y^{eq}$.
(In    the  alternative  method presented in the textbook~\cite{weinbergs08},    which we  adopted  in our 
previous  work~\cite{chemtob16}, $ \l  $   is  denoted   $B_W $.) 
The  interval    of $ x$ near the  freeze out  point
$ x_F = m_K / T_F $,    where the asymptotic series  in $1/\l $  
breaks down,   is defined  by the implicit equation  for $x_F$,
\bea &&  x_F  \simeq \ln (2  a   \l ) - (n+ \ud ) \ln
x_F , \ [a = 0.145 g_K   /   g_\star ]   \label{eq.FO} \eea
which   relates  $\l $  to   $x_F$. 
The  boundary-layer line interval around  $x_F$  
delimits the    region where the  solution for $ Y(x)$ features a fast
variation      between the  two  regimes.
The unknown parameters  are  determined by matching 
the analytic solution   in   this   $x$-interval   
to   the     solutions (in the appropriate limits)  in the outer
intervals.   The  inner  solution in the  post freeze out  limit,  
$ T\to 0  $,   is  then  given  by  
$ Y( \infty ) \simeq  {(n+1) x_F^2   \over \l (n+1+ x_F) }$,  where $ n=0 $ 
for  the   $s$-wave annihilation case at hand.  The  present day     abundance
can then  be evaluated   from the formula 
\bea && \O _A   ={T_{\g , 0} ^3 \over 3 M_\star ^2 H_0  ^2} 
{4  \over 11} 2 m_K Y(\infty )   
= {8 T^3 _{\g , 0} \over  33 H_0 ^2 M_\star } ({g_\star \pi ^2
  \over 90} )^{1/2}   {w^2 z^2 \over \eta ^{12} x_m^2    \xi ^4  \hat \l _3 ^4 } 
{ x_F^2   \over x_F +1} ,  \label{eq.AB} \eea  
where $H_0 = 1.44 \ 10^{-42} GeV ,\ T_{\g 0} = 2.34 \ 10^{-13} \ GeV
,\ \O _A \simeq 0.11$.
We    consider at this point  the interesting possibility    that the parameters    $ z, \ \eta $    could be evaluated by  solving  the  pair of  cold relic  constraint and abundance   equations in Eqs.~(\ref{eq.FO})   and~(\ref{eq.AB}), which we rewrite  in the more  convenient forms 
\bea &&   {\eta ^4  \over  z } =  ({e ^{x_F} x_F^{1/2} w \over  
2  A '})^{1/3} ,\  
\O  _A  = {A w ^2 \over M_\star } ({z \over \eta ^6 })^2  { x_F^2
  \over x_F +1} ,\cr &&   
[A = {8 T _{\g , 0}^3 \over 33 H_0 ^2 } ({g_\star   \pi ^2  \over  90  } )^{1/2}   
{1\over x_m^2 \hat \l _3 ^4 \xi ^4 } \simeq 0.49 \  10^{7}   {M_\star    \over  \hat \l _3 ^{4} } ,\  A' =    a   ({90     \over  g_\star     \pi ^2 } )^{1/2}  
{\xi ^4  \hat \l _3 ^4 x_m^3}  \simeq  0.38 \  10^{20} \hat \l _3 ^{4} ] .  \label{eq.abunX1} \eea 
The     resulting solutions  for the parameters  $\eta $ and $z$  are given    as a function of $ x_F  $     by 
\bea &&  \eta = ({A \over M_\star \O _A }) ^{1/4} ({ 2 A' \over e^{x_F}
  x_F^{1/2}})^{1/6} w ^{1/3} \simeq   { 1.17\   10^{4.92 }  w ^{1/3}   \over
  ( \hat \l _3 \O _A   ^{1/2}   (e ^{x_F } \sqrt {x_F } ) ^{1/3} ) ^{1/2}} 
,\cr &&   z= \eta ^4 ({ 2 A' \over e^{x_F} x_F^{1/2}
  w })^{1/3}  ={A \over M_\star \O _A}   { 2 A' \over e^{x_F}
  x_F^{1/2}}  w  \simeq  {3.7 \  10^{26} w \over  \O _A e ^{x_F } \sqrt {x_F }  } , \eea
where we  have   set  for simplicity,  $  x_F^2  / (x_F +1 ) \to x_F  $ in $\O _A$.  The  proportionality  relation $ z\propto w $  satisfied by the  solution implies   that  the  predicted    mass for the relic particle,   $m_K =  x_m w M_\star  / z $, is  warp factor independent.  (The proportionality   $\eta \propto w ^{1/3} $  is  compatible with the    relation,  $\eta ^3 / z \sim 1/(N g_s )$.)   It is  useful  to    compare  the present    result for $\O _A$ 
with   that     derived  within  the  Chen-Tye approach~\cite{chen06},  where  the pair annihilation rate is  set   by    the  $ 2\to 1 $
(instead of the  $ 2\to 2 $)   process. With the    reaction rate, 
$\D \G _{2\to 1} = n _K (T) \s _{2\to 1} \sim   \l _3 ^2  T  ^{3} $,
one finds  that the  resulting    abundance at  the
radiation domination to matter domination   (RDMD) time,
$\O _A  ^I  = 2 w^2 M_\star / (g_\star ^{1/4} (\xi \hat \l _3 )^2
\rho _{RDMD} )  \simeq  2.5  \ 10^{23}   w^2 /\eta ^6 $,
is  orders of magnitude  larger (with our input parameters) 
 than the  present   prediction  in  Eq.~(\ref{eq.abunX1}).

The  solutions   for $\eta ,\ z $  obtained in our approximate solution to Boltzmann equation  for the   warped modes   abundance 
are plotted   in  Fig.~\ref{therm.inf}  as a  function of  the freeze out  temperature for values in the  interval  $ x_F = T_F / m_K \in (1, 80)$.
The selected   values cover  wide  ranges
owing to the exponential dependence on $ x_F$.
Settling on  natural  values for  the  $z,\ \eta $ parameters of $ O(10) $, favours 
 $ x_F = T_F / m_K \sim 50 $  with  relic masses $ m_K \sim 10^{12} $ GeV.
Going down     to  $ x_F= 40 $,      selects 
larger parameters $ \eta \sim 10 ,\ z \sim 10^{5} $ and a  reduced 
mass $ m_K \sim 10^9 $ GeV,   while allowing masses  as  small as
$ m_K \sim 100 $ GeV, for $ x_F= 20 $ and $ \eta \sim 10^2 $    requires  the unrealistically  large parameter  value  $z \sim 10^{15} $.

\subsection{Double throat case}
\label{subT2} 

Since compactifications with multiple  throats   are    not  excluded, 
we  here consider the   case   involving  an additional $S$-throat  of warp  factor   $ w= w_S \sim 10^{-14}$ hosting TeV scale Standard Model  $D3$-branes.  The  parameters  $w,\ \eta ,\  z $ defined   in Subsection~\ref{sect3sub2}  are    now   attached      suffix labels  $ A,\ S$.
Thermalization  in the  $S$-throat  is    accomplished by  the 
fraction of    warped modes    tunneling   from the  $A$-throat.
We  shall consider the tunneling  rate   described   by the  $ D3$-brane model
in the resonant  bulk   case~\cite{chen06},  $\D  \G _{tun} (A\to bulk) \approx  \D  \G _{tun} (A\to S )  \simeq  w_A ^{9}/\calr _A $.  (The non-resonant  bulk  case   gives a  much  smaller rate, $ \D  \G _{tun} (A\to S )  \simeq  w_A ^{17}/\calr _A$.)        
Recall that   $ A\to S $ tunneling      takes place   provided
its rate exceeds the decay rate, 
$ H  ^{-1} (t_{tun}) \sim t_{tun} <  \D t _{1\to 2}    $,  and  that   
inverse  tunneling  from  long   to short throats is  suppressed.  
Until tunneling  terminates at  time  $t_{tun}\sim w _A^{-9} \calr _A
$,  the cosmic bath is stuck in  a   matter dominated (MD)  regime with
the cosmic evolution satisfying the  scaling  laws,  $T\propto  a^{-2}\propto  t^{-4/3}  ,\ H   = 2/(3 t  ) ,\  \rho = M_\star ^2 / t ^2 $.

The level matching   condition, necessary   for the 
 tunneling from  $ A\to S$ throat   to  take place,  requires that
 the  modes level spacing   be small compared    to the tunneling width,  $ \d m_K  \simeq  w_S / \calr _S < \G _{tun} \simeq w_A^9 /\calr _A   $. This   gives the upper bound $w_S  \leq  w_A^9 \calr _S / \calr _A  \simeq w_A^9  (N_S / N_A )^{1/4} $,  favouring  tunneling to longer throats~\cite{chen06}.
For $ w_S << w_A$, the  low-lying  $A$-throat modes  
change  after tunneling to highly   excited    $S$-throat  modes.
Another constraint stems from  the condition  that the   back-reaction from inflation in  the $A$-throat   does not produce   closed    strings  in the $S$-throat.   This  is expressed   by the   bound on Hubble rate~\cite{frey05,chen06},
  $ H^4 (t)   <  N _S\tau _3  w_S ^4 \sim  N_S^2 w_S ^4 / \calr _S^4$,  which 
imposes a minimal   value for the warp factor, $w _S > H  (t) \calr _S / N_S^{1/2} = {H  (t) \over M_\star } \eta _S ^{3/2} z_S ^{1/2} $ (where we used
 $ N_S^{1/2}  \simeq \calr _S^2 m_s ^2 \simeq z_S^{1/2} / \eta _S ^{3/2} $). 
Setting    the  Hubble rate value  at inflation exit  $ H (t_I ) \simeq  10^{-4} M_\star $  or at  tunneling time   $ H (t_{tun} ) \simeq \G _{tun} $,  gives
the lower bounds,  $ w_S >  10^{-4} \eta _S ^{3/2} z_S ^{1/2} $ or $ w_S  > w_A^9 \eta _S^{3/2} z_S ^{1/2} / z_A $.

 One can   use the   time-temperature and energy-temperature relations,
$ {1\over 2 t } \equiv  H = 0.33 g_\star ^{1/2}  (T)  {T ^2   \over  M_\star },\
\rho (T)  \equiv  3 M_\star ^2 H^2  = {\pi ^2 \over  30 }
g_\star (T)  T  ^{4}  , \ [g_\star (T)  = ({T  \over \a  })^{\g },\ \a  = {w  \over  \calr } ] $  to obtain  the   temperature at tunneling time in
a generic throat $X$,
$T _{X}  (t _{tun} )= ( ({M_\star \G _{tun} / 0.66  })^{2 }    \a _{X} ^{\g } )^{1 /(4  +\g )} $.   Adapting this result  to    the   $ A $- and $S$-throats  with
$\g = 5$  gives the temperatures 
 \bea  &&
 {T _{S}  (t _{tun} ) \over GeV } = 2.67 \   10^{18} w _A^{2} ({w _S^{5} \over z_A^2 z_S ^5})^{1/9} ,\    {T _{A}  (t _{tun} ) \over GeV }= 2.67 \   10^{18}
 {w _A^{23/9} \over z_A ^{7/9} } . \label{eq.TSTA} \eea
 The  wide  gap   between the  respective warp factors  entails that
 $ T_S << T_A $.
The  thermal equilibrium   condition  at tunneling time
can be described  for simplicity    by forming  the ratio of the  $ 2\to 1 $  annihilation   rate to  Hubble rate, $ R _X  (t_{tun} ) =  {\D \G _{2\to 1 } /  H(t_{tun} ) } $,   where  $\D \G _{2\to 1} = n  (t) \s _{2\to 1} \simeq  \a  ^{-\g } T _{tun} ^{3+\g } \l _3 ^2
 \simeq \a _X ^{ -\g / (\g     +4) } ({M_\star \G _{tun} / 0.66 })^{2   (\g +3) /  (\g +4) }  $. Adapting the general  formula  
\bea &&  R _X  (t_{tun} ) ={ (\xi \hat \l_3 )^2\ \eta _X ^6     \over  (0.66) ^ {2 (\gamma +3) / (\gamma +4)} }     {z_X^{\gamma /(\gamma +4)} 
\over w_X ^{(3 \gamma  +8) / (\gamma +4)}    (M_\star t_{tun} )^{(\gamma
  +2) / (\gamma +4)} } ,   \eea    to the $A$- and $S$-throats,  gives    
$ R_S (t_{tun} ) =  1.82 \  10^{8} \eta _S^6 {z_S ^{5/9}  w_A ^7
 /(   z_A ^{7/9}w_S ^{23/9} ) } ,\    R_A (t_{tun} ) =  1.82 \  10^{8} \eta _A^6
{w_A  ^{40/9} / z_A ^{2/9} } ,$   where we used 
 the inputs  values $\g = 5,  \  \xi =7.3 \ 10^4 ,  \hat \l _3 =  0.128 , \  x_m=3.83$. We see that both ratios lie comfortably above unity
and that  thermalization sets in more rapidly in the $S$-throat. 

In order to  avoid  disrupting the   primordial   abundance of light nuclei  the  universe temperature at  tunneling  time  must  lie  above   the  nucleosynthesis  threshold. Imposing the  condition    $T_{S,A}  ( t_{tun} ) >   O(1) \ MeV  $ in   Eqs.~(\ref{eq.TSTA})  yields the   lower bounds  on  warp factors, $ w_S > 10 ^{-117 /5} z_A^{2/5} z_S $ and   $ w_A > 10 ^{-189/ 23} z_A ^{7/23} .$   A  useful  constraint also arises~\cite{chen06}   by substituting  the  above    condition
$ \G _{tun} = H _{tun}  \leq w_S N_S ^{1/2} / \calr _S \simeq 
w_S \calr _S  m_s^2 $ into  Eq.~(\ref{eq.TSTA}). The resulting  upper   bound  on the $S$-throat temperature,   $T_S (t_{tun}) = ( (w_S /\calr _S)^\g (M_\star \G  _{tun} )^2 ) ^{1/(\g +4 )}   <  ( w_S^{\g +2} z_S ^{(2-\g /4)}\eta _S ^{3 \g /4} ) ^{1/(\g +4 )}  m_s $,  lies below  the  warped mass  mass $ w_A/\calr _A $ (upon replacing $ m_s  \to N_A^{1/4} / \calr _A $) but  could dangerously  exceed the warped string  mass  scale $  w_S m_s $ unless one invokes    some fine tuning of  the parameters.

We     consider    next the  cold thermal relic abundance  using  the 
above $ 2\to 1 $  pair  annihilation  rate as in    Chen-Tye approach~\cite{chen06}.
In this  description, the  relationship  $t  \simeq 1 /  (2 H  (t   ) ) = M_\star /(0.66 \ g_\star ^{1/2}  T ^2 ) $    sets the decoupling time  when  modes   become non-relativistic  ($ T_{NR}  \simeq m_K$)  at  $ t _{NR}\simeq  M_\star    / (0.66   \a _A ^2 x_m  ^{(4+\g )/2} ) $.
The  modes abundance at  decoupling in the $A$-throat,
$\rho _K / m_K^4 = n_K( t_{NR} )/ m_K^3  $, is set by substituting the modes number  density,  
$ n_K( t)   \equiv   H  (t) /  D \G  _{ann} = 1/(2 t \l _3^2 ) $,  
and inserting     the softening factor $(t_{tun} / t _{NR} )^{-1/2}$
to  account for the  extended
matter domination  period in the $A$-throat until  tunneling  is completed.
The resulting     general formula for the cold  relic abundance,   $\O _X^{II} = {1 \over (0.33 \   3^{1.5} )^{1/2}   x_m^{\gamma / 4} }  {M_\star \over  \rho _{RDMD}^{1/4} }      ({t_{NR} \over  t_{tun} })^{1/2}   {w^2_X \over (\xi  \hat \l_3 )^2       \eta _X  ^6 } $,  yields   the  abundance in the $A$- and $S$-throats
\bea && \O _S^{II} =  2.38 \  10^{17}  {w_S  w_A ^{9/2}   z_S  \over
\eta _S^{6}  z_A ^{1/2} }  ,\ \
\O _A^{II} =  2.38 \  10^{17}  {w_A^{11/2}  z_A^{1/2}\over  \eta _A^{6} } . \label{eq.abunDT} \eea
It is  clear that the  $A$-throat contribution to the abundance
is  largely  dominant. 
Note   that the  present   prediction using our input parameters   is  considerably  suppressed relative to   Chen-Tye   estimate~\cite{chen06}
$\O _A ^{II} \simeq 10^{24}   w_A^{11/2}  z _A^{1/2}  \eta _A^{-6} $,
which is helpful   in  relaxing the  bounds on   parameters. 


\section{Summary and main conclusions} 
\label{sect5}

We discussed in this work Kaluza-Klein theory for type $ II\ b$
supergravity on Klebanov-Strassler background  based on a large radial distance
approximation  that   preserves    the non-singular geometry near the
conifold apex. The WKB method was used to obtain predictions for  masses,
wave  functions and interactions of warped graviton, axio-dilaton and scalar
(4-form) modes.  The most robust  results    are for  graviton modes.
The mass splittings for  radial, orbital and string excitations
cluster  around $ 1 \ - \ 1.5 $  in units $ g_s M \e ^{2/3}
/ m_s ^2 $. In comparison  to graviton modes, the dilaton modes  are slightly  heavier, due  to  the 3-fluxes  contributions,
while the non-singlet scalar modes
are  significantly  lighter,  due to  attractive contributions from
mixing the   4-form and internal metric supergravity  fields.
The possibility that the scalar mode  $ S_- (100) $     be
a natural candidate for the LCKP~\cite{berndsen07} is mitigated by  the substantial  mixings   between    
scalar modes of  diverse origins in the  classical    background
anticipated from   studies of  mass spectra 
for   multidimensional field spaces~\cite{berg06,benna07}.
The background  deformations from compactification effects
have a strong impact on  the  interactions of warped modes,
modifying the selection rules  imposed by the  throat
isometry and imposing large hierarchies on  decay rates of lightest   modes~\cite{freydacline09}.
The sensitivity of warped modes couplings to the infrared geometry is made manifest   by  strongly suppressed value  for   the    effective  gravitational mass scale $\L _{KK} \simeq M_\star / \Phi _n (r_0)$
relative to the estimates in  models using  softened  warp profiles. 

The thermal evolution of a cosmic component of
warped Kaluza-Klein   modes produced in  the  throat hosting $D3-\bar D3$-brane inflation provides   useful constraints on the    compactification and  warped throat dimensionless  parameters $\eta = L_W /\calr ,\ z= M_\star \calr  $ and $ w $. We  pursued an   analysis along same lines as~\cite{chen06} using
updated values for the pair annihilation and two-body decay
reactions  of warped modes.    Both the    initial  temperature  from   
branes annihilation or the  reheat  temperature from decays
of massive closed strings  lie   well below the  warped string theory  mass scale.  The throat and brane  thermalizations  take easily  place at
temperatures of same orders as the lightest  warped modes. The  empirical
value  for the cold thermal relic  abundance   can be reproduced in a wide interval of  the warped modes mass including the TeV  range, but 
a robust conclusion would  require  a   quantitative  analysis of Boltzmann  equation.

\begin{appendix} 
\label{appendices}  

\section{Review of warped deformed conifold} 
\label{appwdc}

We  provide in this appendix an  introductory review on the deformed
conifold. After a summary  of algebraic  and differential  geometry 
properties we discuss the harmonic analysis  witin the group theory
approach of~\cite{pufu10}.  We      consider      next   the  modified
Klebanov-Strassler  solution for the   metric tensor  replacing 
the  conifold  base $ T^{1,1} $  by  the 
direct product   manifold of geometry   $S^2\times S^3$~\cite{firouz06} and
finally  describe a  simple  construct   that  makes contact 
with  the   analytic  type   formalism   in the singular conifold limit. 

\subsection{Algebraic properties}
\label{appwdcsub1}
  
The    deformed conifold  $\calc _6$~\cite{candelas90} is
part  of the family of  Stenzel spaces  $\calc _{2d-2} $~\cite{cveticpope00}, defined  as    non-compact manifolds of complex
dimension $(d-1) $  satisfying  the quadratic embedding  equation  in $ C^{d} $,
\bea && \sum_{a=1} ^d w_a ^2 = \e ^2 ,\ [\e \in R
  ,\ w_a \in C] \label{app1.eq0} \eea 
where  $\e  \in   C $  is   the complex structure deformation modulus.  
The   invariance   under  $ SO(d)$ orthogonal matrix rotations 
of the complex variables $ w_a ,\ [a=1,\cdots , d] $ 
entails the existence of   the isometry group   
$  G= SO(d) \times Z_2^R $.   At $\e =0$,  the   $Z_2^R$  parity 
$w_a \to - w_a $   is   promoted to  the conserved   charge $ U(1) _R:\ w_a \to e^{i\a }
w_a $, where the  suffix label  for $ U(1) _R $ is a reminder  that  this
corresponds to   the  R-symmetry  of the 4-d  $\caln =1 $  supersymmetric 
dual gauge theory. The    radial    sections $\S _\tau $   at constant  $\tau $ lie   at the intersections of       the  conifold    
$ \sum _a   w _a ^2 = \e ^2  $ with the locus $\sum _a  \vert  w _a \vert
^2 =  \vert \e \vert ^2  \cosh \tau $. 
The fact that  the   undeformed  Stenzel spaces   (at $\e =0$)  are 
real  cones  over    a   compact base  manifold    suggests the   convenient parameterization of the $ w_a $
at finite $\e $ combining     the real radial variable $\tau \in [0,\infty ] $   with  the $d$ pairs of complex conjugate variables $ y_a ,\ \bar y _a $
 parameterizing the   $\S _\tau $ satifying  the  two  conditions 
\bea && \sum _a y_a ^2 = 0,\ \sum _a \vert
  y_a \vert ^2 = 1,\ \ [w_a = {\e \over \sqrt 2} (e^{\tau /2} y_a + e^{-\tau /2} \bar
y_a )   \ \Longrightarrow \ y_a = { e ^{\tau /2} w_a / \e -  e ^{-\tau /2} (w_a /
  \e )^\star  \over  \sqrt 2  \sinh \tau }] . \label{app1.eq1} \eea 
The radial   sections      are compact    manifolds      homeomorphic to 
  Stiefel    coset   spaces $ V_{d,2} = SO(d)/ SO(d-2) $
whose   elements are   organized into  equivalence classes  
$ g\sim  g h ,\ [ g\in SO(d) ,\ h \in SO(d-2) ]$ 
corresponding   to orbits of the isometry group $SO(d)$ acting on
the   element $ y^{(0)} _a = (0,0,\cdots ,  i \sinh (\tau /2) , \cosh
(\tau /2) )$   
fixed     under the stabilizer group $ SO(d-2) $.
Near $\tau =0, \  y^{(0)} _a =  (0,0,\cdots , 0, 1)$  
and   $ \S _\tau \sim SO(d)/ SO(d-1) \sim S^{d-1} $.   
Since   the $\S _\tau  \sim  V_{d,2}$    have   the  same integration measure
$\mu _{SO(d) } / \mu _{SO (d-2)} $    up to an overall $\tau $-dependent
normalization, the    representation  vector space  for the group $SO(d)$   action on $\S _\tau  $ consists of  square normalizable
functions  with  the   integration measure  given by the  group $G$
invariant measure times a    function of $\tau $.    

The spaces $\calc _{2d-2} $   are K\"ahler manifolds
equipped with      a  Hermitean metric  generated  by the
isotropic  K\"ahler  potential $ \calf (\tau )$,
\bea && d  \tilde s ^2 ({\calc _{2d-2}}) =g_{a\bar b} dw ^a d \bar w ^{\bar
  b} ,\ [g_{a\bar b} = {\dh ^2 \calf \over \dh w _a \dh \bar
  w _{\bar b} } =  2 F_1 \d _{ab}  + 4 F_2 \bar w_{a} w_{b}
= {\calf ' \over \e ^2 \sinh \tau } \d _{ab} + {\calf
  '' - \calf ' \coth \tau \over \e ^4 \sinh ^2 \tau } \bar w_{a} 
w_{b}   ]. \label{app1.eq2} \eea 
The  Ricci tensor flatness  condition (on Calabi-Yau manifolds),  $ R_{a\bar
  b} =0 $,       imposes  a differential equation on  
$ \calf ' (\tau )= \dh _\tau \calf (\tau )$ which  can be solved  in terms
of Hypergeometric functions~\cite{pufu10,cveticpope00} 
\bea && ({\calf ' (\tau ) \over \e ^2
  \sinh \tau }) ^{d-2} \calf '' = {d-2\over d-1} \ \Longrightarrow
\ \calf ' (\tau ) =\e ^{4\over d-1}  R^{1 \over d-1} (\tau ),\cr &&
        [R(\tau ) =  
  {d-2\over \e ^2} \int _0 ^\tau d v(\sinh (v) )^{d-2} = { (d-2)\over \e
    ^2 (d-1)} (2 \sinh (\tau /2)) ^{d-1} {_2}F_1 ({d-1\over 2} ,
  {3-d\over 2} ; {d+1\over 2} ; - \sinh ^2 (\tau /2) )
        ].\label{app1.eq3} \eea
        
 We specialize hereafter  to the conifold case ($ d=4$)
 where the   complex variables $ w_a ,\ [a=1,\cdots , 4]  $  and
 their linear  combinations $ z_a$,   can be conveniently packaged  within 
 the  $2\times 2 $ matrix $W$
which   allows defining  the    conifold   and  its     fixed-$\tau $  sections   $\S _\tau $   by  the  pair of  algebraic  equations   
\bea && Det \  (W) = - \sum _a w _a^2  =  z_1 z_2 - z_3 z_4 
 = - \e ^2 ,\ 
\ud Tr (W^\dagger W )  = \sum _a \vert w _a \vert ^2  
= \sum _a \vert z _a \vert ^2  = \hat r  ^3 
= ({2 \over 3})^{3/2} r^{3} =  \vert \e  \vert ^2 \cosh \tau , \cr && 
[W= \pmatrix{ w^3 + i w^4 & w^1-i w^2 \cr w^1+i w^2 & -w^3 + i w^4 }
  \equiv   \pmatrix{ z_3   &   z_1  \cr  z_2  &   z_4 } ,\
  {w_1 \choose  - iw_2 }  =  \ud (z_1 \pm z_2),\ 
{w_3\choose    iw_4} =  \ud (z_3 \mp z_4) ] . \label{app1.eq5} \eea  
(Our  normalization  conventions for  $ w_a  $   and $W$   
coincide    with~\cite{krishtein08,benini09} 
and    would agree   with~\cite{gimon02} 
if  one       substitutes $ w_a  \to  \sqrt {2} w_a ,\ W \to \sqrt {2} W$.)  The    metric  tensor  can be    evaluated     by means  of the   formula  
\bea && d \tilde s^2 ({\calc _6 })  =  {1 \over 2} {\dh \calf   \over  \dh
  \hat r  ^3 } 
Tr (dW^\dagger  d W ) + {1 \over 4}  {\dh ^2 \calf \over
  \dh (\hat  r  ^3  ) ^2  } ]  \vert Tr (W^\dagger d W ) \vert ^2 
=  ( 2 F_1 \d _{ab}  + 4 F_2 \bar w_{a} w_{b} ) d  w^{a}  \bar  dw^{b}
, \cr &&  [F_1 \equiv \ud {\dh \calf  \over  \dh \hat \rho ^2 } = {\e ^{-2/3}K(\tau )
  \over 2} ,\  F_2 \equiv {1 \over 4}  {\dh ^2 \calf \over
  \dh (\hat  \rho ^2 ) ^2  } = {\e   ^{-8/3}  K' (\tau ) \over 4 \sinh
      \tau } 
\equiv {\e ^{-8/3}  K (\tau  ) \over  4  \sinh  ^2  \tau }
  ( {2  \over 3 K ^3(\tau  ) } - \cosh \tau  ) ] 
\label{eq.mconf1}  \eea  
where 
$\e ^2  \sinh \tau d\tau  =  w _a d \bar w _a+   \bar w _a d  w _a
$  and 
\bea && \calf '  (\tau ) \equiv {\dh \calf \over \dh \tau }   
= \e ^{4/3} (\sinh (\tau ) \cosh (\tau ) - \tau)^{1/3} =\e ^{4/3}
\sinh (\tau ) K(\tau ) ,\  
\calf ''\equiv {\dh ^2 \calf \over \dh \tau ^2 }  
= {2\over 3} ({\e ^2 \sinh \tau \over  \calf' }) ^2   . \eea  

At  large $\tau $, the above  metric  is    asymptotic to   the   undeformed conifold   metric, 
\bea && ds ^2 (\calc _6)  \simeq dr ^2 + r
^2 d s^2 ( V_{4,2} ) ,\ [r ^2 \equiv  {3\over 2}  \hat r^2  
= {3 \over 2 ^{5/3} } \e ^{4/3} e   ^{2\tau /3 } ] \eea
where  the sections at fixed   values of the
conic   radial  variable $r$ have the coset space structure
$\S _\tau  =  SU(2) _1  \times SU(2) _2  /U(1)_H \sim T^{1,1} $.
The   invariance of the  embedding equations in Eq.(\ref{app1.eq5})
under $SO(4)$ rotations of the $ w_a $,
modulo  the  $Z_2^R  $-parity $ w_a \to - w_a $,  follows from 
the determinant and  trace    invariance
under    left and right multiplication   
by unitary matrices, $W \to g_1 W g_2 ^\dagger ,\ [g _i \in SU(2) _i
  ,\ i=1,2] $, modulo  $ W\to - W $.   The Klebanov-Strassler
standard solution  for the   metric~\cite{klebstrass00} in Eq.~(\ref{app2.eq1})    follows from   the   parameterization  of $ W$~\cite{minasian99}    
  \bea &&  W = g_1W_\e \s _3\s _1  g_2 ^\dagger   =  {\e }  g_1 e
  ^{\tau \s _3 /2} \s _1 g_2 ^\dagger  = {\e }   
g_1 \pmatrix{ 0 & e ^{\tau /2 } \cr  e ^{-\tau /2 } & 0}  
g_2 ^\dagger ,\ [W_\e  = \e  \pmatrix{e^{\tau /2 }  & 0  \cr    
0 & - e ^{-\tau /2 }}  ,  \cr &&  g_i =  e ^{+i {\phi _i \over 2} \s _3} 
e ^{-i {\t _i \over 2} \s _2} e ^{+i {\varphi  _i \over 2} \s _3} = 
\pmatrix{ c_i e ^{i (\varphi _i +\phi _i  ) /2} & 
- s_i e ^{i (-\varphi _i +\phi _i  )/2}  \cr  s_i e ^{-i (-\varphi _i +\phi _i  )/2} 
 & c_i e ^{-i (\varphi _i +\phi _i  )/2} }  ,\ c_i = \cos {\t _i\over
    2} ,\  s_i = \sin  {\t _i\over 2 }]   
\label{app1.eqKSp}  \eea
where the pair of   Euler angles  $(\t _i, \ \phi _i, \ \varphi _i )  $ 
of  $SU(2)_i \sim S^3 _i $    are   subject  to
the equivalence  $ \varphi _1  =  \varphi _2 = \psi  /2 $.
 
The  complex    coordinates  $ y_a $ in Eq.~(\ref{app1.eq1})
can be expressed as      quadratic products   of  the  two   $2$-spinors 
$ a = (a_i )  \in (\ud , 0) ,\  b = (b_i ) \in (0, \ud ) , \ [i=1,2] $ of the 
isometry   group $SU(2)_{1} \times SU(2)_{2} $~\cite{candelas90}      
\bea && { y_{1} \choose - iy_{2} } =
{a_1 b_1 \mp a_2 b_2 \over \sqrt 2 } ,\ { y_{3} \choose iy_{4} } = -
{a_1 b_2 \pm a_2 b_1 \over \sqrt 2 } , \   [a _{(1,2)} = 
{\cos {\t _1\over 2}     \choose \sin {\t _1\over 2}  } 
e ^{{i\over 2} (\pm \phi _1 + {\psi \over 2} ) } ,\ b _{(1,2)} = 
{\cos  {\t _2\over 2} \choose   \sin  {\t  _2\over 2} } e
  ^{{i\over 2} (\pm \phi _2 + {\psi \over 2} ) } ]. \label{app1.eq4}
\eea
The     $ a_i ,\  b_i $   correspond    in the Klebanov-Witten  dual gauge
theory~\cite{klewit98}   to the pair of chiral supermultiplet  fields  $ A_i
,\ B_i  ,\ [i=1,2]$   carrying  the quantum numbers    
 $ A_i \sim (N_1, \bar N_2 ; 2,1 )_{\ud , 1} ,\
B_i  \sim  (\bar N_1, N_2 ; 1,2 )_{\ud , -1} $ with respect to the
gauge  and global  symmetry groups $ \calg  = SU(N_1) \times
SU(N_2)  $  and  $G  = SU(2)_1 \times SU(2)_2 \times U(1) _r \times
U(1)_b  $.   The  complex variables $ z_a $  in
Eq.~(\ref{app1.eq5}) correspond      then to     the   gauge theory  
composite  fields    $  M_{ij} = B_i A_j  $    with
$( z_1  ,\   z_2 ,\  z_3, \  z_4 ) \sim (B_2
 A_1,\   B_1 A_2 ,\   B_1  A_1,\   B_2 A_2 )  $ 
  
 The deformed conifold embedding   in $ C^4$
is  conveniently   defined by   the parameterization of the  four complex coordinates~\cite{herzog01,mcguirk12},     
\bea && w_1= \e  [\cosh {S }  \cos \t _+ \cos \phi _+ + i\sinh
   {S }     \cos \t _- \sin 
  \phi _+ ] ,\  w_2= \e [- \cosh {S }  \cos \t _+ \sin \phi _+ +
  i\sinh {S }  \cos \t _- \cos \phi _+ ] ,\ \cr && w_3=\e [ - \cosh
  {S }  \sin \t _+ \cos \phi _- + i\sinh {S }  \sin \t _-
  \sin \phi _- ] ,\ w_4= \e [- \cosh {S }  
\sin \t _+ \sin \phi _- - i\sinh
     {S}     \sin \t _- \cos \phi _- ] ,\cr && [S={\tau + i\psi \over 2} ,\
\t _\pm =  \ud (\t _1 \pm \t _2),\ \phi _\pm = \ud (\phi _1 \pm \phi _2)]
    \label{app1.eq6} \eea
derived by substituting Eq.(\ref{app1.eq4}) into Eq.(\ref{app1.eq1}). 
Evaluating  Eq.~(\ref{eq.mconf1})   in  this   parameterization  
yields the  Klebanov-Strassler metric in 
Eq.~(\ref{sect1.eq4}),  which  we rewrite below for convenience, 
  \bea && d\tilde s^2 (\calc _6) = {\e ^{4/3} K(\tau ) \over 2} (
{1\over 3 K ^3(\tau ) } (d \tau ^2 + (g^{(5)} )^2 ) + \cosh ^2 ({\tau \over
2})  ((g^{(3)})^2 + (g^{(4)} )^2 ) + \sinh ^2 ({\tau \over 2})
((g^{(1)})^2 + (g^{(2)})^2 ),\cr && 
[g^{(1,3)} = {1\over \sqrt 2} (e^{(1)}\mp e^{(3)}
),\ g^{(2,4)} = {1\over \sqrt 2} (e^{(2)}\mp e^{(4)} ),\ g^{(5)} =
e^{(5)} ,\cr && {e^{(1)} \choose e^{(2)} } = { - \sin \t _1 d \phi
    _1 \choose d\t _1 },\ {e^{(3)} \choose e^{(4)} } = \pmatrix{ \cos
    \psi & -\sin \psi \cr \sin \psi & \cos \psi } {\sin \t _2 d \phi
    _2 \choose d\t _2 } ,\  e^{(5)} = d\psi + \sum _{i=1}^2\cos \t _i
  d\phi _i ]  \label{eqKSmet1} \eea  
where $ e ^{(a)} $  and $ g ^{(a)} $ are two  bases of 
left  invariant 1-forms   of the compact base    whose 
volume   form  integral yields the base manifold  volume  
\bea &&  V (T^{1,1})  =  {1\over \sqrt { 9 \cdot  6^4 } }  \int  g^{(1)} \wedge \cdots \wedge g^{(5)}  = {1\over 108} \int _0^{\pi }   \sin \t _1
\int _0^{\pi } \sin \t _2  \int _0^{2\pi } d\phi _1 
\int _0^{2\pi } d\phi _2 \int _0^{4\pi } d\psi  = {16\pi ^3 \over 27}
]  . \eea 
An alternative   formula  for the metric~\cite{kuperstein04,kupsonn08}
is  sometimes  used  in terms  of the bases of left invariant  1-forms   $ h_a ,  \ \tilde  h _a ,\ [a=1,2,3]$    associated to the  Lie algebra  generators   $  g_1 ^\dagger  d g _1 = {i\over 2} h_a  \s  _a,
   \ g_2 ^\dagger  d g _2 = {i\over   2} \tilde h_a  \s  _a $   of   the  
$SU(2)_{a} $ groups, $  g _{a} = e ^{+i {\phi _i \over 2} \s _3} 
e ^{-i {\t _i \over 2} \s _2} e ^{+i {\psi  \over 4} \s _3}   , \ [a=i=1, 2]$.   
The 1-forms $ h_a ,  \ \tilde  h _a $  satisfy  the $ SU(2)$  Maurer-Cartan   relations,  $ d h_a =   {i \over 2} \e _{abc}  h_b \wedge h _c,\
 d  \tilde   h_a =   {i \over 2}  \e _{abc}   \tilde   h_b \wedge
 \tilde   h _c  $
   and  are  related to    the basis $ g ^{(a)} $  in Eq.~(\ref{eqKSmet1}) by   the 2-d rotations,  
\bea && {h_1 \choose h_2} ={1\over \sqrt 2}  \calr _{\psi / 2} 
{g^{(1)} + g^{(3)} \choose g^{(2)} + g^{(4)} },\ 
 {\tilde h_1 \choose \tilde  h_2} ={1\over \sqrt 2}  \calr ' _{\psi / 2} 
{g^{(3)} - g^{(1)} \choose g^{(4)} - g^{(2)} },\  h
_3 + \tilde h _3 = g^{(5)} ,\cr &&  [\calr _{\psi /2}   = 
\pmatrix{\cos {\psi \over      2} & \sin {\psi \over 2} \cr  
\sin {\psi \over 2}  & -\cos {\psi \over 2} },\    
\calr  '_{\psi  / 2} = \pmatrix{-\cos {\psi \over 2} & -\sin {\psi
    \over 2} \cr \sin {\psi \over 2}  & -\cos {\psi \over 2} } ,
\label{app1.eq7}\eea
and    to the basis   $ (e_{1,2},\   \e _{1,2,3} )$ 
of~\cite{papado00}  as,    $ h_{1,2} =    e_{1,2},\ \tilde h_{1,2} = \e _{1,2},\
(h_3 + \tilde h_{3} )  = \e _3.$
Note   that  $ h_1 ^2 = (g^{(1)} + g^{(3)} ) ^2  ,\   
h_2 ^2 = (g^{(2)} + g^{(4)} ) ^2  ,\ \tilde h_1 ^2 = (g^{(1)} - g^{(3)} ) ^2  ,\ 
\tilde h_2 ^2 = (g^{(2)} -  g^{(4)} ) ^2  $. 
Substitution in   Eq.~(\ref{eqKSmet1})
leads to  the    alternative equivalent   expression  for the Klebanov-Strassler
metric solution~\cite{kuperstein04},  
\bea && d \tilde s^2 ({\calc _6 })  = \e ^{4/3}  [ B^2 (\tau ) ( d \tau  ^2 +
  (h _3 +  \tilde  h _3  ) ^2 )    + A^2 (\tau )
  ( h_1 ^2 + h_2 ^2 +  \tilde   h_1 ^2 +  \tilde   h_2 ^2 
+ {2\over \cosh \tau }  ( h_2  \tilde   h_2 - h_1  \tilde   h_1 ) )
],\cr &&  [A^2 (\tau ) = 2 ^{-7/3} \coth \tau (\sinh 2\tau - 2\tau
  )^{1/3}  = {1\over 4} \cosh \tau K (\tau )  =  
{\calf ' (\tau ) \over  4  e ^{4/3}  A^2(\tau )  \tanh \tau } 
,\cr && B^2 (\tau )  = 2 ^{-1/3} 3 ^{-1} \sinh ^2 \tau (\sinh
  2\tau - 2\tau )^{-2/3}   = {1\over 6 K^2 (\tau ) }].   \label{appeqdefco1} \eea 

Another    useful  parameterization  of the  deformed   conifold metric~\cite{gimon02}  is  obtained by   representing the  matrix $W$ by   the  product of  $2\times 2 $   unitary matrices $S \in {SU(2)\over  U(1)} \sim S^2 ,\ T \in SU(2)  \sim S^3 $,    associated  to the  collapsing 2-sphere  and the   blown-up 3-sphere    near the  apex, of angle coordinates  
$\t , \phi  $  and   $ (\t _2, \phi  _2, \psi  )  \sim \o _a $, 
\bea && {1\over \sqrt 2} W   = i X P  
\equiv {\e }  T S e ^{\tau \s _3  \over 2}
 S ^\dagger  \s _3 ,\ [X= -i T \s _3 ,\ P =
{\e } U \pmatrix{e ^{\tau /2 } & 0 \cr 0 & e ^{-\tau /2 }}
U^\dagger ,\ U = \s _3 S \s _3 , \cr &&   S = e^{i \phi \s _3 /2} e^{-i \t \s _1/2}
,\  S^\dagger d S = {i \over 2}  s ^a \s  _a 
= {i\over 2} (- d\t \s _1 +\sin \t d\phi \s _2
+ \cos \t d\phi \s _3 ) ,\  T^\dagger d T = {i \over 2} \o ^a \s  _a
,\ [a=1,2,3]  ] .    \label{app1.eq7p}  \eea
The   resulting   expression of the  metric    solution is given by 
\bea && d \tilde s ^2 (\calc _6) = {\e ^{4/3} \over 6
  K^2 (\tau ) } (d\tau ^2 + H_3 ^2 ) + {\e ^{4/3} K (\tau ) \over 4 }
     [\cosh ^2({\tau \over 2 }) (H_1 ^2 + H_2 ^2 ) +
 \sinh ^2 ({\tau \over 2 })
  ( ( 2d\t  +H_2 )^2 +(2 \sin \t d\phi - H_1 )^2 ) ] ,\cr &&
[\pmatrix{H_1 \cr H_2\cr H_3 } = \pmatrix { 0 & \cos \t & -\sin \t \cr
    0 & 1 & 0 \cr 0 & \sin \t & \cos \t } \pmatrix { \sin \phi & \cos
    \phi & 0 \cr \cos \phi & - \sin \phi & 0 \cr 0 & 0 & 1 }
  \pmatrix{\o _1 \cr \o _2\cr  \o _3 } ,\
\pmatrix{\o _1 \cr \o _2\cr  \o _3 } =- \pmatrix{\cos \psi d \t _2 +  \sin \psi \sin  \t _2 d \phi _2  \cr  -\sin \psi d \t _2 +  \cos \psi \sin  \t _2 d \phi _2  \cr  d\psi + \cos \t _2 d \phi _2 } ]    \label{app1.eq8} \eea 
where   we have displayed  in the second line    the  rotations
transforming  $ \o _a \to H_a $, with    $\calr  _{\t }^{(i,j)} ,\  \calr _{\phi }^{ ' (i,j)}    $     being $SO(3)$  representation  matrices for $ S$
associated  to rotations    in the planes  $ (i,j)$  of the space $ R^3$ embedding   $S^2 (\t ,\phi )$.  The bases of 1-forms $ s _a$ and $ \o _a $  
are  related by $ \o ^a S^\dagger  \s _a S= (H_1 \s _1 -H_2 \s _2 +H_3\s _3 )
   \ \Longrightarrow \   H_a =   (\calr _{\t }^{(2,3)} \cdot  
   \calr _{\phi  }^{ (1,2) '} )_{ab} \o _b  .$ 
   Identifying    the     expressions   of $  Tr (d W^\dagger d W )
,\ \vert  Tr ( W^\dagger  d W )  \vert ^2 $   in   the 
parameterizations of $W$ in  Eqs.~(\ref{app1.eqKSp})  and
(\ref{app1.eq7p})  allows expressing the bases of 1-forms
$ g^{(a)} $   as  linear combinations of the 1-forms $s_1, \ s_2 $ and      $ H_a$ ($S$-conjugates  to    the    $\o _a$), 
 \bea && (g^{(1)} + ig^{(2)}) = {e ^{i
    \psi } \over \sqrt 2 }  ( (H_1 -  2 \sin \t d\phi )  + i (H_2 + 2  d\t ) ) ,\   (g^{(3)} + ig^{(4)}) =  {e ^{i \psi } \over \sqrt 2 }
 (H_1 + i H_2),\ g^{(5)} =  H_3.  \label{app1.trsf}  \eea
The  $\tau \to 0$  limit  of  the  metric in Eq.~(\ref{app1.eq8}),    
\bea &&   d \tilde s ^2(\calc _6)  \simeq {\e ^{4/3}  \over
  2^{5/3} 3 ^{1/3} }  (d\tau ^2 + \sum _{a=1}^3 \o _a ^2 +  \tau ^2 
( (d\t + \ud  H_2 )^2 + (\sin \t d\phi - \ud H_1 )^2 ) )  \cr && \Longrightarrow
  \ \tilde r(S^3)=  \tilde  r(S^2) /\tau  = \e ^{2/3} 2 ^{-5/6} 3 ^{-1/6} ,\ 
[r^2 (S^3) = h ^{1/2 } (0) \tilde r^2(S^3)= 
  2^{-4/3} 3^{-1/3}  a_0^{1/2} g_s M \a ']  \label{app1.eq9} \eea
shows that    the       geometry   near the
apex    involves  the collapsing $S^2 (\t , \phi ) $  fibred over  the
blown-up $S^3 (\o ^a)$,  where the     formulas for  the unwarped
and  warped     radii  $ \tilde r (S^k)  ,\ r (S^k)$ were 
inferred   by    means of the familiar method~\cite{minasian99}.  

We  observe in conclusion  that the  deformed conifold   stands out
as  the  prototype    for   conic  Calabi-Yau  threefolds realizing
AdS/CFT     string-gauge   theory duality by quiver gauge theories
with a  renormalization group  flow   of Seiberg cascading  type.
Meanwhile,   several  families  of 6-d conic  Calabi-Yau   throats  
with  horizons   given by Sasaki-Einstein  bases and similar duality  properties  were  discovered. 
One  example  is the  infinite family of  5-d manifolds $
Y^{p,q}$ of  topology $ S^2 \times S^3 $  providing horizons of
conic  Calabi-Yau   manifolds  labelled  by relative prime integers $ p, \ q $,  which  arise  as    partial toric resolutions
of  $ C^3 / (Z_p \times Z_p)$ orbifolds.  The string theory compactifications on the asymptotic
$AdS _5 \times Y^{p,q}$ spacetimes~\cite{martelli04}  are dual to superconformal
quiver gauge theories~\cite{benvsparks04}.     The   supergravity
solution at large  radial distances from the apex  region is discussed
in~\cite{herzov04}. An iterative  construction of the quiver gauge theories  on  $ N  D3$-brane    probes is presented 
in~\cite{benvenuti04}  and the embedding  of  supersymmetry  preserving   
flavour $Dp$-branes  for $ p=3, 5, 7 $   is  discussed in~\cite{canoura05}. 

\subsection{Harmonic  analysis}
\label{appwdcsubII1}

Harmonic analysis   is the branch of mathematics dealing with the  decomposition of   smooth fields on manifolds   using  orthonormal bases of   square integrable functions    given by eigenfunctions of the  manifolds  Laplace-Beltrami  operators.    We  here   summarize    the
decomposition of scalar fields on  harmonic   functions  of
Stenzel  spaces $\calc _{2d-2} $   developed   in~\cite{pufu10}  in terms 
 of the  constrained systems of  coordinates  $ y_a  $  for  the  radial   sections   $ V_{d,2} =SO(d)/ SO(d-2)$.

The     wave functions in fixed representations of $SO(d)$ are   given by 
linear   combinations  with  $\tau $-dependent  coefficients 
of   homogeneous polynomials    of    the  constrained systems of  coordinates  $ y_a  ,\ \bar y_a $  for  the  radial   sections   $ V_{d,2} =SO(d)/ SO(d-2)$  of  fixed  degrees  $n_1, \ n_2 ,\ [n=n_1+  n_2]$, 
\bea && F _{n_1, n_2 } (y_a,\bar y _a) =  \sum _{a _k, b _l} M_{a_1 \cdots a _{n_1}
} ^{b_1 \cdots b_{n_2} } y _{a _1 } \cdots y _{a _{n_1} } 
\bar y _{b_1 } \cdots \bar y _{b _{n_2} } , \label{eqapp.Poly}  \eea 
where  the  constant  coefficients $ M _{ [a
    _k]} ^{ [b_l]}$  are     fully   symmetric under
separate  permutations   of the  upper  or   lower indices,
$ a_k  $  or $ b_l \in (1,\cdots , d)$,  and traceless   under  contractions  of     pairs of these    indices. The  tensors $M$    transform    as   direct    products of the $ SO(d)$
representations  associated to the (single array) Young   diagrams,  
$ (n_1, 0^{d-1} ) $ and    $ (n_2 , 0^{d-1})$. 
For  tensors  which  are   also     traceless     under 
contractions  between        upper     and   lower indices,   
the allowed    direct    products  of $SO(d)$ representations
are those  associated to  (double  array) Young   diagrams  $ (p, q , 0^{d-2}),\ [p+q=n] $~\cite{pufu10}.  

The  metric tensor of  $\calc _{2d
  -2}$   can  be      constructed by   eliminating, say, $ w_1, \ \bar w_1$
in  Eq.(\ref{app1.eq2})   via    the conditions   
$ \sum _a w _a ^2 =\e ^2 ,\ \sum _a \vert w _a
\vert^2 =\vert  \e \vert ^2\cosh \tau $,
\bea && g_{\a \bar \b } = { \calf ' \over 2 w_1 \bar w _1 \s } {(w _{\a }
\bar w _{\bar \b } + w_1 \bar w _1 \d _{\a \bar \b }) } + {\calf ''
- \calf ' \coth \tau \over 2 w_1 \bar w _1 \s ^2 } {(w_{\a } \bar w_1
  - \bar w_{\a } w _1 ) (\bar w_{\b } w _1 - w_{\b } \bar w _1 ) }
,\ [\s = \e ^2 \sinh \tau ]  \eea 
where we have  split up the coordinates indices   as 
$ a = (1, \a ) ,\ [\a =2, \cdots , d]$. 
Grouping    the  complex conjugate indices
into   a single  index  $ A = (\a , \bar \a  ) $, 
yields  the compact matrix notation  of  the   metric
\bea && d\tilde s ^2 (\calc _{2d-2} ) = G_{AB} d w ^A d w ^B ,
\ [G_{AB}= \pmatrix{0 &   G_{ \a \bar \b } \cr G_{\a \bar \b } & 0} ,\
Det ( G_{AB} ) = - \vert Det ( G_{\a \bar \b } ) \vert ^2 = (-
1 )^{1+d} ( {\calf ' \over 2 \s } )^{2 (d-2)} ( {\calf ''\over 2 w_1
  \bar w _1 } )^2] \cr && 
 G ^{\a \bar \b } = {2 \s \over \calf ' }
\d _{\a \bar \b } - {2 \over \calf '' } [ - \bar w _{\a } w _{\b } +
  {\calf '' - \calf ' \coth \tau \over \s \calf ' } \sum _{c=1} ^d {
    (w_{\a } \bar w _{c} - \bar w_{\a } w _c ) (\bar w_{\b } w _c -
    w_{\b } \bar w _c )} ]\eea   
with the familiar  notation   for   the  inverse  metric,  $ G ^{\a \bar
  \b } = (G^{-1})_{\a \bar \b }.$  The  scalar Laplacian  
$ \sqrt {G  ^{-1} } \dh _A G^{A B} \sqrt {G }\dh _B$ is  then   given by  
\bea && \tilde  \nabla ^2 ({\calc  _{2d-2} }) = {4 \over \calf
  ''} \bigg [{1\over 2} P_{ab } \bar P_{ab} + {\calf '' - \calf '
    \coth \tau \over \sinh \tau \calf ' }  (\e ^2\sinh ^2
  \tau \d _{ab} - \cosh ^2 \tau (w _a \bar w_b + \bar w _a w_b ) + (w
  _a w_b + \bar w _a \bar w_b ) ) {\dh  ^2\over \dh w _a \dh \bar w _b} 
  \bigg ] \eea
where $P_{ab} = (w_a {\dh \over \dh \bar w _b}  -  w_b {\dh \over \dh
    \bar w _a} )  ,\  [\bar  P_{ab} = (P_{ab} ) ^\star  ].$ 
It  is convenient to express the   differential operators 
in terms of  the radial derivative $\dh _\tau
$  and the   isometry group  $ SO(d)$  Killing vectors,   
$ \xi ^{(A)} = T ^{(A)} _{ab } (y_a \dh _b + \bar y_a \dh _{\bar b}  )
,\ [a= 1,\cdots , d,\   A= 1, \cdots , d (d-1)/2]  $   for 
a suitable  basis of  the Lie algebra generators  $T ^{(A)} \in so(d)$. 
Using    the  reverse  chain rules  to    relate   $ {\dh / \dh w _b } $
to  $ {\dh /  \dh y _b } $  and $\dh   _\tau $ yields 
the compact   formula for the  Laplacian in terms of  
one   radial operator $  \calT $    and three      angular operators   
$  \calc ,\     \calr, \  \call $~\cite{pufu10},  
\bea && \tilde \nabla ^2   ({ \calc  _{2d-2} })   = \calT
+ g _\calc \calc + g _\calr \calr + g _\call \call , \ \
[g _\calc = - {2 \coth \tau \over \calf ' (
    \tau ) } ,\ g_\calr = - {1 \over \calf '' ( \tau ) } - g _\calc
  ,\ g_\call = {4 \over \calf ' ( \tau ) \sinh \tau } ,\cr &&  
\calT = \nabla  ^2 _\tau = {4 \over \calf '' \calf ^{'d-2} } \dh _\tau
\calf ^{'d-2}   \dh _\tau ,\ \calc
= - 2 \sum _{A=1} ^{d (d-1)/2 } \xi ^{(A)} \xi ^{(A)} =
  y _a y _b \dh ^2_{ab}  +  (\bar y_a y_b - \d _{ab } \bar y_c y_c) \dh
  ^2_{a\bar b} + (d-1) y_a \dh _a + H.\ c.  ,\cr &&  \calr = (y_a \dh _a -
  \bar y_a \dh _{\bar a} )^2 ,\   \call = \ud (\bar y_a y_b + \bar
  y_b y_a -\d _{ab }\bar y_c y_c) \dh ^2_{a\bar b} + {d-2\over 2} \bar y_a
  \dh _a+ H.\ c. ]  \label{eq.lapfu} \eea 
where  $\dh _b = {\dh  / \dh y_b } ,\\dh _{ \bar b} = {\dh  / \dh \bar y_b } ,\ 
\dh ^2 _{a \bar b} = {\dh  ^2   / \dh y_a  \dh  \bar y_b }   $  while $\calc $
coincides   with    the quadratic Casimir  operator  of the   $ SO(d)$ group. 

In the  conifold  case $d =4$, the scalar Laplacian    splits  up into
radial and  angular  parts, 
$ \tilde  \nabla ^2 ({\calc _6}) = \tilde  \nabla ^2 _{\tau } + \tilde 
\nabla ^2 _{5 } , \ [\tilde  \nabla ^2 _{\tau } = \tilde g^{\tau \tau }
  {G ^{-1} (\tau ) }  \dh _\tau G(\tau  ) \dh _\tau ,\  
\tilde  \nabla ^2 _{5 } = - \tilde g^{\tau \tau } V_5  ,\ G = \sqrt {\tilde
  g}   \tilde g^{\tau \tau } ] $.  The     $SO(4)= SU(2)_L \times SU(2)_R$
group   generators $ (T _{L,i } ,\ T _{R,i } ) \sim (  -i \xi ^{(A)} )$
separate  into     the two  commuting   sets of operators,
$  (J_{L,i}  ,\   J_{R, i}) $. In the action  on  symmetric-traceless polynomials,   the  operator   $\calc = 2 (J_L^2 + J_R^2 )$ identifies  to   the   $ SO(4)$ Casimir  operator   while $\calr ,\ \call $      can be expressed  in terms  of quadratic products of  the number  operator $N$  and
 the generators   $\tilde J_i $ of an auxiliary $\tilde  SU(2)$ group,
 \bea && \calr = 4 \tilde J_3 ^2  ,\ \call = -2i \tilde J_2\tilde J_3 +
N \tilde J_1 ,\ [\tilde J _+ =  y^a  \dh _{\bar a}  ,\ 
\tilde J _-  =\bar y^a \dh _a ,\  N= (y^a \dh _a  + \bar
y^a \dh _{\bar a} ),\ \tilde J _3 = \ud (y^a \dh _a  - \bar
y^a  \dh _{\bar a} ) ]  \label{eq.auxsu2}\eea  
where  $\tilde J_\pm =\tilde J _1 \pm   i \tilde J _2 ,\ [
[\tilde J _i,  \tilde J_i ] = i\e _{ijk}     \tilde J_k ]$
are  step  operators   exchanging   $ y_a \leftrightarrow \bar y_a $
and    $ \tilde J^2 = J_{L} ^2 +  J_{R} ^2 - N(N+2) /4 $. 
The operators   $ [N,\  J_L^2 ,\ J_R ^2 , \  J_{L,3}  ,\ J_{R,3} ,
\ \tilde J_3 ]   $    constitute a  maximal    commuting  set  of operators   
admitting  the simultaneous eigenvalues  $ [n=  n_1 + n_2
  ,\  j(j+1),\ l(l+1) ,\ m_L,\ m_R, \ \tilde  m  = r/2 ]
$   with  $ \tilde j (\tilde j +1) =  j (j+1) +  l (l+1) - n(n+2) /4 $. 
The  scalar Laplacian  $\tilde  \nabla ^2 ({\calc _6}) $ decomposes on the operators  $(\calT , \  \calc , \  \calr , \ \call ) $ as in   Eq.~(\ref{eq.lapfu})   with the      coefficients given by  
\bea && \calT = \nabla _\tau  ^2 = {6\over \e
    ^{4/3} \sinh ^2 \tau } 
{\dh_ \tau } (\cosh \tau \sinh \tau-\tau ) ^{2/3} {\dh _\tau } = {96 \over  \e ^{4/3} \sinh ^2 \tau }
\dh _\tau A^4 \tanh ^2 \tau  \dh _\tau ,   \cr &&  
g _\calc = - {\coth ^2 \tau \over  2  \e ^{4/3} A ^2 (\tau ) }
,\  \ g_\calr = -  g _\calc  - {1 \over 4 \e ^{4/3} B ^2 (\tau )  } 
,\ g_\call = {\coth \tau  \over  \e ^{4/3} \sinh  ^2 \tau A ^2 (\tau )}
  \label{app1.Yeq10}  \eea
where  $ A,\ B $ were defined in Eqs.~(\ref{appeqdefco1}). 
One can use the identity \bea &&  g_\calc \calc + g_\calr \calr = g_\calc (\calc - \calr ) + (g_\calc  +  g_\calr ) \calr   = + {\coth ^2 \tau \over \e   ^{4/3} A^2 }
( \nabla _1 ^2 + \nabla _2 ^2 -  \nabla _\psi ^2 )  -  \e ^{4/3} B^2
\nabla _\psi ^2 , \eea    where  the    first term  identifies to    the
coset  space Laplacian,   $  \nabla _{G/H} ^2 =\nabla _{G} ^2   - \nabla _{H} ^2 $,
to obtain the explicit form  of  the scalar Laplacian~\cite{krishtein08}     in the  Euler  angular  coordinates  of  $ T^{1,1}$ in Eq.~(\ref{eqKSmet1}), 
\bea &&  \tilde  \nabla ^2 ({\calc _6})   = \nabla ^2 _\tau + 
{1\over \e ^{4/3} B^{2}}  \dh ^2 _{\psi } + {\coth ^2\tau \over
  \e ^{4/3} A^{2}} (\nabla ^2 _{1} +\nabla ^2 _{2} )+ 
{\cosh \tau \over   \e ^{4/3} A^{2}  
\sinh  ^{2} \tau  }  \nabla ^2 _{m } ,\cr && 
[\calT = \nabla ^2 _\tau ,\  
\calc =-2 (\nabla ^2 _1 +\nabla ^2 _2) - 4 \nabla ^2 _\calr  ,\ 
\calr = -  4 \nabla ^2 _\calr =-\dh^2 /\dh \psi ^2 , 
\ \call = \nabla ^2 _m ,  \cr && 
\nabla ^2 _{i}   = (\sin \t _i )^{-1}\dh _{\t _i } \sin \t _i \dh _{\t
  _i } + ((\sin \t _i )^{-1}\dh _{\phi _i } - \cot   \t _i \dh _{\psi
} )^2]   \eea
where   $\nabla ^2 _{1,2} $ are the Laplacians of $ SU(2) _{j,l}  $ in the 
metric $ ds^2 (S^3_i)  = d \t _i^2 + \sin ^2 \t_i  d\phi _i ^2 +
(d\varphi _i  + \cos \t _i d \phi_i  )^2 $.  

 In the representation space   of the  symmetry  group
 $ SU(2) _j \times  SU(2) _l $,  the   symmetric products   $  (y_a)
^{n_{1} } \sim ([\ud , \ud ]) ^{n_{1 } }  $  and $ (\bar y_a)  ^{n_{2}
} \sim ([\ud , \ud ]) ^{n_{1} } $    carry the representations,   $ [{n_1 \over 2}  ,{n_1 \over 2} ] \times  [{n_2 \over 2},  {n_2 \over 2} ] $  
while  the homogeneous  polynomials  $ F_{n_1,  n_2 } $ in   Eq.~(\ref{eqapp.Poly}), 
of   fixed  total  degree  $ n= n_1   + n_2$,   span
reducible representations  which  decompose  into  sums over  $ k$  increasing  by  2 unit steps,  $F_{n_1,  n_2 }(y,\bar y) \sim   \sum _{k=0} ^{\min (n_1, n_2)}     
([{n\over 2} , {n\over 2} -k ] \oplus [{n\over  2} -k , {n\over 2} ] ) $. 
The irreducible representations  $ [j_L, \ j_R]  $, 
appear  with successive   (-2)    incremental   shifts
from       $ n_1 + n_2 +1  $  down to  $\vert n_1 - n_2 \vert  +1 $   (or from $ [j, j]  $   down to  $ [j,l] + [l,j]$)   with  multiplicities  $ n+1 -2k $  (or  $ 2 \tilde j +1,\    [\tilde j = \min  (j_L, j_R)] $),  where   the multiple   copies  of   each  irreducible representation are labelled by  
the  eigenvalue of $\tilde J_3   = \tilde m  \in 
( -\tilde  j ,\cdots , \tilde  j  )$.
The  polynomials associated  to the states  $\vert n; (j_L, m_L), (j_R, m_R); (\tilde  j,   \tilde  m  ) >$  form a basis  of the direct  product
subspaces   of angular momenta   $(j,  l) $ and $\tilde j$ 
and  magnetic  components $  (m_j,  m_l), \ \tilde m $. For  given 
$ n ,\ \tilde j $, there occur $(2 \tilde j +1 )$   representations
of  $(j,  l) $ type   labelled  by  $ \tilde m $,
in which the   operator $\calc $  is    represented
by  $2 (j (j+1) + l(l+1) )  $ times the unit  matrix,   $
\calr  $   by      a diagonal matrix of entries $ r^2= (2 \tilde m )^2 $  
and       $\call $    by  an off-diagonal matrix
whose  non-vanishing      entries   are given by
$ <...;  \tilde m  \pm 1 \vert  \call \vert ...; \tilde m > =
(  ({n\over 2} \mp \tilde m ) (\tilde j \pm \tilde m  + d ) )^{1/2}  $. 
  The  Laplace operator  eigenfunctions
  are  then  given  by   the   linear combinations 
\bea &&  \calf  ^{j l ( \a ) } (\tau , \T ) = \sum ^{\tilde j  }
_{\tilde  m  =  -\tilde j  }   
f ^{(\a )}  _{\tilde j }   (\tau )  F _{n_1, n_2 } ^{j l} ( y, \bar y ) ,\  
[\a  = 1,\ \cdots , 2   \tilde j +1 , \  \tilde j = min (j ,   l ) ]  \label{app1.eqI13}  \eea
where the     radial  wave functions   $ f ^{(\a )} _{ \tilde j } (\tau ) $ satisfy a  $ (2 \tilde j +1)$-dimensional     system  of second-order differential equations in $\tau $. 
Owing to  the invariance  under the  $ Z_2$ parity,   changing 
$ y _a \to \bar y_a  $ (or $\tau \to -\tau $),   the  wave functions 
of  fixed $(j, \ l)$   decompose into  two  decoupled  sets  of same
$r$-parity    polynomials. For illustration, we quote  below the      
equations~\cite{pufu10}      in  the two  lowest     representations 
$ (j_L,j_R) =(j, l)$  
\bea && \bullet \  (j, 0) ,\ \tilde j=0 :\ [\calT 
+ 2 j(j+1) g_\calc + E_m ^2 h  (\tau ) ] f ^{(0)}  =0,\cr &&  \bullet
\ (j=\ud , l= \ud  )
,\ \tilde j=  \ud    :\ [\calT + 3  g_\calc + g_\calr \pm g_\call + E_{m
    \pm }  ^2    h (\tau ) ]  f _1^ {(\pm  )}  (\tau )  =0,\  
  [f _1^ {(\pm  )}   = (f  _{1} \pm f _{-1} ) ]. \eea

The  polynomial  basis    $F^{j, l} (y,\bar y)  $     is   equivalent
to  the  basis of the  coset space $ T^{1,1} \sim
SU(2)_L \times SU(2)_R / U(1)_q$  harmonic  functions, 
\bea && (\nabla ^2 _{ T^{1,1}}   + H_0 ^\nu  ) Y ^{\nu }
_{  (m_L, m_R) } (\T )  = 0,\  [H_0 ^\nu  = 6 (j(j+1) +l(l+1) -{r^2\over 8 } ) ,\
\nu = ( j l r) ].  \eea
In the  Young tableau  for the direct product  representations,  $ D^ {j_L}
\times D^ {j_R}  $,  the   $( j_L, j_R)$ tableaux
differing by the positions of $ \pm 1/2 $ states  in  the  boxes
appear  in  $ 2 \tilde j +1  ,\ [\tilde j = \min  (j_L, j_R)] $  copies 
associated to the     basis of harmonics $ Y^{j_L, j_R}$   of $ SU(2)_L
\times  SU(2)_R$~\cite{ceresole99,ceresoII99}.   These are 
labelled by  the $ r= 2  \tilde  m $ charge,   eigenvalue of  the Cartan generator $ -i( T_{L,3} -  T_{R,3} )$. The relations between 
the polynomials and harmonic  functions   bases
can be  determined by  acting with  the  ladder  operators~\cite{pufu10},
$ \tilde J_ \pm $    and   $ J_{L, \pm } - J_{R, \pm } $. 

The   10-d    wave  operator for
graviton fields $ \Psi $,   decomposed on  harmonics of     $\calc _6$,
is given by  the scalar Laplacian
 \bea && 0=  ( \nabla
^2 _{10} -\mu ^2 ) \Phi  =  h^{-1//2}  (\tau )  (h   (\tau ) \tilde \nabla _4 ^2 +
 \tilde \nabla _6 ^2  -h^{1/2}  (\tau ) \mu ^2 ) \Psi (X),\cr && 
 [\Psi  (X) =  \sum _{jl ,\a }    \phi ^{jl , (\a )} (x)   \calY ^{jl ( \a ) }     (\tau, \T ) ,\ \caly ^{j l ( \a ) } (\tau , \T ) = \sum ^{\tilde j
  }  _{r =  -\tilde j  }  f ^{(\a )}  _r (\tau ) Y ^{j l r} (\T ) ,\
\a  = 1,\ \cdots , 2
  \tilde j +1 , \  \tilde j = min (j ,   l ) ].   \label{app1.eq13}   \eea
The second-order differential  wave equations for the  wave functions $ \calY ^{jl ( \a ) }  (\tau , \T ) $,   
\bea && 0= (h^{1/2} (\tau ) \tilde \nabla _4 ^2 + h^{-1/2}  \tilde \nabla _6 ^2
-\mu ^2 ) \calY ^{jl ( \a ) }  (\tau , \T )  = g ^{\tau
    \tau } ( G^{-1} \dh _\tau G \dh _\tau + \tilde g_{\tau \tau } (h
  \tilde \nabla _4 ^2 - h^{1/2} \mu ^2 + \tilde \nabla _5 ^2 )  )\calY
  ^{jl ( \a ) }(\tau , \T ) ,    
\cr  &&     [\tilde \nabla _5 ^2 = g _\calc (\tau ) \calc + g _\calr (\tau ) \calr
  + g _\call (\tau ) \call    ,\  \ \  {1\over G}  \dh _\tau G \dh _\tau 
= {1  \over  S^{2/3} } \dh _\tau S^{2/3} \dh _\tau  =  
{\e ^{4/3} \sinh ^2 \tau \over 6 S^{2/3} } \calT = \tilde g _{\tau \tau } \calT  
,\cr && S ( \tau ) \equiv  (\sinh  \tau \ - \tau  )  ,\ 
\tilde g _{\tau \tau } = {\e ^{4/3} \over 6 K ^2(\tau ) } ,\  
\sqrt {\tilde \g _5} G (\tau ) 
= \sqrt { g_{10} } g  ^{\tau \tau } = \sqrt {\tilde g_{10} } \tilde g
^{\tau \tau } = {\e ^{8/3}\over 16} K^2 (\tau ) \sinh ^2 \tau \sqrt {\tilde \g _5}  ]  . \eea
can  be  transformed  to  a  system of  coupled 
 second-order radial differential  equations  
 by substituting   the angular  Laplacian $ \tilde \nabla _5 ^2$
 by   a  $\tau$-dependent matrix in the  vector space  of the
radial wave functions   $f ^{(\a )}  _r (\tau ) .$ 
In the large distance limit, where  $\calc \to 2 (j (j+1) + l(l+1) ) ,\ \calr  \to  r^2 ,\ \calr  \to 0,$  the latter simplifies to  the
diagonal matrix, 
\bea && - V_5 = \tilde g_{\tau \tau } \tilde \nabla _5 ^2 =  {e
  ^{4/3} \over 6 K^2 (\tau ) } [ (g _\calc \calc +  g _\calr \calr  ) 
  + g _\call \call  ] \simeq  - {M_5 ^2 \over 9} = - {H_0 \over 9}.\eea

\subsection{Approximate  separable   conifold  geometry}  
\label{appwdcsub2}

The technical   complexity     of  harmonic   analysis
on the  deformed conifold  case  can be   circumvented    by    using
an approximate separable  ansatz  for  the metric.
The change   proposed   by  Firouzjahi and Tye (FT)~\cite{firouz06} involves   
a slight   modification of    Klebanov-Strassler  metric replacing
the component along   $ g ^{(5)}  $  as,
$1/ (3 K^{3}(\tau ) ) \to \ud \cosh ^2 ({\tau \over 2})  $. (The
replacement   $ K (\tau  ) \approx  K \vert _{FT}   = 
({2\over 3})^{1/3} / \cosh ^{2/3}  (\tau /2) $  is  accurate    near
the  origin but  not  at large  $\tau $, 
where $ \lim _{\tau \to \infty }  (K (\tau  ) /  K \vert  _{FT} )
\to  (3/4)^{1/3} .$)  The  geometry in this approximation  is that of a cone
over a  compact  base $ S^3 \times S^2 $,    
\bea &&  d\tilde s ^2  _{KS} \to  d\tilde s^2_{FT}  = {\e ^{4/3} K (\tau )
  \over 2} [ {d\tau ^2 \over 3 K^3 (\tau ) } + \cosh ^2 ({\tau \over
    2} ) d \O ^2 (S^3)  + \sinh ^2 ({\tau \over 2} ) d \O ^2 (S^2)]
,\cr && [d \O ^2 (S^3) = \ud g ^{(5) 2} + g ^{(3) 2} + g ^{(4) 2}
  ,\ d \O ^2 (S^2)  = g ^{(1) 2} + g ^{(2) 2} ,\ Vol (S^{3}) =
8 \pi ^2,\ Vol (S^{2}) = 4\pi ] \label{app2.eq3} \eea
with  an   $ S^3 $     of finite  radius and  an $ S^2 $  that   collapses    near the apex,  as  seen on the limiting  formula at $ \tau \to 0 $,     
\bea && d\tilde s ^2 _6 \simeq { \e ^{4/3} \over 2
  ^{2/3} 3 ^{1/3} } ( {d\tau ^2 \over 2 } + d \O ^2 (S^3)  + {\tau ^2
  \over 4 } d\O ^2 _{S^2} )   , \ [\tilde r^2 (S^3)= {\e ^{4/3} \over 2 ^{2/3}
    3^{1/3} } ,\ \tilde r^2 (S^2)={\e ^{4/3} \tau ^2 \over 2 ^{8/3}
    3 ^{1/3} }] . \label{app2.eq7}   \eea   
The   integration  measure     and the auxiliary       function 
$G(\tau )$,   analogous to those   in Eq.(\ref{sect2.eq4}),   are   now given    by  
\bea && \sqrt {\tilde g_6} = {\e ^4
  \over 2^3 \ 3 ^{1/2} } K^{3/2} (\tau ) \cosh ^3 {\tau \over 2 }  \sinh ^2
{\tau \over 2 } \sqrt {\tilde \g _5} ,\ G ( \tau ) = { \sqrt { \tilde g_{10} } \over \tilde g  
  _{\tau \tau } } = {\e ^{8/3} 3 ^{1/2} \over 2^4} K^{7/2} (\tau )
\sinh ^2 \tau  \cosh {\tau \over 2 }  \sqrt { \tilde \g _5} .\eea

Since the   metric is  factorizable,  the   Kaluza-Klein reduction  of the  metric tensor components   along $ M_4$  introduces graviton modes   described by products of radial wave functions   times harmonic functions of $S^{3} $ and $ S^{2} $.  For the   round   $S^{3} $ metric  in hyperspherical  coordinates,
\bea &&  ds ^2 ( S^3 ) = d   \chi  ^2 +
\sin ^2  \chi \ ( d\t ^{2} + \sin ^{2} \t d \phi ^{2} ) ,\ [\t \in [0,\pi ],\ \phi \in [0, 2\pi ] ,\ \chi  \in [0,\pi ] ] \eea 
the basis of  harmonic    scalar  Laplacian eigenfunctions,
$ (\nabla ^2 _{S^3}  + j^2 -1 ) Q ^j _{L, m} = 0 $,   is given  by   Fock polynomials $\Pi _L^j  (\chi )  $~\cite{gerlachgupta73} times   spherical harmonic (Legendre) polynomials  $Y^L_m (\t , \phi ) $,
\bea && Q ^j _{L, m} (\chi  , \t , \phi ) = \Pi ^j _L (\chi ) Y ^L _m ( \t , \phi ) ,\ [\Pi ^j _L (\chi ) = (\sin \chi  )^L  {d ^{L+1} \cos (j \chi ) \over d
  (\cos \chi ) ^{L+1} } ,\ Y^ {L }
_m (\t , \phi ) = P_L (\cos \t ) e ^{i m \phi }   ,\cr &&
j=1,  2 , \cdots  ,   \   L=0,  \cdots , j-1 ,\ 
m =(-L, \cdots ,   L )  ,\   (\nabla ^2 _{S^2}  + L(L+1) ) Y ^L_m (\t , \phi )  =0 ]  \label{app2.eq11} \eea 
labelled by  the  principal integer  angular  momentum    $ j \geq 1   $   and the  secondary integer  angular  momentum $L\in [ 0,\  (j-1)] $, 
setting the  total  $j$ quantum number degeneracy at 
$\sum  _{L=0} ^{j-1} (2L+1) = (1 + 3 + 5 +\cdots + (j-1)) =j^2$. 
The    scalar Laplace operator eigenfunctions for the    $S^{2} $ round metric  are  the  familiar spherical harmonics, 
$(\nabla ^2 _{S^2} +   l(l+1) ) Y^l _m (\t ,\ \phi ) =0$,
expressed in terms of  Legendre polynomials.

The graviton modes   wave functions  in representations  $(j,l)$
factorize   into  products of radial  functions by angular functions 
labelled  by the  radial excitation index  $m$  and  the     harmonic  basis degeneracy index $r$,  
$\Psi_{m, r}  (\tau , \T )  = R_{(j,l), m, r}   (\tau )  \Phi _{(j,l) ,r} (\T ) $.
The   rescaled  radial wave functions,
$B_m  (\tau )  = R_m  (\tau )   \sqrt  {G(\tau )  } $,  are governed by    a Schr\"odinger  type    wave equation
$(\dh _\tau ^2 - V_{eff}  ) B _{m}  (\tau ) \Phi _{(j,l) ,r}  \T ) =0 $  with 
effective potential    depending  on  the mode   mass $ E_m$ and
 charge  quantum numbers $(j, l)$, 
\bea &&  V_{eff} (\tau , \T ) = -  g_{\tau \tau } ( E_m
  ^2 h^{1/2} (\tau ) -\mu ^2   +h ^{-1/2}   {K^{ab} (\tau ) O^{ab} (\T )
  \Phi _m \over \Phi _m } ) + G_1 (\tau )
\cr  && = -  { \e ^{4/3} h ^{1/2} \over 6  K^2
(\tau ) } ( E_m ^2 h ^{1/2} -\mu ^2 ) - {1 \over 3
K^3 (\tau ) }  ( { \tilde \nabla ^2 _{S^3} \over 
\cosh ^2 ({\tau \over 2 } ) }  +  { \tilde \nabla ^2 _{S^2}
\over   \sinh ^2 ({\tau \over 2 } ) }  )  + G_1 (\tau )
\cr  && = -  {\hat E_m ^2   \over   2 ^{1/3}  3 } { I(\tau )  \over  K^2
(\tau ) }  +   \hat \mu ^2  {I ^{1/2} (\tau ) \over K ^2 (\tau
  ) }  + {1 \over 3 K^3 (\tau ) }  ( { j^2 -1 \over 
\cosh ^2 ({\tau \over 2 } ) }  +  { l (l+1) \over   
\sinh ^2 ({\tau \over 2 } ) }  )  + G_1 (\tau ) , \cr &&
[\hat E_m^2 =  \e ^{-4/3} (g_s M \a   ' ) ^2    E_m^2 ,\ \hat \mu ^2 =   
{2 ^{1/3} \over 6}  g_s M \a ' \mu  ^2 ,\ G_1 = {(G^{1 \over  2} )'' 
\over  G^{1 \over  2}   } ,\   \mu ^2 =   {4\over \a ' } (N_\star -1 ) ,\    [N_\star -1= 0, 1, \cdots ] ].      \label{app2.eq8} \eea
The repulsive  centrifugal  barrier term, $\d V_{eff} = {l(l+1) /(3 K^3 (\tau ) \sinh ^2    (\tau /2)  ) } $  (finite for $l\ne 0 $), arising from 
the contribution  of the  $ S^2$-sphere which  collapses   near the origin,  
   produces    an  inner classical turning point  at   $ V_{eff}  (\tau _0' )=0 $.  

\begin{figure} [t]  
\epsffile{   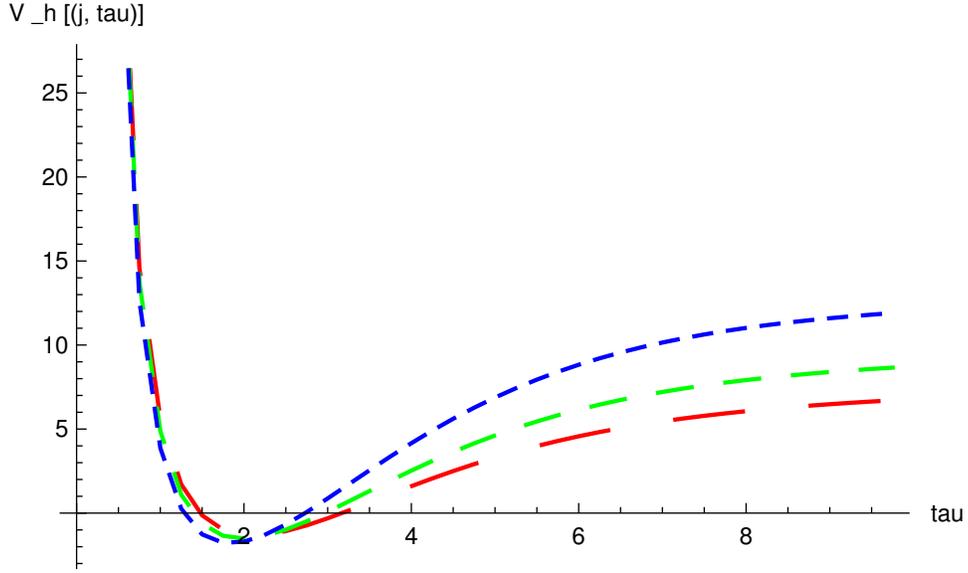}
\caption{\it \label{wcofb1}  
  Plots  versus   $\tau $  of  the effective radial  potentials  $ V_{eff}   (\tau   ) $   of  graviton  modes in the  approximate metric~\cite{firouz06} of the deformed conifold. The   red-green-blue coloured  curves
drawn   with  dashes of decreasing   sizes
refer to the states  of angular momenta  $ l=2,\  j =1,2, 3$.}   
\end{figure}

We  have evaluated the mass spectra   for  a set of charged  and radially excited  graviton   modes   by  means of   the  WKB   
prescription   described  in Eq.~(\ref{sect2.eq10}). 
The    phase integral   with  $\d _W = \d _{l,0}$ 
sets  at $ (n-{1\over 4}) \pi $  for $l=0$ and 
$ (n-{1\over 2}) \pi $  for $l > 0$. 
The effective potential  $\tau $-profiles  for $ l=0$ modes are similar    to
those    found  in  Section~\ref{sect2sub1}, but those for  $ l> 0$ modes  
have  centrifugal barriers rising at  turning points   located at     $\tau _0'  =1.2 - 1.5 $,  as  is apparent on   Fig.~\ref{wcofb1}. The     
weak dependence on $j$  of the potentials  in the  well regions  becomes stronger   at larger $j$ in the  tail regions.
We  display in  Table~\ref{tabFT2}   results for   the  masses  of   a set of charged, radially  and string excited graviton   modes.
The mass gaps   $\d \hat E_m $  for the  first  radial and string   oscillator
excitations  lie  both  at $  1$, while   that  for the first  $l$-excitation  lies at $2$.
Our  results   for $ \hat E _m (l, j)= [ 2.03, 3.62, 5.09, 3.97 ]
$ at $  (l,j )=  [(0,1) , (0,2) , (0,3), (1,1) ]  $   are fairly close  to those quoted in~  \cite{firouz06}, $[ 2.02, 3.61, 5.49, 4.0] $. 
To  speed up   the numerical solution convergence, one  could  for expediency impose a hard  wall    boundary    at $  \tau _0'   \simeq =0.5  $, 
but   the iterative   resolution  procedure fails to converge  for $ l>2 $.   

\begin{table}  \begin{center}  
\caption{\it \label{tabFT2}  
Masses    $ \hat E^{(N_\star ,l,j)} _m $  of the  ground state and first
two  radial  excitations ($  m=1,2,3   $)  for graviton modes
in singlet and charged   representations of   angular momenta   ($l= 0, 1, 2,\ j=1,2,3$    in the approximate $ S^2\times  S^3 $  geometry of  the  deformed conifold~\cite{firouz06}
and   in ground state and  first  string   oscillator excitation 
($ N_\star-1  = 0, \ 1) $.} 

\begin{tabular}{|c| c|c|c|} \hline
 $\ \hat E^{(N_\star -1 ,l , j)} _m \ $ & $ j=1 $ & $ j=2 $ & $j=3 $ \\  \hline
$ m $ & $ 1 \ \ \ \  2 \ \ \ \ \  3   $ & $ 1 \ \ \ \  2 \ \ \ \ \  3  $
  &$1 \ \ \ \  2 \ \ \ \ \  3  $   \\ \hline    
$ \hat E^{(  0, 0  , j)} _m  $&$ {2.03\ \ \  3.07\ \ \  4.11}  $&
  $ {3.62\ \ \  4.72\ \ \  5.78} $&$   {5.09\ \ \  6.23\ \ \  7.31}  $    \\  
$ \hat E^{(  0, 1 , j)}_m   $&$ {3.97\ \ \  5.05\ \ \  6.23}  $&${4.52\ \ \  6.25\ \ \  7.48}   $&$
{6.38\ \ \  7.66\ \ \  8.95}$          \\   
$ \hat E^{(  0, 2  , j)}_m $&$ {6.09\ \ \  7.17\ \ \  8.23} $&$ {6.89\ \ \  7.96\ \ \  9.07}  $&$ {7.96\ \ \  9.05\ \ \  10.2}  $         \\   \hline \hline 
$ \hat E^{(  1, 0  , j)} _m $&$ {3.23\ \ \  4.40\ \ \  5.52}  $&$ {4.28\ \ \  5.44\ \ \  6.55} $& ${5.54\ \ \  6.71\ \ \  7.83}   $ \\  
$ \hat E^{(  1, 1  , j)} _m  $&$ {4.78\ \ \  5.90\ \ \  6.98}  $&$ {5.66\ \ \  6.77,
  7.85} $&$ {6.80\ \ \  7.90\ \ \  9.00} $  \\   \hline
$\hat E^{(  1,  2, j)} _m   $&$ {6.65\ \ \  7.76\ \ \  8.84} $&$  {7.35\ \ \  8.45\ \ \  9.53} $&$  {8.33\ \ \  9.44 \ \ \  10.5}  $  \\   \hline 
\end{tabular} \end{center}   \end{table}    \vskip 0.3 cm
  
\subsection{Undeformed  conifold  limit}
\label{appwdcsub3} 

We consider the  approximate  construct proposed in~\cite{firouz06}
using a large $\tau $ limit   of the metric  in  the    $ S^2\times S^3 $  geometry  which leads to analytic solutions for the modes  wave funnctioms   
analogous   to those  of the hard wall  model~\cite{chemtob16}. 
The modified effective potential is defined through  the change   of radial variable $ \tau \to \rho $    by the  ansatz
\bea && V_{eff}(\tau )
=  - {4 \rho ^2 \over  9 \nu ^2 } +   V_5 (\tau )+ G_1 (\tau ) ,\ \
[\rho = {\nu \l _1 \over 2} e ^{-{2\tau \over 3 \nu }} = {3 ^{1/2}  \e ^{2/3} \l _1 \over  2 ^{5/6}  r } , \cr &&     G_1 (\tau ) \to G_1 = 4/9  ,\    \l _1 = a \hat E_m  ,\   V_5 (\tau )\to Q^2  = \hat  c\hat \mu ^2 + \hat  d(j^2 -1) + \hat  f l(l+1) + \hat g   ]     \label{app2.eq15} \eea
depending on the free  adjustable  parameters  $a ,\ \nu $ and $\hat  c,\ \hat  d ,\ \hat  f  ,\ \hat g  $  associated to  the radial potential and the angular potential  $V_5$. 
(Compare to the  large $\tau $  limit for $T^{1,1}$ base manifold,
$V_5  (\tau ) = - \tilde g _{\tau \tau }   \tilde \nabla _5 ^2
\simeq  - \tilde \nabla {T^{1,1}}^2 /9   \to   H_0 ^\nu / 9$.) 
The modes   wave functions are  given     by radial  functions $ R_m (\rho ) = B_m  (\rho ) / G^{1/2} (\tau )  $  times  harmonic  functions of $ S^3 $ and $ S^2$,   where the radial  wave equation,   $0=  (\dh  ^2_\tau  - V_{eff} ) B_m  (\tau )$, can   be solved   in terms of   Bessel functions,      
\bea && 0= ( \rho  \dh _\rho  \rho  \dh _\rho  -
{9 \nu ^2 \over 4}  \tilde V_{eff} ) B_m = (\rho ^2 B'' _m + \rho  B ' _m + (\rho ^2 -\tilde \nu ^2 )  B_m (\rho ) ) ,  \cr &&
\Longrightarrow \ B_m (\rho ) = {\rho ^{2}  \over N '_m }  ( J_{\tilde \nu } (
\rho ) + b _m Y_{\tilde \nu  }  ( \rho ) )  ,\ 
     [\tilde \nu ^2 =\nu ^2(1 +{9 Q^2  \over 4}  ) ]. \eea

     In the WKB quantization rule, analogous to   Eq.(\ref{sect2.eq10}),
     the  phase integral  for $  V_{eff} = -{4\rho ^2 \over 9 \nu ^2}  + Q^2 + {4 \over 9}  $     over  the interval between the  origin and  the potential  turning point,  $(\tau _0' =0 , \ \tau _0 )$,  corresponding
to   the variable  $ z= {\rho \over \tilde \nu } $ interval    
 $ ( z_0 = 1,\    z'_0 = {\nu \l _1 / (2 \tilde \nu ) } ) $,  is given by
\bea &&\pi (n -{1\over 2} + {\d _{l,0} \over 4})  =
\int _{\tau '_0}  ^{\tau _0}  d \tau ( -   V_{eff}(\tau ) )  ^{1/2}  
=\tilde \nu \int _{z _0}  ^{z '_0} d  z
{({ z ^2 } -1 )^{1/2} \over z } = \tilde \nu [( z^{'2} _0 -1 )^{1/2} - \tan ^{-1} ( (z^{'2} _0  -1 )^{1/2} ) ]  . \label{app2.eq20}  \eea 
The   fit  to numerical  values of masses for singlet modes in the  warped deformed conifold   case gave~\cite{firouz06},   $\nu  \approx  2.44,\ a   \approx  2.48 .$ The other constant parameters in $Q$ can then  be  determined  by  fitting  the masses of string and  charged modes in Table~\ref{tabFT2}.
Since   our $Q $ is related to $\calq $ of~\cite{firouz06} by $ 2(1 + {9\over 4} Q^2 )^{1/2}  / a =\calq $, which  for  modes of  large charges becomes,  $Q \simeq a \calq /3 $,  our (hatted)  parameters  for $Q$   can be evaluated from      the (unhatted) ones   of~\cite{firouz06}  using
$ [\hat c, \ \hat d, \ \hat f, \ \hat g 
] = (a  /3 )  [c, \ d, \ f, \ g ] \approx  0.83 \times  
[ 3.4 ,\   1.4 ,\   3.4 ,\  0.65] . $

\end{appendix}

\end{document}